\documentclass[twocolumn,10pt,pra,aps,floatfix]{revtex4}

\usepackage{amsmath}
\usepackage{amsfonts}
\usepackage{amssymb}
\usepackage{graphicx}
\usepackage{hyperref}
\usepackage{color}

\newcommand{\ket}[1]{|{#1}\rangle}
\newcommand{\bra}[1]{\langle{#1}|}
\newcommand{\boldrho}{\mbox{\boldmath$\rho$}}

\begin{document}

\title{Cold polar molecules in 2D traps:\\Tailoring interactions with external fields for novel quantum phases}

\date{\today}

\author{A.~Micheli}\email{andrea.micheli@uibk.ac.at}
\author{G.~Pupillo}
\author{H.~P.~B\"uchler}
\author{P.~Zoller}

\affiliation{Institute for Theoretical Physics, University of Innsbruck, A-6020 Innsbruck, Austria,\\
  Institute for Quantum Optics and Quantum Information of the Austrian Academy of Sciences, A-6020, Innsbruck, Austria}

\begin{abstract}
  We discuss techniques to engineer effective long-range
  interactions between polar molecules using external static
  electric and microwave fields. We consider a setup where molecules
  are trapped in a two-dimensional pancake geometry by a
  far-off-resonance optical trap, which ensures the stability of the
  dipolar collisions. We detail how to modify the {\em shape} and
  the {\em strength} of the long-range part of interaction
  potentials, which can be utilized to
  realize interesting quantum phases in the context of cold molecular
  gases.
\end{abstract}

\maketitle

\section{Introduction}\label{sec:secIntro}

The realization of Bose Einstein condensates and quantum degenerate
Fermi gases with cold atoms has been one of the highlights of
experimental atomic physics during the last decade~\cite{Ultracold},
and in view of recent progress in preparing cold molecules we expect
a similarly spectacular development for molecular ensembles in the
coming
years~\cite{SpecialIssue1,Ex1,Bethlem00,Ex04,Crompvoets01,Rempe04,Hinds04,Ex2,Wang,Ex4,Ex5,Hudson31,EX99,DeMille02,Greiner03,Regal03,EX03,Micheli06}.
The outstanding features of the physics of cold atomic and molecular
gases are the microscopic knowledge of the many-body Hamiltonians,
as realized in the experiments, combined with the possibility to
control and tune system parameters via external fields. Examples are
the trapping of atoms and molecules with magnetic, electric and
optical traps, allowing for the formation of quantum gases in 1D, 2D
and 3D geometries, and the tuning of contact inter-particle
interactions by varying the scattering length via Feshbach
resonances~\cite{Fano61,Duine04}. This control is the key for the
experimental realization of fundamental quantum phases, as
illustrated by the superfluid-Mott insulator quantum phase
transition with bosonic atoms in an optical lattice~\cite{Bloch02},
and the BEC-BCS crossover in atomic Fermi
gases~\cite{Regal04a,Bartenstein04,Zwierlein05,Partridge05,Chin04}

As discussed in our recent work~\cite{Buechler07}, polar molecules
prepared in the electronic and vibrational ground state offer new
possibilities to control inter-particle interactions. In fact,
effective interactions with a {\em given
  potential shape} can be engineered under conditions of tight 2D
confinement, by applying static (DC) and microwave (AC) fields. The
engineered potentials can display both repulsive and/or attractive
character. This control of the interactions - in combination with
low-dimensional trapping - opens the way to realizing novel quantum
phases and quantum phase transitions. As an example
Ref.~\cite{Buechler07} discusses a quantum phase transition from a
superfluid to a self-assembled crystal for a gas of polar molecules
in the {\em strongly interacting limit}, where the stability of the
collision processes is guaranteed by the confinement in a 2D
geometry. It is the purpose of the present paper to present in some
detail the molecular aspects
behind this engineering of effective two-body interactions.\\

\begin{figure}[htbp]
  \begin{center}
    \includegraphics[width=\columnwidth]{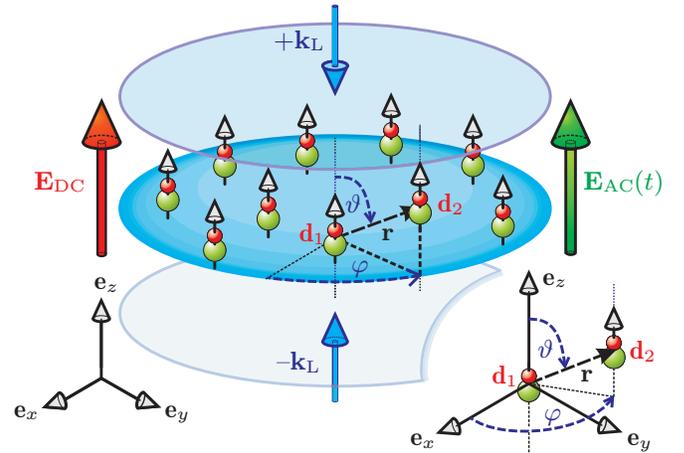}
  \end{center}
  \caption{\label{fig:fig1}(color online) System setup: Polar molecules are
    trapped in the ($x,y$)-plane by an optical lattice made of two
    counter-propagating laser beams with wavevectors $\pm{\bf
    k}_{\rm
      L}=\pm{k}_{\rm L}{\bf e}_z$ (arrows on the top/bottom). The dipoles ${\bf d}_j$ are
    aligned in the $z$-direction by a DC electric field ${\bf E}_{\rm
      DC}\equiv E_{\rm DC}{\bf e}_z$ (arrow on the left). An AC
      (microwave)
    field ${\bf E}_{\rm AC}$ is indicated (arrow on the right).
    Inset: Definition of polar ($\vartheta$) and azimuthal
      ($\varphi$) angles for the relative orientation of the
      inter-molecular collision axis ${\bf r}$ with respect to a
      space-fixed frame with axes $\{{\bf e}_x,{\bf e}_y,{\bf
      e}_z\}$.}
\end{figure}
The interaction potential between atoms, in particular Alkali atoms
in their electronic ground state, is dominated at large distances by
an {\em attractive} $C_6/r^6$ potential. In the many body
Hamiltonian for a dilute quantum gas this gives rise to an effective
two-body short range interaction in the form of a contact
interaction with a scattering length $a_{\rm s}$. Polar molecules
have strong {\em permanent} electric dipole moments in their
electronic-vibrational ground state manifold, and pairs of molecules
aligned by external DC or AC electric fields will interact via
(comparatively strong) dipole-dipole interactions with
characteristic long-range $1/r^3$ dependence
\cite{SpecialIssue1,Krems05,Krems06,Avdeenkov03,Ticknor05}. These
dipole-dipole interactions will be attractive or repulsive,
depending on the relative orientation of the dipoles.

The alignment of the dipoles corresponds to the {\em dressing} of
the lowest energy excitations of the internal molecular degrees of
freedom, which are related to rotations of the molecule. The
rotational dynamics can be manipulated using external electric DC
and AC (microwave) fields. This dressing of rotational states by
external fields together with the dipole-dipole interaction forms
the basis to shape the effective molecular interactions.

\begin{figure}[htbp]
  \begin{center}
    \includegraphics[width=\columnwidth]{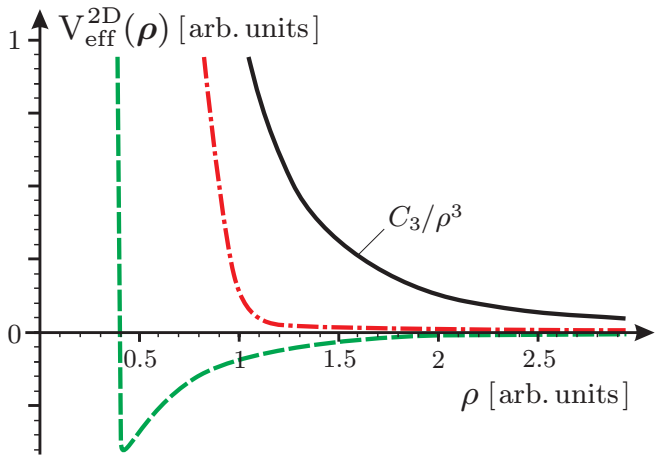}
  \end{center}
  \caption{\label{fig:fig3added}(color online) Qualitative sketch of effective 2D
  potentials
  $V_{\rm eff}^{\rm 2D}(\boldrho)$ for polar
  molecules confined in a 2D (pancake) geometry. Here,
  $\boldrho=r\sin\vartheta(\cos\varphi,\sin\varphi)$ is the 2D coordinate
  in the plane $z=0$ and $\rho=r\sin\vartheta$, (see inset of Fig.~\ref{fig:fig1}).
  Solid line: Repulsive dipolar potential $V_{\rm eff}^{\rm 2D}(\boldrho)=C_3/\rho^3$ induced by
  a DC electric field. Dash-dotted line: ``Step-like'' potential
  induced by a single AC (microwave) field and a weak DC field. Dashed
  line: Attractive potential induced by the
  combination of several AC (microwave) fields and a weak DC field.
  Here, the potentials $V_{\rm eff}^{\rm 2D}(\boldrho)$ and the separation $\rho$ are given in arbitrary
  units. For the ``step-like'' case (dash-dotted line) $\rho=1$ corresponds to the Condon point $r_{\rm C}$ of
  Sec.~\ref{Sec:CouplingDCAC}.}
\end{figure}

\begin{figure}[htbp]
  \begin{center}
    \includegraphics[width=\columnwidth]{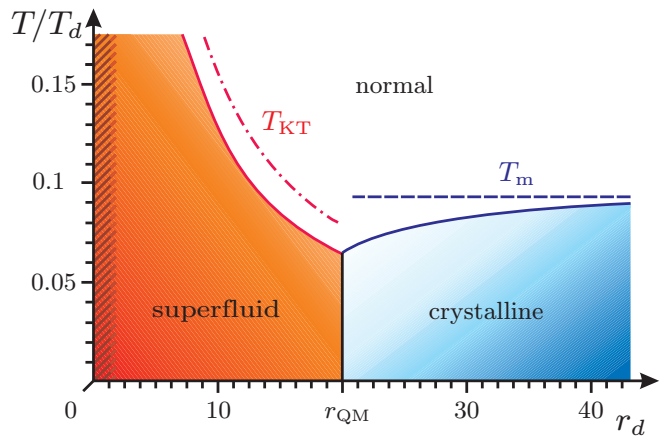}
  \end{center}
  \caption{\label{fig:fig0}(color online) Sketch of the phase diagram for a homogeneous 2D system of polar
  molecules interacting via the effective 2D
  repulsive potential $V_{\rm eff}^{\rm 2D}(\boldrho)=C_3/\rho^3$.
  $T$ is the temperature in units of $T_d\equiv C_3/k_{\rm B} a^3$, with
  $a$ the average inter-particle distance and $k_{\rm B}$ the Boltzmann
  constant. The symbol $r_d\equiv E_{\rm int}/E_{\rm kin}=C_3 m /\hbar^2
  a$ is the interaction ($E_{\rm int}=C_3/a^3$) to kinetic energy ($E_{\rm kin}=\hbar^2/m a^2$)
  ratio. A crystalline phase appears for large ratios $r_d > r_{\rm QM}$
  and small temperatures $T<T_{\rm m}$. The critical ratio $r_{\rm QM}\approx 18\pm4$
  for the quantum melting to a superfluid phase has been determined in
  Ref.~\cite{Buechler07}, while the classical
  melting temperature $T_{\rm m}$ (dashed line) to a normal gas phase has been calculated
  in Ref.~\cite{Kalia81}. The finite-temperature superfluid to normal fluid phase transition
  is of the Berezinskii-Kosterlitz-Thouless
  type~\cite{BKT72,BKT73}
  and it appears below the upper
  bound $T_{\rm KT}=\pi \hbar^2/2 k_{\rm B} m a^2$ (dashed-dotted line). The crossover to an unstable
  regime for small repulsion and finite confinement
  in the $z$-direction (see Fig.~\ref{fig:fig1})
  is indicated by a hatched region (see text, Sec.~III).}
\end{figure}

One example discussed in Ref.~\cite{Buechler07} deals with polar
molecules confined in a 2D (pancake) trap (see Fig.~\ref{fig:fig1}).
The molecular dipoles are aligned perpendicular to the plane by a DC
field. Thus, the effective 2D interactions are {\em repulsive} and
{\em long range} $V_{\rm eff}^{\rm 2D}(\boldrho)=C_3/\rho^3$, with
$\boldrho=r\sin\vartheta(\cos\varphi,\sin\varphi)$ the 2D coordinate
in the plane $z=0$ and $\rho=r\sin\vartheta$ (see
Fig.~\ref{fig:fig3added}, solid line). The interaction strength
$C_3$ is proportional to the square of the induced dipole moment for
the dressed rotational ground state. Depending on the interaction
strength, we find the appearance of a crystalline phase, and an
associated quantum melting to a superfluid phase as a function of
the square of the induced dipole-moment. The corresponding phase
diagram is discussed in Ref.~\cite{Buechler07}, and it is reproduced
in Fig.~\ref{fig:fig0} (see also Ref.~\cite{Astrakharchik07}).

In the present work we present in detail the microscopic molecular
theory underlying this engineering of the interaction potential for
trapped polar molecules offered by DC and AC microwave fields. We
focus both on potentials which are repulsive $1/r^3$ (DC field) and
on potentials which have a marked ``step-like'' character, that is,
the character of the repulsive potentials varies considerably in a
small region of space (an AC plus a DC field). Three example cases
of effective 2D potentials $V_{\rm eff}^{\rm 2D}(\boldrho)$ are
shown in Fig.~\ref{fig:fig3added}. The use of multi-chromatic AC
fields can lead to the realization of interesting potentials (for
example the attractive potential of Fig.~\ref{fig:fig3added}),
however, in this work we focus on monochromatic AC fields only.

In all cases, the derivation of the effective 2D interactions
proceeds in two steps: First, we derive a set of Born-Oppenheimer
(BO) potentials by diagonalizing the Hamiltonian for the relative
motion of two particles for fixed molecular positions. Within an
adiabatic approximation, the corresponding eigenvalues play the role
of an effective 3D interaction potential. Second, we obtain an
effective 2D dynamics by integrating-out the fast transverse motion
of the molecules along the direction of the tight parabolic
confinement.\\

The paper is organized as follows: In Sec.~\ref{sec:secMolecularHam}
we discuss the Hamiltonian for a single rotating polar molecule
dressed by DC and AC (microwave) fields under conditions of strong
optical confinement. The collisions of two polar molecules are
considered in Sec.~\ref{Sec:TwoMolecules}. After reviewing the
molecular collisions in the absence of external fields, in
Sec.~\ref{sec:secDC} we consider the case of interactions in the
presence of a DC electric field. In particular, the stabilizing
effects of a parabolic potential confining the particles to a 2D
plane are analyzed in Sects.~\ref{sec:secDCTrap} and
\ref{sec:secDCStab}, while the effective 2D interaction potential
$V_{\rm eff}^{\rm 2D}(\boldrho)=C_3/\rho^3$ is derived in
Sec.~\ref{sec:secEff2D}. The interactions in the presence of an AC
field are studied in Sec.~\ref{sec:secAC}. In the absence of
external confinement, this case is analogous to the 3D optical
shielding developed in the context of ultracold collisions of
neutral atoms~\cite{Zilio96,Napolitano97,Weiner99}. As in the
latter, we find a strong dependence of the 3D shielding potential on
the polarization of the AC field. The presence of "holes" in the 3D
shielding potential for certain polarizations and of several
degeneracies in the two-particle spectrum for all polarizations
render the pure AC-field case less appealing for realizing stable
collisional setups in 2D. In fact, both the former and the latter
processes open loss channels for the ground-state interaction. In
Sec.~\ref{Sec:CouplingDCAC} we analyze the interactions in combined
DC and AC fields, and we show that the DC field helps to greatly
suppress the presence of possible loss channels at large distances,
while an additional harmonic confinement along $z$ avoids populating
the regions of space where ``holes'' analogous to those of
Sec.~\ref{sec:secAC} occur. Thus, by introducing a {\em tight
optical confinement} in the $z$-direction, in this case {\em it is
possible to realize stable two-dimensional collisional setups.}
Two-dimensional inter-particle interactions can be designed, whose
character varies markedly between long and short distances, allowing
for much greater flexibility in tuning by external fields than the
pure DC case of Sec.~\ref{sec:secDC}.

\section{Molecular Hamiltonians}\label{sec:secMolecularHam}

The purpose of this section, which forms the basis of discussion in
the following sections, is to review the single-molecule rotational
spectroscopy. In particular we are interested in the rotational
excitations of cold $\nu{}^{2S+1}\Lambda(v)$ spin-less ($S=0$) polar
molecules in their electronic ($\nu=0$) and vibrational ($v=0$)
ground-state, with zero-projection ($\Lambda=0$) of the total
angular momentum on the internuclear axis~\cite{Herzberg50,Brown03}.
The spectroscopic notation for the electronic-vibrational
ground-state of these molecules is $X{}^1\Sigma(0)$. Moreover, we
are interested in manipulating the rotational states of these
molecules using DC and AC electric fields and in confining the
particles using a (optical) far-off-resonance trap (FORT). The
application of these external fields will serve as a key element to
engineer effective interaction potentials between the molecules.

Our goal in this section is to derive a low energy effective
Hamiltonian for the external motion and internal rotational
excitations of a single molecule in its electronic-vibrational
ground state of the form
\[ H(t) = \frac{{\bf p}^2}{2m} + H_{\rm rot} + H_{\rm DC} + H_{\rm
  AC}(t) + H_{\rm opt}({\bf r}).\] In the last equation, ${\bf
  p}^2/2m$ is the kinetic energy for the center-of-mass motion of
  a molecule of mass $m$, while
$H_{\rm rot}$ accounts for the rotational degrees of freedom. The
terms $H_{\rm DC}$, $H_{\rm AC}(t)$ and $H_{\rm opt}({\bf r})$ refer
to the interaction with electric DC and AC (microwave) fields and to
the optical trapping of the molecule in the ground
electronic-vibrational manifold, respectively.

\subsection{Rotational excitations of ${}^1\Sigma$ molecules}

We consider spin-less polar molecules with $\Sigma$ electronic
ground-states in their electronic-vibrational ground state,
$X{}^1\Sigma(v=0)$.  The low-energy internal excitations correspond
to the rotation of the internuclear axis of the molecules with total
internal angular momentum ${\bf
J}$~\cite{Herzberg50,Brown03,Judd75}. The corresponding Hamiltonian
$H_{\rm rot}$ is the one of a rigid spherical
rotor~\cite{Herzberg50}
\begin{eqnarray}
  H_{\rm rot}=B{\bf J}^2.\label{eq:rotor}
\end{eqnarray}
Here $B$ is the rotational constant for the electronic-vibrational
ground state, which is of the order of $B\sim h~10~{\rm
GHz}$~\cite{NISTDataWeb}. We denote the energy eigenstates of
Eq.~\eqref{eq:rotor} by $\ket{J,M}$, where $J$ is then quantum
number associated with the total internal angular momentum and $M$
is the quantum number associated with its projection onto a {\em
space-fixed} quantization axis. The excitation spectrum is $E_{J}=B
J(J+1)$, which is anharmonic. Each $J$-level is $(2J+1)$-fold
degenerate.\\

A polar molecule has an electric dipole moment, ${\bf d}$, which
couples its internal rotational levels. This dipole moment gives
rise to the dipole-dipole interaction between two molecules. For
$\Sigma$-molecules the dipole operator is along the internuclear
axis ${\bf e}_{ab}$, i.e. ${\bf d}=d{\bf e}_{ab}$. Here, $d$ is the
``permanent'' dipole moment of a molecule in its
electronic-vibrational ground-state.

The spherical components of the dipole operator on a {\em
space-fixed} spherical basis $\{{\bf e}_{-1},{\bf e}_0,{\bf e}_1\}$,
with ${\bf
  e}_{q=0}\equiv{\bf e}_z$ and ${\bf e}_{\pm 1}=\mp({\bf e}_x\pm i{\bf
  e}_y)/\sqrt{2}$, are given by $d_q={\bf e}_q\cdot{\bf d}=d
  C_q^{(1)}(\theta,\phi)$ ,where $C_q^{(k)}(\theta,\phi)$ are the
  unnormalized spherical harmonics and $\theta$ ($\phi$)
  is the polar (azimuthal) angle for the orientation of the molecule in the
  space-fixed frame~\cite{Herzberg50,Brown03,Judd75}, respectively. We note that for
a spherically-symmetric system, e.g.~in the absence of external
fields, the eigenstates of the rotor have no net dipole-moment,
$\bra{J,M}{\bf d}\ket{J,M}=0$. On the other hand, the component
$d_q$ couples the rotational states $\ket{J,M}$ and
$\ket{J\pm1,M+q}$ according to
\begin{eqnarray}
  \bra{J\pm1,M+q}d_q\ket{J,M} =
  d(J,M;1,q|J\pm1,M+q)\times\nonumber\\
   \times (J,0;1,0|J\pm1,0)\sqrt{\frac{2J+1}{2(J\pm1)+1}},\nonumber
\end{eqnarray}
where $(J_1,M_1;J_2,M_2|J,M)$ are the Clebsch-Gordan-coefficients.\\

In the following we are interested in the interaction of the
molecules with an external DC electric field along ${\bf e}_z$,
${\bf E}_{\rm DC}=E_{\rm DC}{\bf e}_0$, and with AC microwave fields
with either linear polarization ($q=0$) or circular polarization
($q=\pm1$) relative to ${\bf e}_z$, ${\bf E}_{\rm
  AC}(t)=E_{\rm AC}e^{-i\omega t}{\bf e}_q+{\rm c.c.}$. These
fields couple to a molecule via the electric dipole interaction,
\begin{subequations}
  \begin{eqnarray}
    H_{\rm DC} &=& -{\bf
      d}\cdot{\bf E}_{\rm DC} = - d_0 E_{\rm DC},\label{eq:DCelectric}\\
    H_{\rm AC}(t) &=& -{\bf
      d}\cdot{\bf E}_{\rm AC}(t) = - d_q E_{\rm AC}e^{-i\omega
      t}+{\rm h.c.}\label{eq:acelectric},
  \end{eqnarray}
\end{subequations}
which try to align the molecule along the field, while competing
with its rotation, as $[{\bf J}^2,d_q]\neq0$.

\subsection{Coupling of rotational states by DC and AC electric
  fields.}

\begin{figure}[htbp]
  \begin{center}
    \includegraphics[width=\columnwidth]{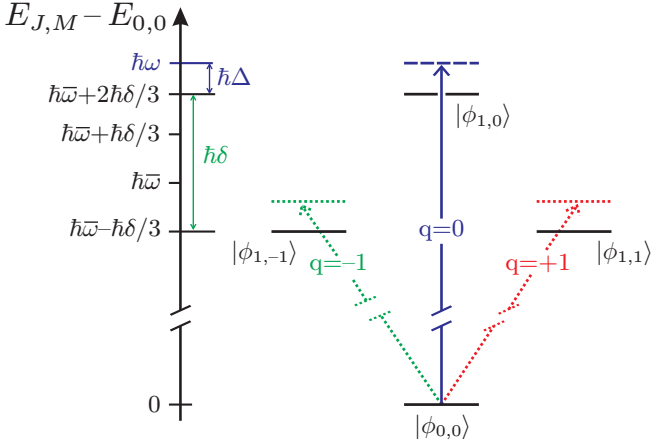}
  \end{center}
  \caption{\label{fig:fig2}(color online) Solid lines: Energies $E_{J,M}$ (left)
    and states $\ket{\phi_{J,M}}$ (right) of Eq.~\eqref{eq:EigStatic} with $J=0,1$,
    for a molecule in a weak DC electric field ${\bf E}_{\rm DC}=E_{\rm DC}{\bf e}_0$
     with $\beta\equiv d E_{\rm DC}/B \ll
      1$. The DC-field-induced splitting $\hbar \delta$
      and the average energy separation
      $\hbar \bar{\omega}$ are $\hbar \delta=3B\beta^2/20$
      and $\hbar \bar{\omega}=2B+B\beta^2/6$, respectively. Dashed
      and dotted lines: Energy levels for a molecule in
      combined DC and AC fields (The AC-Stark shifts of the dressed states are not shown).
      Dashed line: The AC field is monochromatic, with
      frequency $\omega$, linear polarization $q=0$, and detuning $\Delta=\omega-(\bar{\omega}
      +2 \delta/3)>0$. Dotted lines: Schematics of energy levels
      for an AC-field with polarization $q=\pm1$ and frequency $\omega'\neq\omega$.}
\end{figure}

\subsubsection{Coupling to a DC electric field}\label{Sec:CouplingDC}

The effects of a DC electric field, ${\bf E}_{\rm DC}$, on a single
polar molecule are: (a) To split the $(2J+1)$-fold degeneracy in the
rotor spectrum, and (b) to align the molecule along the direction of
the field, which amounts to inducing a finite dipole moment in each
rotational state.\\

\begin{table}[b]
  \begin{tabular}[t]{|c|c|c|}
    \hline
    $g_0$&$\bra{\phi_{0,0}}d_0\ket{\phi_{0,0}}$&$
    (d\beta/3)(1-7\beta^2/360)$ \\
    $g_1$&$\bra{\phi_{1,\pm1}}d_0\ket{\phi_{1,\pm1}}$&$
    (d\beta/10)/(1-3\beta^2/5600)$ \\
    $g_2$&$\bra{\phi_{1,0}}d_0\ket{\phi_{1,0}}$&$
    -(d\beta/5)(1-19\beta^2/350)$ \\
    \hline $f_0$& $\bra{\phi_{1,0}}d_0\ket{\phi_{0,0}}$&$
    (d/\sqrt{3})(1-43\beta^2/360)$ \\
    $f_1$& $\bra{\phi_{1,\pm1}}d_{\pm1}\ket{\phi_{0,0}}$&$
    (d/\sqrt{3})(1-49\beta^2/1440)$ \\
    $f_2$&$\bra{\phi_{1,0}}d_{\mp1}\ket{\phi_{1,\pm1}}$&$
    (3d\beta/20)(1+11\beta^2/1400)$ \\
    \hline
  \end{tabular}\caption{\label{tab:tab1} Permanent ($g_n$) and transition
  ($f_n$) dipole moments of the $4$ states belonging to the
rotational ($J=0,1$)-manifolds, in the presence of a weak polarizing
DC field, ${\bf E}_{\rm DC}=E_{\rm DC}{\bf e}_0$. Here, $\beta=d
E_{\rm DC}/B\ll1$ is the ratio of the electrostatic energy and
rotational constant, and the dipole moments are given up to order
third order in $\beta$.}
\end{table}

We choose the direction of the DC-field as the quantization axis,
${\bf E}_{\rm DC}\equiv E_{\rm DC}{\bf e}_0$. Then, the internal
Hamiltonian is that of a rigid spherical pendulum
\cite{Herzberg50,Townes55}
\begin{eqnarray}
  H = H_{\rm rot}+H_{\rm DC} = B{\bf J}^2 - d_0 E_{\rm
    DC}, \label{eq:pendulum}
\end{eqnarray}
which conserves the projection of the angular momentum $J$ on the
quantization axis, i.e. $M$ is a good quantum number. The energy
eigenvalues and eigenstates of Eq.~\eqref{eq:pendulum} are labeled
as
$E_{J,M}$ and $\ket{\phi_{J,M}}$, respectively. \\

We are interested in weak fields, $E_{\rm DC}\ll B/d$, where the
effects of the electric field are a {\em quadratic} DC Stark shift
of the rotational energy levels and a finite induced dipole moment
along the axis of the field in each rotational state. For a typical
rotational constant, $B\sim h~10~{\rm GHz}$, and a dipole-moment
$d\sim 9~{\rm Debye}$ this corresponds to considering DC fields
(much) weaker than $B/d\sim 2~{\rm kV/cm}$. To lowest order in
$\beta\equiv dE_{\rm DC}/B$ the energy eigenvalues and eigenstates
are \cite{Herzberg50,Townes55}
\begin{subequations}\label{eq:EigStatic}
\begin{eqnarray}
  E_{J,M}/B &=& J(J+1) +
  \frac{\beta^2}{2}\frac{1-3M^2/J(J+1)}{(2J-1)(2J+3)},\\
  \ket{\phi_{J,M}} &=& \ket{J,M} -
  \frac{\beta}{2}\frac{\sqrt{J^2-M^2}}{\sqrt{J^3(2J+1)}}
\ket{J-1,M}+\nonumber\\
&&+
\frac{\beta}{2}\frac{\sqrt{(J+1)^2-M^2}}{\sqrt{(J+1)^3(2J+1)}}\ket{J+1,M}.
\end{eqnarray}
\end{subequations}
Thus, the ground state energy is shifted downwards by
$E_{0,0}=-B\beta^2/6$, while the energies of the lowest excited
states are split by
\begin{eqnarray}
\hbar \delta\equiv E_{1,0}-E_{1,\pm
1}=3B\beta^2/20,\label{eq:eqdelta}
\end{eqnarray}
see solid lines in Fig.~\ref{fig:fig2}.
  The average energy separation of the $(J=0)$ and $(J=1)$-manifolds is
\begin{eqnarray}
  \hbar \overline{\omega} =\sum_{M=-1}^1 (E_{1,M}-E_{0,0})/3= 2B +
  B\beta^2/6.\label{eq:eqomega}
\end{eqnarray}

The induced dipole moments to lowest order in $\beta$ are
\begin{eqnarray}
  \bra{\phi_{J,M}}{\bf d}\ket{\phi_{J,M}} &=&  d\beta
  \frac{3M^2/J(J+1)-1}{(2J-1)(2J+3)}{\bf e}_0.\nonumber
\end{eqnarray}
This equation shows that the ground state acquires a finite dipole
moment $g_0\equiv
  \bra{\phi_{0,0}}d_0\ket{\phi_{0,0}}=d\beta/3$ along the field axis,
while the lowest excited states acquire a dipole moment
$\bra{\phi_{1,M}}d_0\ket{\phi_{1,M}}=d(3M^2-1)\beta/10$. For later
convenience, perturbative values in the small parameter $\beta$ of
the transition and induced dipole moments  are reported in
Table~\ref{tab:tab1} for the four single-particle states
$\ket{\phi_{J,M}}$ with $|M|\leq J\leq 1$. The transition and
induced dipole moments are labeled as $f_n$ and $g_n$, respectively.

\subsubsection{Coupling to an AC electric field}

Similar to the case of a DC electric field, the basic effect of an
AC electric field ${\bf E}_{\rm AC}$ on a single molecule is to
polarize it by dressing its energy levels. The characteristic
time-dependence of the AC field allows for: (a) Addressing
individual rotational transitions by applying one or several
non-interfering microwave fields (multi-mode field); (b) Realizing
dressing fields  that can be not only {\em linearly}, but also {\em
circularly} polarized. In this work, we consider the case of a
single AC microwave field with polarization $q$ and frequency
$\omega$, ${\bf E}_{\rm AC}({\bf r},t)=E_{\rm AC}({\bf
r})e^{-i\omega t}{\bf e}_q+{\rm c.c.}$, and derive the dressed
energy-levels for a molecule in the field. For the sake of
generality - and for later convenience - we consider the case where
the AC field is superimposed to a weak DC field, which provides for
a splitting of the first excited $(J=1)-$manifold, as shown above.\\

Given a polarization $q$, the frequency $\omega$ is chosen close to
the transition from the ground-state to one state of the
$(J=1)$-manifold, $\ket{\phi_{0,0}}\leftrightarrow\ket{\phi_{1,q}}$,
i.e. $\omega\sim \overline{\omega}+\delta(2/3-q^2)$, where the
states $\ket{\phi_{J,M}}$ are those of Eq.~\eqref{eq:EigStatic}. The
corresponding wavelength is of the order of centimeters, which
largely exceeds the size of our system and therefore one can neglect
the position dependence of the microwave field, i.e. recoil effects,
$E_{\rm AC}({\bf r})\approx E_{\rm AC}$. The electric dipole
interaction of Eq.~\eqref{eq:acelectric} reads
\begin{eqnarray}
  H_{\rm AC}(t) &=& - d_q E_{\rm AC}e^{-i\omega t} + {\rm
  h.c.}.\label{eq:acsingle}
\end{eqnarray}
The Rabi frequency $\Omega$ and the detuning $\Delta$ are
$\Omega\equiv E_{\rm
AC}\bra{\phi_{1,q}}d_q\ket{\phi_{0,0}}/\hbar=E_{\rm
AC}f_{|q|}/\hbar$ and
$\Delta\equiv\omega-(E_{1,q}-E_{0,0})/\hbar=\omega-[\overline{\omega}+\delta(2/3-q^2)]$,
respectively (see Fig.~\ref{fig:fig2}).\\

In Sec.~\ref{Sec:TwoMolecules} we consider a specific setup where
the AC field has linear polarization, $q=0$. Here we illustrate how
to obtain the dressed energy levels of a molecule in this field by
diagonalizing the Hamiltonian $H=H_{\rm rot}+ H_{\rm DC} + H_{\rm
AC}(t)$ in a Floquet picture. First, we expand the Hamiltonian on
the basis $\ket{\phi_{J,M}}$, which diagonalizes the
time-independent part of $H$ as $H_{\rm rot}+H_{\rm
DC}=\sum_{J,M}\ket{\phi_{J,M}}E_{J,M}\bra{\phi_{J,M}}$. Then, we
consider the effect of the AC field driving the
$(\ket{\phi_{0,0}}\leftrightarrow\ket{\phi_{1,0}})$-transition with
Rabi-frequency $\Omega\equiv f_0 E_{\rm AC}/\hbar \approx d E_{\rm
AC}/\sqrt{3}\hbar$ and detuning
$\Delta=\omega-(E_{1,0}-E_{0,0})/\hbar= 2B(1+2\beta^2/15)/\hbar$.

A transformation to the Floquet picture is obtained by expanding the
time-dependent wave-function in a Fourier series in the AC frequency
$\omega$. After applying a rotating wave approximation, i.e. keeping
only the energy conserving terms, we obtain the {\em
time-independent} Hamiltonian $\tilde H$, which describes the
coupled two-level system in the basis
  $\{\ket{\phi_{0,0}},\ket{\phi_{1,0}}\}$ as
\begin{eqnarray}
  \tilde{H} &=& -\hbar\left[\begin{array}{cc}0&\Omega\\
      \Omega&\Delta\end{array}\right] +E_{0,0}.\nonumber
\end{eqnarray}
The corresponding dressed energy eigenvalues of ${\tilde H}$ for the
ground state and excited state (minus one photon energy $\hbar
\omega$) are given by
  \begin{eqnarray}
    \tilde{E}_{0,0}-E_{0,0} &=&-\frac{\hbar\Delta}{2}+\frac{\hbar\Delta}{2}\sqrt{1+\frac{4\Omega^2}{\Delta^2}}
    \approx +\frac{\hbar\Omega^2}{\Delta},\nonumber\\
    \tilde{E}_{1,0}-E_{0,0} &=&-\frac{\hbar\Delta}{2}-\frac{\hbar\Delta}{2}\sqrt{1+\frac{4\Omega^2}{\Delta^2}}
    \approx -\hbar\Delta-\frac{\hbar\Omega^2}{\Delta},\nonumber
  \end{eqnarray}
respectively. We note that the AC field induces an AC-Stark shift
$\approx\pm\hbar\Omega^2/\Delta$, on the ground and the excited
state, respectively. Thus, the shift depends on the detuning
$\Delta$, and in particular on its sign, and on the Rabi-frequency
$\Omega$.

\subsection{Optical trap}

An essential ingredient of our setup is the tight confinement of the
molecules in a 2D-plane. This is realized for example by a
far-off-resonant optical trap (see Fig.~\ref{fig:fig1}).  The latter
drive far-off-resonance transitions from $X\Sigma(0)$ to the
electronically excited states, $\nu\Lambda(v)$. The goal of this
section is to obtain the resulting trapping potentials for the
lowest rotational excitations, $J=0,1$.\\

A detailed discussion of the complex nature of molecular electronic
excitations~\cite{Herzberg50,Brown03} is beyond the scope of the
present discussion. For a detailed treatment of an example case we
refer to Ref.~\cite{Kotochigova06}. We consider here a simple model,
where the fine and hyperfine interactions are neglected. Then the
basic molecular structure is obtained as follows: In the adiabatic
approximation one diagonalizes the Hamiltonian for the electrons and
the two nuclei as a function of the internuclear separation
$r_{ab}$, thus obtaining a set of Born-Oppenheimer (BO) potentials,
 $E_{\nu\Lambda}(r_{ab})$. Here $\nu$ is the
main electronic quantum number, while $\Lambda$ denotes the quantum
number associated with the operator for the total angular momentum
component of the molecules along the internuclear axis, ${\bf
e}_{ab}\cdot{\bf J}$. The latter gives rise to a large splitting of
the electronic manifolds $\sim A_\nu|\Lambda|^2$, where
$A_\nu=\hbar^2/I_e$ is the inverse of the (small) moment of inertia
of the electrons~\cite{Herzberg50}. Then, the vibration of the
nuclei in the BO-potentials yields a series of bound-states
$v=0,1,2,\ldots$ with energy $E_{\nu\Lambda(v)}$.

Deep optical traps are obtained as follows: The laser, ${\bf E}_{\rm opt}({\bf r},t)={\bf
  E}_{\rm opt}({\bf r})e^{-i\omega_{\rm L}t}+{\rm c.c.}$,
drives the electronic transitions to the lowest excited states,
labeled $A$ and $B$, with frequency $\omega_{\rm L}$ tuned near the
minima of the BO-potentials.  Since spontaneous emission in the
excited states is typically a few MHz, deep traps on the order of
one MHz with a negligible inelastic scattering rate ($\sim$ a few
Hz) require detunings on the order of hundreds of GHz from the
vibrational resonances.  Since these detuning are much larger than
$B$, one can neglect the rotational structure in the electronic
ground and excited states in deriving the optical potential. The
effective interaction of the molecules with the off-resonant
laser-field is thus described by
\begin{eqnarray}
H_{\rm opt}({\bf r}) &=& {\bf E}_{\rm opt}({\bf
    r})^*\cdot\hat\alpha(\omega_{\rm L})\cdot{\bf E}_{\rm opt}({\bf
    r}),\label{eq:effectiveHopt}
\end{eqnarray}
with the dynamic polarizability-tensor
\begin{align}
  \hat\alpha(\omega_{\rm L})=\alpha_\parallel(\omega_{\rm L}) {\bf e}_0'\otimes{\bf e}_0' +
  \alpha_\perp(\omega_{\rm L}) \sum_{\Lambda=\pm1}(-1)^\Lambda{\bf
    e}_{\Lambda}'\otimes{\bf e}_{-\Lambda}'\nonumber\\
=\alpha_\perp(\omega_{\rm L})\sum_{q=-1}^{+1}(-1)^q{\bf e}_q\otimes{\bf e}_{-q}+
\left[\alpha_\parallel(\omega_{\rm L})-\alpha_\perp(\omega_{\rm L})\right]\times\nonumber\\
\times
\sum_{p,q}(-1)^{p-q}C_{-p}^{(1)}(\theta,\phi)C_{q}^{(1)}(\theta,\phi){\bf
    e}_{p}\otimes{\bf e}_{-q}.\label{eq:polarizabilitybodytensor}
\end{align}
\noindent Here $\{{\bf e}_{-1}',{\bf e}_{0}',{\bf e}_{+1}'\}$
denotes a body-fixed spherical basis with ${\bf e}_0'\equiv{\bf
e}_{ab}=\sum_q(-1)^qC^{(1)}_{-q}(\theta,\phi){\bf e}_{q}$ being the
internuclear axis, and $\alpha_\parallel(\omega_{\rm L})$
[$\alpha_\perp(\omega_{\rm L})$] is the dynamic polarizability at
frequency $\omega_{\rm L}$ in the direction parallel (perpendicular)
to the internuclear axis~\cite{Herzberg50,Friedrich95}.  The
parallel and perpendicular components are given by the
$\Sigma-\Sigma$ ($\Delta\Lambda=0$) and $\Sigma-\Pi$
($\Delta\Lambda=\pm1$) transitions, respectively, and read
\begin{subequations}
  \begin{eqnarray}
    \alpha_\parallel(\omega_{\rm L}) &=&
    \sum_{\pm}\sum_{\nu,v}\frac{|d_{\nu\Sigma(v)-X\Sigma(0)}|^2}{E_{X\Sigma(0)}-E_{\nu\Sigma(v)}\pm\hbar\omega_{\rm L}},\\
    \alpha_\perp(\omega_{\rm L}) &=&
    \sum_{\pm}\sum_{\nu,v}\frac{|d_{\nu\Pi(v)-X\Sigma(0)}|^2}{E_{X\Sigma(0)}-E_{\nu\Pi(v)}\pm\hbar\omega_{\rm L}}.
  \end{eqnarray}\label{eq:polarizabilitycomponents}
\end{subequations}
\noindent Here $d_{\nu\Lambda(v)-X\Sigma(0)}$ denotes the transition
dipole moment from the $X\Sigma(0)$ to $\nu\Lambda(v)$, and the sum
over $\pm$ accounts for the near-resonant and typically far
off-resonant terms. From Eq.~\eqref{eq:polarizabilitycomponents} we
see that the anisotropy in the dynamic polarizabilities,
$\alpha_\perp(\omega_{\rm L})-\alpha_\parallel(\omega_{\rm L})$, is
due both to the different dipole-moments and to the large splitting
of the excited $\nu\Sigma(v)$ and $\nu\Pi(v)$ states.

In our setup we consider a pair of circularly polarized
counter-propagating laser beams, ${\bf E}_{\rm opt}({\bf r})=E_{\rm
  opt}\cos(k_{\rm L}z){\bf e}_+$, with wave-vectors $\pm {\bf k}_{\rm L}=\pm \omega_{\rm L}{\bf
  e}_0/c$ along $z$, trapping the molecules in the $x-y$ plane (see
Fig.~\ref{fig:fig1}). From Eq.~\eqref{eq:effectiveHopt} we obtain
the following Hamiltonian for the optical
trapping~\cite{Friedrich95}
\begin{align}\label{eq:TensorPolarizability}
 H_{\rm opt}({\bf r}) =\alpha_0(\omega_{\rm L})|E_{\rm opt}|^2\cos^2(k_{\rm L} z)C_0^{(0)}(\theta,\phi)+\nonumber\\
+\alpha_2(\omega_{\rm L})|E_{\rm opt}|^2\cos^2(k_{\rm L}
z)C_0^{(2)}(\theta,\phi),
\end{align}
where $\alpha_0(\omega_{\rm L})\equiv[\alpha_\parallel(\omega_{\rm
L})+2\alpha_\perp(\omega_{\rm L})]/3$ and $\alpha_2(\omega_{\rm
L})\equiv[\alpha_\parallel(\omega_{\rm L})-\alpha_\perp(\omega_{\rm
L})]/3$. The first term in Eq.~\eqref{eq:TensorPolarizability},
proportional to $C_0^{(0)}(\theta,\phi)=1$, gives an overall shift,
which is common to all the rotational states. The second term is
responsible for tensor shifts, which split the excited rotational
states according to $|M|$, as
\begin{eqnarray}
\bra{J,M}C^{(2)}_0(\theta,\phi)\ket{J,M}=\frac{J(J+1)-3M^2}{(2J-1)(2J+3)}.\label{eq:Tensor}
\end{eqnarray}
Typical depths of optical lattices are of the order of $\lesssim
h~1~{\rm MHz}$, and thus much smaller $B$. Therefore we may neglect
the far-off resonant Raman coupling between different $J$ manifolds,
i.e. $J\leftrightarrow J\pm2$.

We consider {\em tight} optical traps, such that the molecule in the
ground state are strongly confined at one potential minimum of
$\bra{\phi_{0,0}}H_{\rm opt}({\bf
r})\ket{\phi_{0,0}}=\alpha_0(\omega_L)|E_{\rm
  opt}|^2\cos^2(k_L z)$. For a light field which is
(far) red detuned from the electronic excited states, i.e.
$\hbar\omega_L \ll E_{\nu\Lambda(v)}-E_{X\Sigma(0)}$, the dynamic
polarizabilities $\alpha_\parallel(\omega_{\rm L})$ and
$\alpha_\perp(\omega_{\rm L})$ are negative and the trapping
potential for the ground-state is attractive, since
$\alpha_0(\omega_{\rm L})<0$. We assume the molecule to be strongly
confined near the field anti-node $z=0$. Then the optical trapping
is essentially given by a tight harmonic trap
\begin{eqnarray}\label{eq:eq14}
 H_{\rm opt}({\bf r})&\approx&|\alpha_0(\omega_{\rm
L})||E_{\rm opt}|^2 (-1+k^2 z^2) + \nonumber\\
&&+\alpha_2(\omega_{\rm L})|E_{\rm opt}|^2 (1-k_{\rm
L}^2z^2)C_0^{(2)}(\vartheta,\varphi).
\end{eqnarray}
From the last expression we see that the tensor-shifts induce a
position-dependent splitting for the excited rotational manifolds,
which at $z=0$ is analogous to that induced by a DC field, but it
has a strong modulation in space. The tensor-shifts are thus seen as
position and state-dependent potentials, and the last term in
$H_{\rm opt}({\bf r})$ is (in principle) unwanted for our purposes,
since it gives rise to different trapping frequencies $\omega_\perp$
for the ground and excited states.

However, we note that by applying a second laser, ${\bf E}_{\rm
opt}'({\bf r},t)$, of frequency $\omega_{\rm L}'$ with wavevector
${\bf k}_{\rm L}'$ and polarization ${\bf e}_{\rm L}'$, one can
eliminate the state-dependent potentials - up to a position
independent splitting of the excited states. Given the large number
and variety of available excited electronic-vibrational states
several choices are possible. One choice is, e.g., to apply an
additional laser with the same polarization ${\bf e}_{\rm L}'$ as
the first laser, i.e. ${\bf e}_{\rm L}'={\bf e}_+$, but having a
node at $z=0$ and being blue detuned from the electronic
transitions, i.e.~$E_{\rm opt}'({\bf r},t)=E_{\rm opt}'\sin(k_{\rm
L}'z)e^{-i\omega_{\rm L}'t}{\bf e}_{+}+{\rm c.c.}$ with $\omega_{\rm
L}'\gg (E_{\nu\Lambda(0)}-E_{X\Sigma(0)})/\hbar$ for
$\nu\Lambda=A\Pi,B\Sigma$. This induces an additional state
dependent optical trapping potential given by $H_{\rm opt}'({\bf r})
= |E_{\rm opt}'|\sin^2(k_{\rm L}'z)[\alpha_0(\omega_{\rm
L}')+\alpha_2(\omega_{\rm L}')C_{0}^{(2)}(\theta,\phi)]$. Tuning the
laser-frequency $\omega_{\rm L}'$ with respect to the vibrational
resonances one can force both $\alpha_\perp(\omega_{\rm L}')$ and
$\alpha_\parallel(\omega_{\rm L}')$ to be positive, see
Eq.~\eqref{eq:polarizabilitycomponents}. The additional trapping
potentials are zero at the node $z=0$, in particular for the
ground-state the trapping potential is repulsive (thus enhancing the
trapping given by the first laser), while the excited-state
position-dependent trapping $\propto z^2 C_0^{(2)}(\theta,\phi)$ of
Eq.~\eqref{eq:eq14} can be compensated for by tuning the strength of
the second laser, $E_{\rm opt}'$.

The parabolic trapping potential for our setup is then given by
\begin{eqnarray}
H_{\rm opt}({\bf r}) = \frac{1}{2}m\omega_\perp^2 z^2 - V_0 + V_2
C_{0}^{(2)}(\theta,\phi),
\end{eqnarray}
where the first term is a state-independent harmonic trapping along
${\bf e}_z$ at frequency $\omega_\perp=[2|\alpha_0(\omega_{\rm
L})||E_{\rm opt}k_{\rm L}|^2/m+2\alpha_0(\omega_{\rm L}')|E_{\rm
opt}'k_{\rm L}'|^2/m]^{1/2}$, the second terms gives an overall
Stark-shift, $V_0=|\alpha_0(\omega_{\rm L})||E_{\rm opt}|^2$, and
the last term is a splitting of the excited rotational states,
$J>0$, which is independent of the
position $z$, $V_2=\alpha_2(\omega_{\rm L})|E_{\rm opt}|^2$.\\

Concluding, the Hamiltonian for a single molecule is
\begin{align}
  H(t) = \frac{{\bf p}^2}{2m} + \frac{1}{2}m\omega_\perp^2 z^2 - V_0 + V_2 C_{0}^{(2)}(\theta,\phi)+ \nonumber\\
+B{\bf J}^2 - d_{0} E_{\rm DC} - \left(d_q E_{\rm AC} e^{-i\omega t}
+ {\rm h.c.}\right).\label{eq:eq9}
\end{align}

\bigskip

\section{Two molecules}\label{Sec:TwoMolecules}

We consider the interactions of two polar molecules $j=1,2$ confined
to the $x-y$ plane by a tight harmonic trapping potential of
frequency $\omega_\perp$, directed along $z$. The interaction of the
two molecules at a distance ${\bf r}\equiv{\bf r}_2-{\bf r}_1=r{\bf
e}_{r}$ is described by the Hamiltonian
\begin{eqnarray}
  H(t)  =   \sum_{j=1}^{2} H_j(t) + V_{\rm dd}({\bf r}),\label{eq:eqTwoMol}
  \label{eq:eq10}
\end{eqnarray}
where $H_j(t)$ is the single-molecule Hamiltonian
Eq.~\eqref{eq:eq9}, and $V_{\rm dd}({\bf r})$ is the dipole-dipole
interaction
\begin{eqnarray}
V_{\rm dd}({\bf r}) =  \frac{{\bf d}_1\cdot{\bf d}_2-3\left({\bf
d}_1\cdot{\bf
      e}_r\right)\left({\bf e}_r\cdot{\bf d}_{2}\right)}{r^3}.
\end{eqnarray}
Here, ${\bf d}_j$ is the dipole operator of the molecule $j$, and
${\bf e}_r\cdot{\bf d}_j$ is its projection onto the collision axis
${\bf e}_r$. The projection reads ${\bf e}_r\cdot{\bf
  d}_j=\sum_{q=-1}^{+1} (-1)^q C_{-q}^{(1)}(\vartheta,\varphi)
  d_{q;j}$,
where $C_q^{(1)}(\vartheta,\varphi)\equiv {\bf e}_q\cdot{\bf e}_r$
are unnormalized spherical hamonics with $\vartheta$ and $\varphi$
polar and azimuthal angles relating the orientation of ${\bf e}_r$
with respect to a space-fixed frame ${\bf e}_q$, respectively. The
terms $d_{q;j} \equiv{\bf e}_q\cdot{\bf d}_j$ are the spherical
components of the projection of the dipole operator of molecule $j$
onto the space-fixed frame ${\bf e}_q$.

In the absence of external fields $E_{\rm DC}=E_{\rm AC}=0$, the
interaction of the two molecules in their rotational ground state is
determined by the van-der-Waals attraction $V_{\rm vdW}\sim
C_{6;0}/r^{6}$ with $C_{6;0}\approx-d^{4}/6B$. This expression for
the interaction potential is valid outside of the molecular core
region $r>r_{B }\equiv (d^{2}/B)^{1/3}$, where $r_{B}$ defines the
characteristic length where the dipole-dipole interaction becomes
comparable to the splittings of the rotational levels, see below. In
the following we show that it is possible to \emph{induce} and
\emph{design} interaction potentials which are long-range, by
dressing the interactions with appropriately chosen static and/or
microwave fields. In fact, the combination of the latter with
low-dimensional trapping allows to engineer effective potentials
whose {\em strength} and {\em shape} can be both tuned. The
derivation of the effective interactions proceeds in two steps: (i)
We derive a set of Born-Oppenheimer (BO) potentials by first
separating Eq.~\eqref{eq:eq10} into center-of-mass and relative
coordinates, and diagonalizing the Hamiltonian for the relative
motion for fixed molecular positions. Within an adiabatic
approximation, the corresponding eigenvalues
play the role of an effective 
3D interaction potential in a given state manifold dressed by the
external field. (ii) We eliminate the motional degrees of freedom in
the tightly confined $z$ direction to obtain an effective 2D
dynamics with interaction $V_{\rm eff}^{\rm 2D}(\boldrho)$. In the
following we consider the cases of a static field and a microwave
field, coupling the lowest rotor states.

\subsection{Effective interactions in the presence of a DC electric
field}\label{sec:secDC}

In this section we consider the collisions of two ground-state
molecules in the presence of a DC electric field, ${\bf E}_{\rm
DC}=E_{\rm DC}{\bf e}_0$. The Hamiltonian Eq.~\eqref{eq:eqTwoMol}
now reads
\begin{eqnarray}\label{eq:Hamilt}
  H &=& \sum_{j=1}^{2} \left[\frac{{\bf p}_j^2}{2m} +
    \frac{1}{2}m\omega_\perp^2 z_j^2 + B{\bf J}_j^2 - E_{\rm DC}
    d_{0;j}\right] + V_{\rm dd}({\bf r})\nonumber\\
    &=& \sum_{j=1}^{2} \left[\frac{{\bf p}_j^2}{2m} +
    \frac{1}{2}m\omega_\perp^2 z_j^2 \right] + H_{\rm int}({\bf r}),\nonumber\\
\end{eqnarray}
where $d_{0;j}={\bf e}_0\cdot{\bf d}_j$ and $H_{\rm int}({\bf r})$
is the internal Hamiltonian including the dipole-dipole interaction,
$H_{\rm int}({\bf r}) = \sum_j \left[B{\bf J}_j^2 - E_{\rm DC}
d_{0;j}\right] + V_{\rm dd}({\bf r})$, respectively. In this section
we are interested in ground-state collisions, and thus for
convenience we set $V_2=0$ in Eq.~\eqref{eq:Hamilt}, that is, we
neglect possible tensor-shifts in the excited-state energies of each
molecule. We can further rewrite Eq.~\eqref{eq:Hamilt} by splitting
$H$ in center of mass and relative coordinates as $H=H_{\rm
com}+H_{\rm rel}$ with
\begin{subequations}\label{eq:eq12}
\begin{eqnarray}
  H_{\rm com} &=& \frac{{\bf P}^2}{4m} + m\omega_\perp^2
  Z^2,\\
  H_{\rm rel} &=& \frac{{\bf p}^2}{m}
  + \frac{1}{4}m\omega_{\perp}^2 z^2 + H_{\rm int}({\bf r}).
\end{eqnarray}
\end{subequations}
Here, ${\bf R}=({\bf r}_1+{\bf r}_2)/2$ and ${\bf P}={\bf p}_1+{\bf
  p}_2$ are the center of mass coordinate and momentum of the two
molecules, while ${\bf r}={\bf r}_2-{\bf r}_1$ and ${\bf p}=({\bf
  p}_2-{\bf p}_1)/2$ are the relative coordinate and momentum,
respectively. Equations~\eqref{eq:eq12} show that the dipole-dipole
interaction couples the internal degrees of freedom to the relative
motion, while the latter and the harmonic motion of the center of
mass remain decoupled. Thus, the non-trivial system's dynamics is
entirely determined by $H_{\rm rel}$. In the following we focus our
discussion on this term.

As explained above, in the spirit of the BO-approximation we can
obtain effective interaction potentials for the collision of the two
particles by diagonalizing $H_{\rm rel}$ for fixed particle
positions and zero kinetic energy. In the adiabatic approximation,
the resulting eigenvalues are energy-surfaces which act as effective
potentials in each state manifold. Here we first analyze the case of
collisions in the absence of external fields, that is $E_{\rm
DC}=\omega_\perp=0$, Sec.~\ref{sec:secNoField}. Then, in
Sec.~\ref{sec:secDCField} we add a static electric field of small
strength $E_{\rm DC} \ll B/d$. The effects of finite trapping
$\omega_\perp\neq 0$ are treated in the following section
Sec.~\ref{sec:secDCTrap} for the most relevant case of ground-state
collisions. The stability of ground-state collisions is investigated
in Sec.~\ref{sec:secDCStab}. The effective two-dimensional potential
for ground state collision is derived in Sec.~\ref{sec:secEff2D}.

\subsubsection{Collisions in the absence of external
fields}\label{sec:secNoField}

\begin{figure}
  \begin{center}
    \includegraphics[width=\columnwidth]{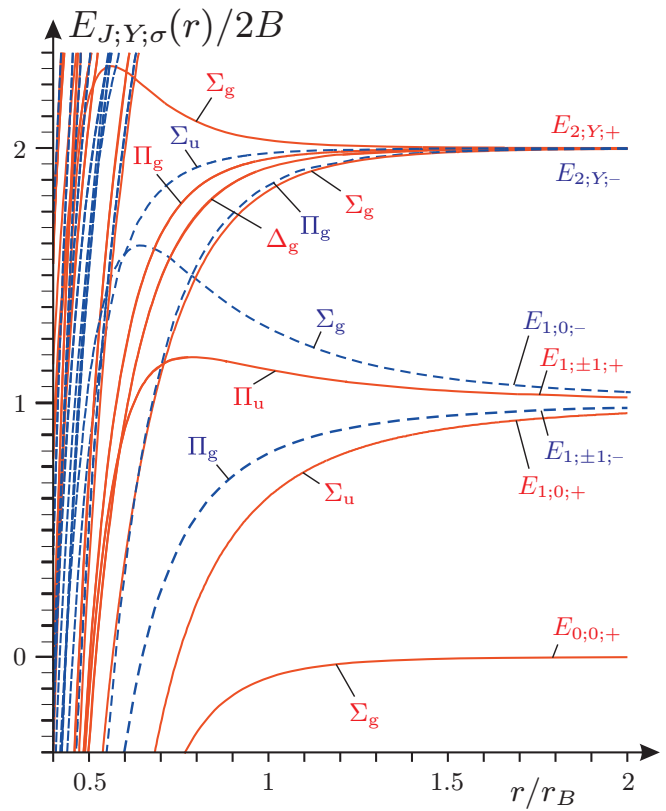}
    \caption{\label{fig:fig3}(color online) BO-potentials $E_{J;Y;\sigma}({\bf r})$
    as a function of the distance
    $r$, for two molecules interacting in the absence of external fields,
    $E_{\rm DC}=\omega_{\perp}=\beta=0$ (see text and
    Tab.~\ref{tab:tab2}). Here, $J=J_1+J_2$, $Y$ and $\sigma$ are the total number of
    rotational excitations shared by the two molecules, the quantum
    number associated with the projection of the total internal angular
    momentum onto the collision axis, $={\bf e}_r\cdot({\bf
J}_1+{\bf J}_2)$,
    and the permutation symmetry under the exchange of the particles,
    respectively.
    The solid and dashed curves
    correspond to symmetric ($\sigma=+$) and antisymmetric ($\sigma=-$) eigenstates,
    respectively.  Each potential energy surface is
    labeled by the corresponding energy and eigenstate (see Tab.~\ref{tab:tab2}). Here, $r_B\equiv(d^2/B)^{1/3}$, with $d$ the permanent
    dipole moment and $B$ the rotational constant of each molecule,
    respectively. Note that the $\Pi$- and the $\Delta$-states are doubly
    degenerate.}
  \end{center}
\end{figure}

\begin{table}[b]
\begin{tabular}[t]{|c|c|c|c||c||c|c|c|}
\hline $n$ & $J$ & $Y$ & $\sigma$ & $Y_{p}$ & $E_{n}^{(0)}$ &
 $C_{3;n}/d^2$ & $C_{6;n}\times6B/d^4$ \\\hline\
$0$ & $0$& $0$& $+$ & $\Sigma_{\rm g}$ & $0$ & $0$ & $-1$
\\\hline
$1$ & $1$& $0$& $+$ & $\Sigma_{\rm u}$ & $2B$ & $-2/3$ & $-22/45$ \\
$2,3$ & $1$& $\pm1$& $-$ & $\Pi_{\rm g}$ & $2B$ & $-1/3$ & $-19/45$ \\
$4,5$ & $1$& $\pm1$& $+$ & $\Pi_{\rm u}$ & $2B$ & $+1/3$ & $-19/45$ \\
$6$ & $1$& $0$& $-$ & $\Sigma_{\rm g}$ & $2B$ & $+2/3$ & $-22/45$
\\\hline
$7,8$ & $2$& $0_{\pm}$& $+$ & $\Sigma_{\rm g}$ & $4B$ & $0$ & $-(48\mp39\sqrt{3})/50$ \\
$9,10$ & $2$& $\pm1$& $-$ & $\Pi_{\rm u}$ & $4B$ & $0$ & $-39/20$ \\
$11,12$ & $2$& $\pm2$& $+$ & $\Delta_{\rm g}$ & $4B$ & $0$ & $-24/25$ \\
$13,14$ & $2$& $\pm1$& $+$ & $\Pi_{\rm g}$ & $4B$ & $0$ & $-51/25$ \\
$15$ & $2$& $0$& $-$ & $\Sigma_{\rm u}$ & $4B$ & $0$ & $-6/25$
\\\hline
\end{tabular}
\caption{\label{tab:tab2}Perturbative expansions of the effective
potentials, $E_n
  \equiv E_{J,Y,\sigma}$, for the three lowest-energy manifolds,
  $J=J_1+J_2=0,1,2$ with $J_j=0,1$, for interactions in the absence of external fields
  $E_{\rm DC}=E_{\rm AC}=\omega_\perp=0$. First column: index $n=0,1,2,\ldots$
  labeling the energy potentials and states. Second column: total number of
  rotational excitations shared by the two molecules $J=J_1+J_2$.
  Third column: the quantum number $Y$ associated with the projection of the
  total internal angular momentum along the collision axis, ${\bf e}_r\cdot({\bf
J}_1+{\bf J}_2)$, which is denoted by
  $\Sigma,\Pi,\Delta$ for $Y=0,1,2$, respectively. For $n=7,8$ the subindex $\pm$ indicates the presence of two states,
  which are split by the Van-der-Waals interaction.
  Fourth column:
  permutation symmetry $\sigma=\pm$. Fifth column: spectroscopic
  notation with parity $p=\sigma(-1)^J$ denoted by ``g'' (gerade) for
  $p=+1$ and ``u'' (ungerade) for $p=-1$, respectively. Sixth column:
  Asymptotic energy $E_n^{(0)}\equiv E_n(r\rightarrow\infty)$.
  Seventh column: dipole-dipole coefficient $C_{3;n}$.
  Last column: Van-der-Waals coefficient
  $C_{6;n}$. Perturbative energy eigenvalues are expressed in the
  form: $E_n({\bf r})=E_n^{(0)}+C_{3;n}/r^3+C_{6;n}/r^6$.}
\end{table}

In the absence of external fields ($E_{\rm DC}=\omega_\perp=0$) and
for zero kinetic energy, diagonalizing $H_{\rm rel}$ amounts to
diagonalizing $H_{\rm int}({\bf r})$ as a function of ${\bf r}$,
\begin{align}\label{eq:eqInt}
  H_{\rm int}({\bf r}) =
\sum_{j=1}^{2} B{\bf J}_j^2  + V_{\rm dd}({\bf r}) = \sum_n
\ket{\Phi_n({\bf r})}E_n({\bf r})\bra{\Phi_n({\bf r})},
\end{align}
where $E_n({\bf r})$ and $\ket{\Phi_n({\bf r})}$ are the $n^{\rm
  th}$-adiabatic energy eigenvalues and two-particle
eigenfunctions, respectively, and $n$ is a collective index for a set
of quantum numbers to be specified below. Each eigenvalue $E_n({\bf
  r})$ plays the role of an effective interaction in a given state
manifold dressed by the external field. At infinite separations of
the molecules, the eigenfunctions $\ket{\Phi_n^{(0)}({\bf
r})}\equiv\ket{\Phi_n^{(0)}(\vartheta,\varphi)}=\ket{\Phi_n(r\rightarrow\infty,\vartheta,\varphi)}$
are symmetrized products of the (rotated) single-particle
eigenstates $\ket{J_j,M_j}_j' \equiv e^{-i\varphi
J_{z;j}}e^{-i\vartheta J_{y;j}}\ket{J_j,M_j}_j$, which are
independent of the distance $r$. For finite $r$ the two-particle
eigenstates are superposition of several single-particle states,
which are mixed by the dipole-dipole interaction $V_{\rm dd}({\bf
r})$.

 A few eigenvalues $E_n({\bf r})$ of
Eq.~\eqref{eq:eqInt} are plotted as a function of ${\bf r}$ in
Fig.~\ref{fig:fig3}. Figure~\ref{fig:fig3} shows that the energy
spectrum behaves quite differently for $r<r_B$ and $r>r_B$, $r_B
\equiv (d^2/B)^{1/3}$. In fact, for $r<r_B$ a large number of level
crossings and anticrossings occurs, which make the fulfillment of
the adiabatic approximation generally impossible. The region $r<r_B$
is the \emph{molecular core} region. In the following we focus on
the region $r> r_B$, where the lowest-energy eigenvalues group into
well defined manifolds, which are approximately spaced by an energy
$\approx 2B$. For ground-state collisions, the adiabatic
approximation is here trivially fulfilled.\\

Since we are interested in ground-state collisions, we restrict our
discussion to the $J_j=0$ and $J_j=1$ manifolds of each molecule,
which amounts to take into account $16$ rotational two-particle
states. The corresponding eigenvalues $E_n({\bf r})$ are clearly
distinguishable in Fig.~\ref{fig:fig3} in the region $r>r_B$. The
manifolds are approximately split by $2B$, according to the number
of rotational excitations $J_1+J_2$ shared by the two molecules. The
two-particle energy eigenstates and eigenpotentials can be
classified according to the following symmetries of $H_{\rm
int}({\bf r})$: (a) The projection of the total internal angular
momentum along the collision axis, ${\bf e}_r\cdot({\bf J}_1+{\bf
J}_2)$, is conserved and associated with a quantum number $Y$; (b)
The Hamiltonian is invariant under the exchange of the two
particles, which is associated with the permutation symmetry
$\sigma=\pm$ under the exchange of the two particles. This implying
that symmetric (antisymmetric) states couple to symmetric
(antisymmetric) states only; And (c) the parity
$p=\sigma(-1)^{J_1+J_2}$ is conserved. The spectroscopic notations
labeling the eigenstates, $\ket{\Phi_n({\bf r})}$, and potentials,
$E_n({\bf r})$, with $n\equiv(J=J_1+J_2;Y;\sigma)$ in
Fig.~\ref{fig:fig3} are explained in the caption of
Tab.~\ref{tab:tab2}.

Analytic results for the energy eigenvalues and eigenstates of
$H_{\rm int}({\bf r})$ for large enough inter-particle distances $r$
can be derived using a perturbative expansion in $V_{\rm dd}({\bf
r})/B$. Our results for the energy eigenvalues $E_n({\bf r}) \equiv
E_{J_1+J_2;Y;\sigma}({\bf r})$ of $B\sum_j {\bf J}_j^2 + V_{\rm
dd}({\bf r})$ are summarized in Tab.~\ref{tab:tab2}. There, the
asymptotic energy $E_n^{(0)}\equiv E_n(r \rightarrow \infty)$, the
dipole-dipole coefficient $C_{3;n}$ and the Van-der-Waals
coefficient $C_{6;n}$ are reported, so that the perturbative
expression for $E_n({\bf r})$ takes the form $E_n({\bf
r})=E_n^{(0)}+C_{3;n}/r^3+C_{6;n}/r^6$. The Table shows that the
ground-state energy $E_{0}({\bf r})\equiv E_{0;0;+}({\bf r})$ is
shifted downwards by an amount $E_{0,0,+}({\bf r})= - d^4/6Br^6$,
which is the usual Van der Waals shift due to off-resonant
dipole-dipole interactions. The first excited manifold,
($J_1+J_2=1$), consists of $6$ states, of which $3$  are symmetric
and $3$ are antisymmetric. These states are split by the resonant
dipole-dipole interaction according to their angular momentum along
the collision axis, $|Y|=0,1$ and $\sigma=\pm$, as reported in
Table~\ref{tab:tab2}. Finally, the second excited manifold,
($J_1+J_2=2$), consists of $9$ states, of which $6$ are symmetric,
$Y=0_\pm,\pm1,\pm2$ with $\sigma=+$, and $3$ are antisymmetric,
$Y=0,\pm 1$ with $\sigma=-$.

\subsubsection{Collisions in a DC field: Effective 3-D
interaction}\label{sec:secDCField}

\begin{figure*}[htbp]
  \begin{center}
    \includegraphics[width=\textwidth]{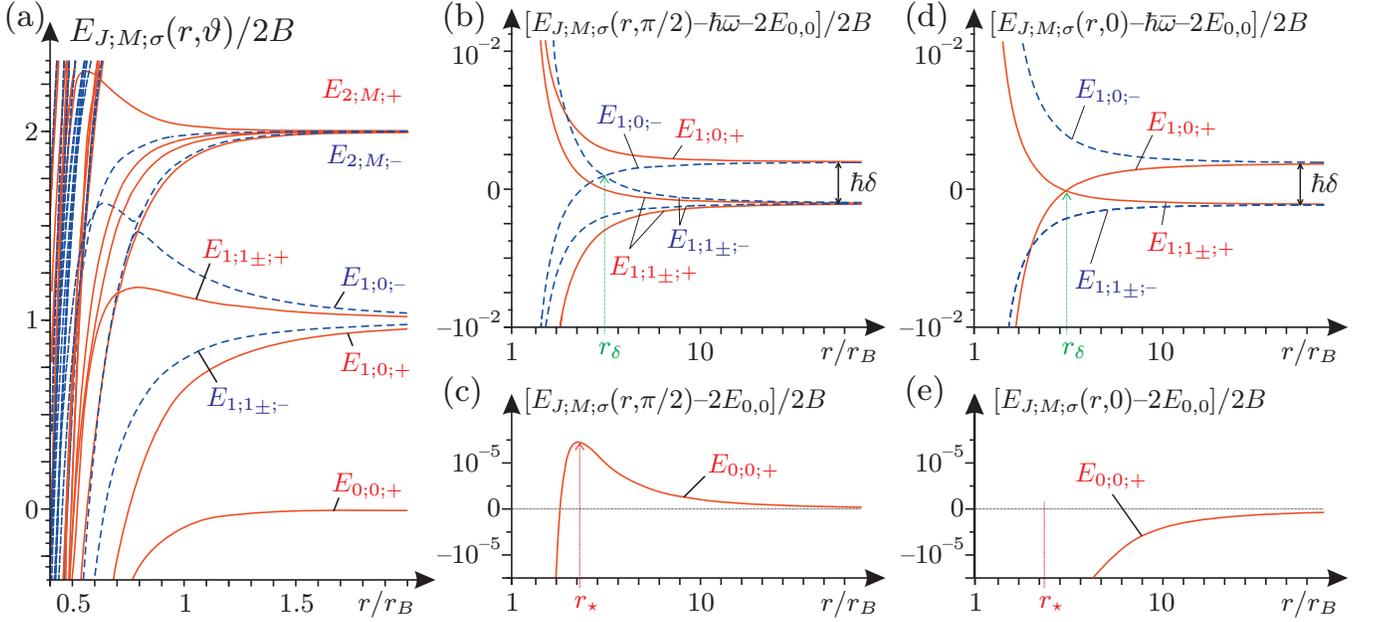}
    \caption{\label{fig:fig4}(color online) BO-potentials $E_{J;M;\sigma}(r,\vartheta)$ for two molecules colliding
      in the presence of a DC field, with
      $\beta\equiv d E_{\rm DC}/B=1/5$ and $(J;M;\sigma)$ the quantum numbers of
      Tab.~\ref{tab:tab3}. The solid and dashed curves
      correspond to symmetric ($\sigma=+$) and antisymmetric ($\sigma=-$) eigenstates,
      respectively. (a): BO-potentials for the 16 lowest-energy eigenstates
      $E_n(r,\vartheta)$. The molecular-core region is
      identified as the region $r<r_B=(d^2/B)^{1/3}$, while for $r\gg r_B $ the eigestates group into manifolds separated by one
      quantum of rotational excitation $2B$. (b) and (e): Blow-ups
      of the first-excited energy manifold of panel (a) in the region $r \gtrsim
      r_B$ for $\vartheta=\pi/2$ and $\vartheta=0$, respectively. Note the
      electric-field-induced splitting $\hbar \delta\equiv 3 B \beta^2 /20$ (see Sec.~II~B~1).
      The distance $r_\delta$ where the dipole-dipole interaction
      becomes comparable to $\hbar \delta$ is $r_{\delta}=(d^2/\hbar\delta)^{1/3}$. (c) and (e): Blow-ups
      of the ground-state potential $E_{0,0;+}(r,\vartheta)$ of panel (a) in the region $r \gtrsim
      r_B$ for $\vartheta=\pi/2$ and $\vartheta=0$, respectively. The distance
      $r_\star$, cf.~Eq.~\eqref{eq:rstar},
      where the dipole-dipole interaction becomes comparable to the Van-der-Waals attraction
      is indicated. Note the repulsive (attractive) character of the potential for $\vartheta=\pi/2$
      ($\vartheta=0$) and $r>r_\star$.}
  \end{center}
\end{figure*}

\begin{table*}[htbp]
  \begin{tabular}[t]{||c||c|c|c||c|c|c||c||}
\hline\hline
$n$&$J$&$M_\mu$&$\sigma$&$E_n^{(0)}-2E_{0,0}$&$C_{3;n} h_n(\vartheta)$&$C_{6;n}(\vartheta)6B/d^4$&$\ket{\Phi_n^{(0)}(\vartheta,\varphi)}$\\
\hline\hline
$0$&$0$&$0$&$+$&$0$&$g_0^2\Upsilon$&$-1$&$\ket{\phi_{0,0};\phi_{0,0}}$\\
\hline
$1$&$1$&$1_-$&$+$&$\hbar(\overline{\omega}-\delta/3)$&$(g_0g_1-f_1^2)\Upsilon-f_1^2$&$-A_1-(21+\Upsilon)/45$&$\sum_\pm{\pm}e^{\mp i\varphi}\ket{ \phi_{0,0};\phi_{1,\pm1}}/2+{\rm perm.}$\\
$2$&$1$&$1_-$&$-$&$\hbar(\overline{\omega}-\delta/3)$&$(g_0g_1+f_1^2)\Upsilon+f_1^2$&$-A_1-(21+\Upsilon)/45$&$\sum_\pm{\pm}e^{\mp i\varphi}\ket{\phi_{0,0};\phi_{1,\pm1}}/2-{\rm perm.}$\\
$3$&$1$&$1_+$&$+$&$\hbar(\overline{\omega}-\delta/3)$&$g_0g_1\Upsilon+f_1^2$&$-19/45$
&$\sum_\pm e^{\mp i\varphi}\ket{\phi_{0,0};\phi_{1,\pm1}}/2+{\rm perm.}$\\
$4$&$1$&$1_+$&$-$&$\hbar(\overline{\omega}-\delta/3)$&$g_0g_1\Upsilon-f_1^2$&$-19/45$
&$\sum_\pm e^{\mp i\varphi}\ket{\phi_{0,0};\phi_{1,\pm1}}/2-{\rm perm.}$\\
\hline
$5$&$1$&$0$&$+$&$\hbar(\overline{\omega}+2\delta/3)$&$(g_0g_2+f_0^2)\Upsilon$&$+A_1-(20-\Upsilon)/45$
&$\ket{\phi_{0,0};\phi_{1,0}}/\sqrt{2}+{\rm perm.}$\\
$6$&$1$&$0$&$-$&$\hbar(\overline{\omega}+2\delta/3)$&$(g_0g_2-f_0^2)\Upsilon$&$+A_1-(20-\Upsilon)/45$
&$\ket{\phi_{0,0};\phi_{1,0}}/\sqrt{2}-{\rm perm.}$\\
\hline
$7$&$2$&$2$&$-$&$2\hbar(\overline{\omega}-\delta/3)$&$g_1^2\Upsilon$&$-3(46+19\Upsilon)/100$
&$\sum_\pm \pm\ket{\phi_{1,\pm1};\phi_{1,\mp1}}/\sqrt{2}$\\
$8$&$2$&$2_-$&$+$&$2\hbar(\overline{\omega}-\delta/3)$&$g_1^2\Upsilon$&$-3(22-5\Upsilon)/100$
&$\sum_\pm\pm e^{\mp 2i\varphi}\ket{\phi_{1,\pm1};\phi_{1,\pm1}}/\sqrt{2}$\\
$9$&$2$&$2_0$&$+$&$2\hbar(\overline{\omega}-\delta/3)$&$g_1^2\Upsilon$&$-3(A_2+A_3)/100$
&$\sum_\pm(c_\xi\ket{\phi_{1,\pm1};\phi_{1,\mp1}}-s_\xi e^{\mp2i\varphi}\ket{\phi_{1,\pm1};\phi_{1,\pm1}})/\sqrt{2}$\\
$10$&$2$&$2_+$&$+$&$2\hbar(\overline{\omega}-\delta/3)$&$g_1^2\Upsilon$&$-3(A_2-A_3)/200$
&$\sum_\pm(s_\xi\ket{\phi_{1,\pm1};\phi_{1,\mp1}}+c_\xi
e^{\mp2i\varphi}\ket{\phi_{1,\pm1};\phi_{1,\pm1}})/\sqrt{2}$\\
\hline
$11$&$2$&$1_-$&$+$&$2\hbar(\overline{\omega}+\delta/6)$&$(g_1g_2-f_2^2)\Upsilon-f_2^2$&$-3(13+2\Upsilon+2\Upsilon^2)/100$
&$\sum_\pm {\pm}e^{\mp
    i\varphi}\ket{\phi_{1,0};\phi_{1,\pm1}}/2+{\rm perm.}$\\
$12$&$2$&$1_-$&$-$&$2\hbar(\overline{\omega}+\delta/6)$&$(g_1g_2+f_2^2)\Upsilon+f_2^2$&$-39/20$
&$\sum_\pm \pm e^{\mp
    i\varphi}\ket{\phi_{1,0};\phi_{1,\pm1}}/2-{\rm perm.}$\\
$13$&$2$&$1_+$&$+$&$2\hbar(\overline{\omega}+\delta/6)$&$g_1g_2\Upsilon+f_2^2$&$-3(27+5\Upsilon)/100$
&$\sum_\pm e^{\mp i\varphi}\ket{\phi_{1,0};\phi_{1,\pm1}}/2+{\rm perm.}$\\
$14$&$2$&$1_+$&$-$&$2\hbar(\overline{\omega}+\delta/6)$&$g_1g_2\Upsilon-f_2^2$&$-3(27-19\Upsilon)/100$
&$\sum_\pm e^{\mp i\varphi}\ket{\phi_{1,0};\phi_{1,\pm1}}/2+{\rm perm.}$\\
\hline
$15$&$2$&$0$&$+$&$2\hbar(\overline{\omega}+2\delta/3)$&$g_2^2\Upsilon$&$-3(34-14\Upsilon-\Upsilon^2)/100$
&$\ket{\phi_{1,0};\phi_{1,0}}$\\
\hline\hline
  \end{tabular}
  \caption{\label{tab:tab3} Perturbative expressions for the $16$ lowest-energy BO-potentials
    $E_n({\bf r})=E_n^{(0)}+C_{3;n} h_n(\vartheta)/r^3+C_{6;n}(\vartheta)/r^6$ of two
    molecules interacting in the presence of a DC electric field ${\bf E}_{\rm
      DC}=(B\beta/d){\bf e}_0$. First column: The collective quantum number
    $n\equiv(J=J_1+J_2;M\equiv|M_1|+|M_2|;\sigma=\pm)$, labeling the eigenstates $E_n({\bf r})$.
    Second column: The number $J=J_1+J_2$ of rotational excitations
    shared by the two molecules. Because of the presence of the DC
    field, parity is not conserved and $J$ is a simple index that
    labels the various energy manifolds for $r\gg r_B,r_\delta$. Third
    column: The quantum number $M\equiv|M_1|+|M_2|$. The additional subindex $\mu$ for $M>0$ labels superposition of states with the same $(Y;M;\sigma)$, which depend on the azimuthal angle $\varphi$ (see eigth column).  Fourth column:
    The quantum number $\sigma=+$ ($\sigma=-$) denoting symmetric (antisymmetric)
    states under permutation of the two molecules. Fifth column:
    Asymptotic energies $E_n^{(0)}$ for infinite separation.
    The quantities $\hbar\delta$ and $\hbar\overline{\omega}$ are defined in
    Eq.~\eqref{eq:eqdelta} and Eq.~\eqref{eq:eqomega}, respectively.
     Sixth column: The $C_{3;n}$ coefficient and the angular dependence
    $h_n(\vartheta)$.
    The dipole moments $g_n$ and $f_n$ are defined in
    Table~\ref{tab:tab1}, while the angular
    distribution $\Upsilon$ is $\Upsilon\equiv 1-3\cos^2\vartheta$.
    Seventh column: The contributions to the $C_{6;n}(\vartheta)$ coefficient up to order ${\cal O}(\beta^2)$. The values
    $A_1$, $A_2$ and $A_3$ are $A_1\equiv40(2-\Upsilon-\Upsilon^2)(f_0f_1+f_2g_0)^2/d^4\beta^2$,
$A_2\equiv 33+6\Upsilon-\Upsilon^2/2$,
$A_3\equiv13(1+\Upsilon)/\cos\xi$, respectively. Here, $\xi$
      is defined by the relation
      $\tan\xi=(14-\Upsilon)(2+\Upsilon)/26(1+\Upsilon)$.
    Last column: Eigenstates
$\ket{\Phi_n^{(0)}(\vartheta,\varphi)}\equiv\ket{\Phi_n(r\rightarrow\infty,\vartheta,\varphi)}$
   valid at infinite separation. Here,
``perm.'' denotes the
  permuted state, e.g.~$\ket{\phi_{1,2};\phi_{3,4}}\rightarrow\ket{\phi_{3,4};\phi_{1,2}}$.}
\end{table*}

We now turn to study the collision of the two molecules in the
presence of a weak static electric field applied in the
$z$-direction but in the absence of optical trapping, that is, ${\bf
E}=E_{\rm DC}{\bf e}_0$ with $E_{\rm DC} \ll B/d$ and
$\omega_\perp=0$. As explained in Sec.~\ref{Sec:CouplingDC}, the
effects of a DC electric field on each molecule are to partially
split the $(2J+1)$-fold degeneracy in the rotor spectrum (the
modulus of the projection $M$ is conserved), and to align the
molecule along the direction of the field, which amounts to inducing
a finite dipole moment
${}_j\bra{\phi_{J_j,M_j}}d_{0;j}\ket{\phi_{J_j,M_j}}_j$ in each
rotational state.

Analogous to the discussion above, the effective interaction
potentials for the collision of the two particles can be obtained in
the adiabatic approximation by diagonalizing the Hamiltonian $H_{\rm
rel}=H_{\rm int}({\bf r})= \sum_{j=1}^{2} \left[ B{\bf J}_j^2 -
E_{\rm DC} d_{0;j}\right] + V_{\rm dd}({\bf r})= \sum_n
\ket{\Phi_n({\bf r})}E_n({\bf r})\bra{\Phi_n({\bf r})}$, where now
the asymptotic energy eigenstates
$\ket{\Phi_n^{(0)}(\vartheta,\varphi)}$ are symmetrized products of
the single-particle states $\ket{\phi_{J_j,M_j}}_j$ of
Eq.~\eqref{eq:EigStatic}. The quantity $n\equiv(J;M;\sigma)$ is the
collective quantum number labeling the eigenvalues $E_n({\bf r})$,
with $J=J_1+J_2$, $M\equiv|M_1|+|M_2|$, and $\sigma=\pm$. We note
that, because of the presence of the DC field, here $J$ is a simple
label for the various energy manifolds, and not a quantum number.
The energies of the eigenvalues $E_n({\bf r})$ and the associated
eigenvectors are tabulated in Tab.~\ref{tab:tab3}.

Similar to the zero-field discussion, in the weak field limit $\beta
\ll 1$ and for $r>r_B$ we expect the eigenvalues of $H_{\rm
int}({\bf r})$ to group into manifolds, which are approximately
separated by the rotational spacing $2B$. On the other hand, because
of the finite induced dipole moments
${}_j\bra{\phi_{J_j,M_j}}d_{0;j}\ket{\phi_{J_j,M_j}}_j$, for the two
molecules can now interact resonantly {\em via} the dipole-dipole
interaction $V_{\rm dd}({\bf r})$ in each state manifold. This has
important consequences for ground-state collisions. In fact, the new
effective ground-state potential $E_0({\bf r})$ derived in
perturbation theory in $V_{\rm dd}({\bf r})/B$ reads
\begin{eqnarray}\label{eq:eqVeff3D}
    V_{\rm eff}^{\rm 3D}({\bf r}) \equiv  E_0({\bf
      r})\approx \frac{C_{3;0}}{r^3}\left(1-3\cos^2\vartheta\right)
+\frac{C_{6;0}}{r^6},
  \end{eqnarray}
where a constant term $2E_{0,0}=-\beta^2 B/3$ due single-particle DC
Stark-shifts has been neglected. The constants $C_{3;0}\approx
d^2\beta^2/9$ and $C_{6;0}\approx - d^4/6B$ are the dipolar and Van
der Waals coefficients for the ground-state, respectively
(see~Tab.~\ref{tab:tab3}). Equation~\eqref{eq:eqVeff3D} is valid for
$r\gg r_B$, and it shows that for distances $r \gg r_\star$ with
\begin{eqnarray}\label{eq:rstar}
r_\star\equiv
\left(\frac{2|C_{6;0}|}{C_{3;0}}\right)^{1/3}\approx\left(\frac{3d^2}{B\beta^2}\right)^{1/3}
\end{eqnarray}
the dipole-dipole interaction dominates over the Van der Waals
attractive potential, and $V_{\rm eff}^{\rm 3D}({\bf r})\sim
C_{3;0}(1-3\cos^2\vartheta)/r^3$~\cite{Buechler07}. In fact, the
potential has a local maximum in the plane $z=r \cos \vartheta = 0$
at the position $r=r_\star$, where the dipole-dipole and Van der
Waals interactions become comparable. The height of this maximum is
\begin{eqnarray}\label{eq:Vstar}
V_\star=\frac{{C_{3;0}}^2}{4|C_{6;0}|}\approx \frac{B\beta^4}{54},
\end{eqnarray}
and the curvature along $z$ is $\partial_z^2 V(r=r_\star,z=0) =
-6C_{3;0}/r_\star^5 \equiv - m\omega_{\rm c}^2/2$, which defines a
characteristic frequency
\begin{eqnarray}\label{eq:omegac}
\omega_{\rm c}\equiv
\left(\frac{12C_{3;0}}{mr_\star^5}\right)^{1/2},
\end{eqnarray}
to be used below. The latter has a strong dependence $\beta^{8/3}=(d
E_{\rm DC}/B)^{8/3}$ on the applied electric field.

We notice that if it were possible to confine the collisional
dynamics to the $(z=0)$-plane, purely repulsive long-range
interactions with a characteristic dipolar spatial dependence $\sim
1/r^3$ could be attained. In the following sections, we analyze the
conditions for realizing sufficiently strong confinements to the
$(z=0)$-plane by employing a tight harmonic optical trap in the $z$-direction.\\

Figure~\ref{fig:fig4} shows the eigenvalues $E_n({\bf r})$ as a
function of the interparticle distance $r$, for $\beta=1/5$. The
vector ${\bf r}$ is expressed in spherical coordinates ${\bf
r}=(r,\vartheta,\phi)$, with $z=r \cos\vartheta$.
Figure~\ref{fig:fig4}(a) shows the different behavior of the energy
spectrum for $r<r_B$ and $r>r_B$, analogous to the zero-field case.
Even at finite $\beta$ we can clearly distinguish the molecular core
region $r<r_B$ where the adiabatic approximation breaks down. In
this plot, the continuous and dashed lines correspond to the cases
$\vartheta=\pi/2$ and $\vartheta=0$, respectively, which are almost
indistinguishable on the scale of the graph.
Figures~\ref{fig:fig4}(b,c) and Figs.~\ref{fig:fig4}(d,e) are
blow-ups of the two lowest-energy manifolds of
Fig.~\ref{fig:fig4}(a), for $\vartheta=\pi/2$ and $\vartheta=0$,
respectively. Different from the zero-field case,
Fig.~\ref{fig:fig4}(b) and Fig.~\ref{fig:fig4}(d) show that the
excited state manifold with one quantum of rotation ($J_1+J_2=1$) is
asymptotically split into two sub-manifolds. This separation
corresponds to the electric-field-induced splitting of the $J_j=1$
manifold of each molecule, and it is thus given by $\hbar \delta = 3
B \beta^2/20$ of Eq.~\eqref{eq:eqdelta}. More importantly,
Fig.~\ref{fig:fig4}(c) and Fig.~\ref{fig:fig4}(e) show that the
effective ground-state potential has a very different character for
the cases $\vartheta=\pi/2$ and $\vartheta=0$, respectively. In
fact, for $\vartheta=\pi/2$, corresponding to collisions in the
$(z=0)$-plane [see Fig.~\ref{fig:fig4}(c)], the potential is
repulsive and decaying at large distances as $1/r^3$ in agreement
with the discussion above. On the other hand, for $\vartheta=0$ [see
Fig.~\ref{fig:fig4}(e)] the potential is purely attractive, with
dipolar character. As mentioned above, in the next section we show
that the probability to sample this attractive part of the potential
during the collision can be largely suppressed in the case
$\omega_\perp\neq0$, for a sufficiently tight transverse trapping.

\subsubsection{Parabolic confinement}\label{sec:secDCTrap}

The presence of a finite trapping potential of frequency
$\omega_{\perp}$ in the $z$-direction provides for a
position-dependent energy shift of Eq.~\eqref{eq:eqVeff3D}. The new
potential reads
\begin{eqnarray}\label{eq:eqE0r}
  V({\bf r})&\equiv&  V_{\rm eff}^{\rm 3D}({\bf r}) + \frac{1}{4}m \omega_\perp^2
  z^2 \nonumber\\
  &=&  \frac{C_{3;0}}{r^3}\left(1-3\cos^2\vartheta\right)
  +\frac{C_{6;0}}{r^6} + \frac{1}{4}m \omega_\perp^2z^2.
  \end{eqnarray}

As noted before, for $z=0$ the repulsive dipole-dipole interaction
dominates over the attractive Van der Waals at distances $r\gg
r_\star$ given in Eq.~\eqref{eq:rstar}. In addition, for
$\omega_\perp
> 0$ the harmonic potential confines the particle's motion in the
$z$ direction. Thus, the combination of the dipole-dipole
interaction and of the harmonic confinement yields a repulsive
potential which provides for a three-dimensional barrier separating
the long-distance from the short-distance regime. If the collisional
energy is much smaller than this barrier, the particle's motion is
confined to the long-distance region, where the potential is purely
repulsive.

\begin{figure}[htbp]
  \begin{center}
    \includegraphics[width=\columnwidth]{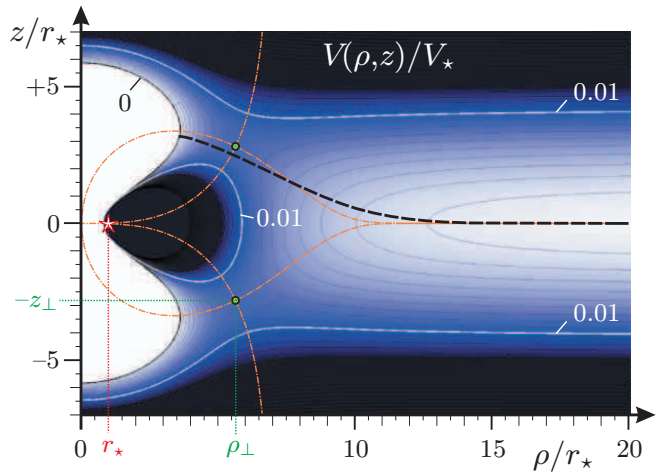}
  \end{center}
  \caption{\label{fig:fig5}(color online) Contour plot of the effective potential $V(\rho,z)$
  of Eq.~\eqref{eq:eqE0r}, for two polar molecules interacting
  in the presence of a DC field $\beta>0$, and a confining harmonic potential in the
  $z$-direction, with trapping frequency $\omega_\perp=\omega_{\rm
  c}/10$, where $\omega_{\rm c}\equiv (12C_{3;0}/mr_\star^5)^{1/2}$ of Eq.~\eqref{eq:omegac} and
  $r_\star=(2 |C_{6;0}|/C_{3;0})^{1/3}$ of Eq.~\eqref{eq:rstar}. The contour lines are shown for $V(\rho,z)/V_\star \geq 0$, with
    $V_\star=B \beta^4/54$.
  Darker regions represent stronger repulsive interactions. The
    combination of the dipole-dipole interactions induced by the
    DC field and of the harmonic confinement leads to
    realizing a 3D repulsive potential. The
  repulsion due to the dipole-dipole interaction and of the harmonic
  confinement is distinguishable at $z\sim 0$ and $z/r_\star\sim
  \pm 7$, respectively. Two saddle points (circles) located at
  ($\rho_\perp,\pm z_\perp$) separate the long-distance region where
  the potential is repulsive $\sim 1/r^3$ from the attractive
  short-distance region. The gradients of the potential
  are indicated by dash-dotted lines. The thick dashed line
    indicates the instanton solution for the tunneling through the
    potential barrier.}
\end{figure}

Figure~\ref{fig:fig5} is a contour plot of $V({\bf r})$ in units of
$V_\star$, cf.~Eq.~\eqref{eq:Vstar}, for $\beta>0$ and
$\omega_\perp=\omega_{\rm c}/10$, with ${\bf r}\equiv(\rho,z)=r(\sin
\vartheta,\cos\vartheta)$ (the angle $\varphi$ is neglected due to
the cylindrical symmetry of the problem). Darker regions correspond
to a stronger repulsive potential. The repulsion due to the
dipole-dipole and harmonic potentials is clearly distinguishable at
$|z|/r_\star \sim 0 $ and $7$, respectively. The lesser-dark regions
located at $(\rho_\perp,\pm z_\perp)\equiv
\ell_\perp(\sin\vartheta_\perp,\pm\cos\vartheta_\perp)$ correspond
to the existence of two saddle points, see circles in
Fig.~\ref{fig:fig5}. Here, $\ell_\perp$ and $\cos\vartheta_\perp$
are $\ell_\perp=(12C_{3;0}/m\omega_\perp^2)^{1/5}$ and
$\cos\vartheta_\perp=\sqrt{1-(r_\star/\ell_\perp)^3}/\sqrt{5}$,
respectively, while the barrier at the saddle point is
$V(\rho_\perp,\pm
z_\perp)=C_{3;0}/\ell_\perp^3+C_{6;0}/\ell_\perp^6$. The figure
shows that for distances $r \gg \ell_\perp \geq r_\star,r_B $ the
effective interaction potential Eq.~\eqref{eq:eqE0r} is purely
repulsive. The existence of two saddle points at distances $r\sim
\ell_\perp$ separating the long- from the short-distance regimes is
a general feature of systems with $\beta>0$ and
$\omega_{\perp}/\omega_{\rm c}<1$. Thus, $\ell_\perp$ defines the
characteristic length-scale for attaining purely repulsive 3D
potentials in the presence of a static electric field. Actually, we
show below that for collisional energies smaller than
$V(\rho_\perp,|z_\perp|)$ the dynamics of the particle can be
reduced to a {\em quasi} two-dimensional (2D) one, by tracing over
the fast particle motion in the $z$-direction.

For strong trapping $\omega_\perp{\geq}\omega_{\rm c}$ the two
saddle points collapse into a single one located at $z=0$, and $\rho
= \ell_\perp \sim r_\star$. In this limit the dynamics is purely 2D,
with the particles strictly confined to the $(z=0)$-plane. The
long-distance regime is separated from the short-distance one by the
potential barrier of height $V(\ell_\perp,0)=V_\star=B\beta^4/54$.
The amount of harmonic confinement required to achieve this pure 2D
regime increases rapidly with $\beta$ as $\omega_{\rm
c}\propto\beta^{8/3}$. While for a typical rotational constant
$B/h\sim~5~{\rm GHz}$ and a weak DC field $\beta=1/10$, $\omega_{\rm
c}$ is of order of $\omega_{\rm c}/2\pi \sim 10~{\rm kHz}$, for a
(reasonable)
 electric field $\beta=1/3$ we find $\omega_{\rm c}/2\pi
\sim~1~{\rm MHz}$. This value of $\omega_{\rm c}$ exceeds the
tightest experimental optical traps $\omega_\perp^{\rm max}/2\pi\sim
150~{\rm kHz}$. Thus, in general the
dynamics should be considered {\em quasi} 2D.\\

When an {\em ensemble} of polar molecules is considered, inelastic
collisions and three body recombination may lead the system to a
potential instability, associated with the attractive character of
the dipole-dipole
interaction~\cite{Ticknor05,Bortolotti06,Santos03}. In our
discussion, this instability is associated with the population of
the short-distance region $r < \ell_\perp$, which can be efficiently
suppressed. In fact, for collisional energies smaller than the
potential barrier $V(\rho_\perp,\pm z_\perp)$ the particles are
mostly confined to the long-distance regime, where they scatter
elastically. That is, when a cold ensemble of molecules is
considered the barrier provides for the stability of the system by
``shielding'' the short-distance attractive part of the two-body
potential. In this limit, residual losses are due to the tunneling
through the potential barrier. In the next section we estimate the
tunneling rate $\Gamma$ associated with this process, and we show
that it can be efficiently suppressed for reasonable values of
$\beta$ and $\omega_\perp$. Thus, it is possible to realize stable
2D configurations of strongly interacting polar molecules
interacting {\em via} dipole-dipole interactions~\cite{Buechler07}.

\subsubsection{Stability of long-range
collisions}\label{sec:secDCStab} In the following we calculate the
rate $\Gamma = \Gamma_0 e^{-S_{\rm E}/\hbar}$ of particle tunneling
through the barrier $V(\rho_\perp,\pm z_\perp)$ using a
semi-classical/instanton approach~\cite{Coleman77}. In particular,
we focus on determining the quantity $S_{\rm E}$, the euclidian
action of the semiclassical trajectory~\cite{Coleman77}, which is
responsible for the exponential suppression of the tunneling. The
constant $\Gamma_0$ is related to the quantum fluctuations around
the semiclassical trajectory, and its value is strongly
system-dependent. For the crystalline phase of
Ref.~\cite{Buechler07} (see also Fig.~\ref{fig:fig0}), it is the
collisional "attempt frequency", proportional to the characteristic
phonon frequency $\Gamma_0\sim \sqrt{C_{3;0}/m a^5}$, with $a$ the
mean interparticle distance.

\begin{figure}[b]
  \begin{center}
    \includegraphics[width=\columnwidth]{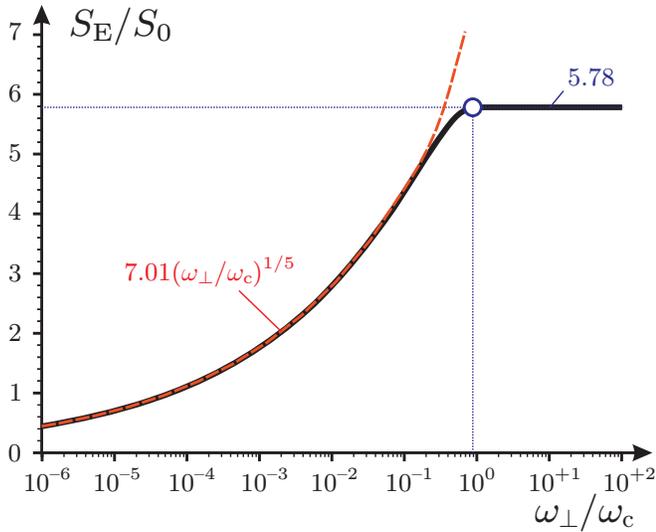}
    \end{center}
  \caption{\label{figs:fig5}(color online) The euclidian action $S_{\rm E}$ of Eq.~\eqref{eq:eq22}
    as a function of $\omega_\perp/\omega_{\rm c}$ (solid line). For $\omega_\perp<\omega_{\rm c}'\approx
0.88~\omega_{\rm c}$ ($\omega_\perp>\omega_{\rm c}'$) the "bounce"
occurs for $z(0)\neq0$ (within the plane $z(0)=0$), see text. The
point $\omega_{\rm c}'$ is signaled by a circle. The dashed line
    is the $C_{6;0}$-{\em independent} expression
    $S_{\rm E}\approx 7.01 S_0(\omega_\perp/\omega_{\rm c})^{1/5}$ (see text), with
    $S_0=\sqrt{m |C_{6;0}|}/\hbar r_\star^2$. For
    $\omega_\perp>\omega_{\rm c}'$ the action is $S_{\rm E}\approx5.78 S_0$,
    which is $\omega_\perp$-{\em independent}, consistent with the
    "bounce" occurring in the ($z=0$)-plane.}
\end{figure}

The relative motion of the two particles in the effective potential
$V({\bf r})$ of Eq.~\eqref{eq:eqE0r} is analogous to that of a
single (fictitious) particle with reduced mass $m/2$, and dynamics
determined by the Hamiltonian $H = {\bf p}^2/m+V({\bf r})$. The
associated euclidian action, that is the action in imaginary time
$\tau$, is given by
\begin{eqnarray}\label{eq:guidoginopino}
  S_{\rm E}\left[{\bf r}(\tau)\right] =
  \int_{-\infty}^{+\infty}
  d\tau\left[\frac{m}{4}\left(\frac{\partial{\bf r}}{\partial
        \tau}\right)^2+V({\bf r})\right],
\end{eqnarray}
where ${\bf r}(\tau)$ is the particle's trajectory. We remark that
Eq.~\eqref{eq:guidoginopino} corresponds to the action in real time,
with an inverted potential $-V({\bf r})$. The classical trajectories
are found by minimizing the action Eq.~\eqref{eq:guidoginopino},
yielding the following equation of motion
\begin{eqnarray}\label{eq:eqofmotion}
  \frac{m}{2}\frac{d^2 {\bf r}}{d\tau^2}=+{\bf \nabla}V({\bf
    r}).
\end{eqnarray}
The ``energy'' $\overline E={\bf p}^2/m-V({\bf r})$ is conserved
along each classical trajectory. The instanton solution is then the
trajectory with the smallest action $S_{\rm E}$, which approaches
${\bf r}(\tau\rightarrow\pm\infty)=(\rho\rightarrow\infty,0)$
asymptotically at time $\tau\rightarrow\mp\infty$, see dashed line
in Fig.~\ref{fig:fig5}. The energy of the particle along this
trajectory is zero. The action $S_{\rm E}$ reads
\begin{eqnarray}
  S_{\rm E} = 2\int_0^\infty d\tau 2V[{\bf r}(\tau)]=
  2\int_{{\bf r}(0)}^{{\bf r}(\infty)} ds \sqrt{m V({\bf
      r})},\label{eq:eq22}
\end{eqnarray}
where ${\bf r}(0)$ is the ``bouncing point'' reached at $\tau=0$~\cite{Coleman77}.\\

We solve Eq.~\eqref{eq:eqofmotion} numerically for the classical
trajectories with zero energy, for a generic value of $\beta$ and
$\omega_\perp$. The obtained action $S_{\rm E}$ is plotted in
Fig.~\ref{fig:fig5} as a function of $\omega_\perp/\omega_{\rm c}$,
in units of
$S_0=\sqrt{m|C_{6;0}|}/r_\star^2=(2Bd^4m^3\beta^8/3^7)^{1/6}$. We
notice that the action shows different behaviors for
$\omega_\perp\ll \omega_{\rm c}$ and $\omega_\perp\gg\omega_{\rm
c}$. In particular, for $\omega_\perp\ll \omega_{\rm c}$ the action
increases with increasing $\omega_\perp$, while for
$\omega_\perp\gg\omega_{\rm c}$ it is $\omega_\perp$-independent.
The transition between the two different regimes mirrors the change
in the nature of the underlying potential $V({\bf r})$ as a function
of $\omega_\perp/\omega_{\rm c}$ described following
Eq.~\eqref{eq:eqE0r}, as explained below.

We find numerically that for $\omega<\omega_{\rm
c}'\approx0.88~\omega_{\rm
  c}$ the ``bouncing point'' ${\bf r}(\tau=0)=[\rho(0),z(0)]$
  of the instanton solution occurs for $z(0)\neq0$, see dashed line in
Fig.~\ref{fig:fig5}. This is consistent with the existence of two
saddle points located at $V(\rho_\perp, \pm z_\perp)$, with $z_\perp
>0$. Since the saddle points appear approximately at a length $r\sim
\ell_\perp \gg r_\star$, it is expected that in this regime the
action is independent of the short distance behavior of the
potential, that is of the $C_{6;0}$-coefficient of the Van der Waals
attraction. Accordingly, Fig.~\ref{fig:fig5} shows that $S_{\rm E}$
is well approximated by $S_{\rm E}\approx
7.01S_0(\omega_\perp/\omega_{\rm c})^{1/5} =
5.86(C_{3;0}^2m^3\omega_\perp/8)^{1/5}=1.43\hbar(\ell_\perp/a_\perp)^{2}$
(dotted line), which only depends on the $C_{3;0}$-coefficient of
the dipole-dipole interaction and the confinement along $z$, {\em
via} $a_\perp=(\hbar/m \omega_\perp)^{1/2}$.

For $\omega\geq\omega_{\rm c}'$ we find numerically that the
``bounce'' takes place in the plane $z=0$. This is consistent with
the existence of a single saddle point located at $V(\ell_\perp,0)$
for $\omega_\perp > \omega_{\rm c}$, as discussed in the previous
section. The ``bouncing point'' is at
$\rho(\tau=0)=\ell_\perp=r_\star/2^{1/3}$ and the action is $S_{\rm
E}=S_0 2^{5/3}\sqrt{\pi}\Gamma(7/6)/\Gamma(5/3)\approx 5.78~S_0$.
The latter is independent of $\omega_\perp$, which is again
consistent with the collisional dynamics being purely 2D.

The reason why the transition between the two behaviors of the
instanton solution happens at a value of $\omega_\perp$ which is
slightly different from $\omega_{\rm c}$ is that the instanton
solution accounts for the kinetic energy of the fictitious particle.
Thus, the particle is not always forced to follow the gradient of
the potential, see dash-dotted (red) lines in Fig.~\ref{fig:fig5}.
This results in the ``bounce'' occurring in the plane $z=0$ even for
values of $\omega_\perp$ slightly
smaller than $\omega_{\rm c}$.\\

From the discussion above it follows that in the limit of strong
interactions and tight transverse confinement $\Gamma$ rapidly tends
to zero. We illustrate this for the example of SrO, which has a
permanent dipole-moment of $d\approx 8.9~{\rm Debye}$ and mass
$m=104~{\rm amu}$. Then, for a tight transverse optical lattice with
harmonisc oscillator frequency $\omega_\perp=2\pi\times150{\rm kHz}$
and for a DC-field $\beta=dE_{\rm DC}/B=1/3$ we have
$(C_{3;0}^2m^3\omega_\perp/8\hbar^5)^{1/5}\approx 3.39$ and obtain
$\Gamma/\Gamma_0\approx e^{-5.86\times3.39}\approx 2\times10^{-9}$.
Even for DC field as weak as $\beta=dE_{\rm DC}/B=1/6$ we still
obtain a suppression by five order of magnitudes, as
$\Gamma/\Gamma_0\approx e^{-5.86\times1.94}\approx 10^{-5}$.

\subsubsection{Effective 2-D interaction}\label{sec:secEff2D}

\begin{figure}
  \begin{center}
    \includegraphics[width=\columnwidth]{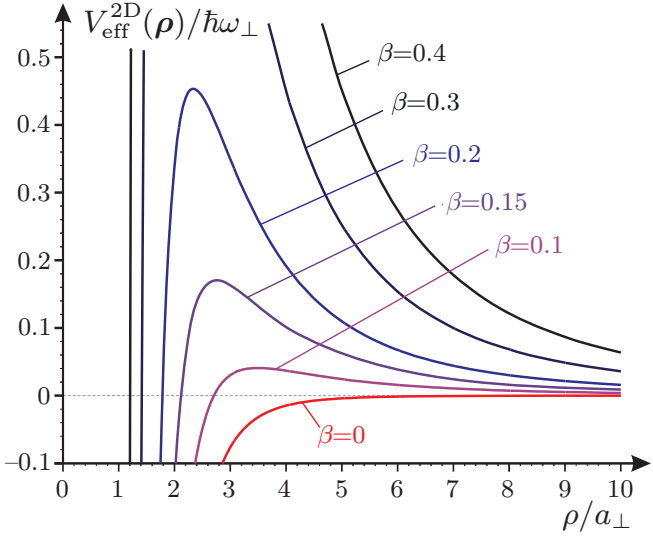}
  \end{center}
  \caption{\label{figs:fig8}(color online) The ground-state effective 2D
    BO-potentials $V_{\rm eff}^{\rm 2D}(\boldrho)$ of
    Eq.~\eqref{eq:eff2dpotential}
    as a function of the molecule separation
    $\rho$ in the ($z=0$)-plane for various strengths of the DC electric field
    $\beta=0,0.1,0.15,0.2,0.3,0.4$. The quantity $a_\perp$ is the harmonic oscillator length in the $z$-direction. The molecular parameters are chosen as
    $(d^4m^3B/\hbar^6)^{1/2} = 1.26\times10^6$ and the frequency of the harmonic potential in the $z$-direction is $\omega_\perp=15B/10^6\hbar$.
    This corresponds to the case of SrO with a mass $m=104~{\rm amu}$, a rotational constant $B\approx h~10~{\rm GHz}$ and a permanent dipole moment $d\approx 8.9~{\rm Debye}$ in a tight confining potential with $\omega_\perp=2\pi\times150{\rm kHz}$, where $a_\perp\approx 25{\rm
    nm}$.}
\end{figure}
In the limit of strong interactions and tight optical confinement,
it is possible to derive effective two-dimensional potentials by
integrating out the fast particle motion in the transverse direction
$z$.

For $r > \ell_\perp \gg a_\perp$, the two-particle eigenfunctions in
the $z$-direction approximately factorize into products of
single-particle harmonic oscillator wave-functions
$\psi_{k_1}(z_1)\psi_{k_2}(z_2)$. In first order perturbation theory
in $V_{\rm eff}^{\rm 2D}/\hbar\omega_\perp$, the effective 2D
interaction potential $V_{\rm eff}^{\rm 2D}$ reads
\begin{eqnarray}\label{eq:eff2dpotential}
V_{\rm eff}^{\rm 2D}(\boldrho) &\approx& \int dz_1dz_2
  \psi_0(z_1)^2\psi_0(z_2)^2 V_{\rm eff}^{\rm 3D}({\bf r})\nonumber\\
  &=& \frac{1}{\sqrt{2\pi}a_\perp}\int dz e^{-z^2/2a_\perp^2}V_{\rm eff}^{\rm 3D}({\bf
  r}).
\end{eqnarray}
Expression Eq.~\eqref{eq:eff2dpotential} is valid for large
separations $r > \ell_\perp \gg a_\perp$ where the potential is
(much) smaller than the harmonic oscillator spacing, i.e. $|V_{\rm
eff}^{\rm 2D}(\boldrho)|\ll\hbar\omega_\perp$. When this condition
breaks down, more harmonic oscillator states should be considered in
addition to the ground states $\psi_{0}(z_1)\psi_{0}(z_2)$ in
deriving $V_{\rm eff}^{\rm 2D}$ from Eq.~\eqref{eq:eqVeff3D}. In any
case, for large separations $\rho\gg \ell_\perp$ the 2D potential
reduces to
\[
V_{\rm eff}^{\rm 2D}(\boldrho)\approx V_{\rm eff}^{\rm
3D}(\rho,0)=\frac{C_{3;0}}{\rho^3}+\frac{C_{6;0}}{\rho^6}.
\]
\\

Finally in the adiabatic approximation we obtain the effective 2D
Hamiltonian $H_{\rm eff}^{\rm 2D}$
\begin{eqnarray}
H_{\rm eff}^{\rm 2D} = \sum_{j=1}^{2} \frac{\overline{{\bf
p}}_j^2}{2m}+V_{\rm eff}^{\rm 2D}(\boldrho),
\end{eqnarray}
where $\overline{{\bf p}}_j\equiv(p_{x;j},p_{y;j})$ is the
(two-dimensional) momentum in the plane $z=0$ of molecule $j=1,2$
and $\boldrho\equiv(x_2-x_1,y_2-y_1)$ is the (two-dimensional)
separation of the molecules in the plane $z=0$. The derivation of
$H_{\rm eff}^{\rm 2D}$ is the central result of Sec.~\ref{sec:secDC}.\\

\subsection{Effective interactions in the presence of an AC microwave
field}\label{sec:secAC}

In this section we consider the interactions of two polar molecules
in the presence of an AC microwave field of frequency $\omega$ and
polarization $q$, with respect to the direction of transverse
trapping ${\bf e}_z$, i.e. ${\bf E}_{\rm AC}(t) = E_{\rm
AC}e^{-i\omega t}{\bf
  e}_q + {\rm c.c.}$. The spatial dependence of ${\bf E}_{\rm AC}(t)$
is neglected, in accordance to the discussion of
Eq.~\eqref{eq:acsingle}. The field is blue-detuned from the
($J_j$=0-1) transition of the single-particle rotor spectrum by
$\Delta=\omega-2B/\hbar>0$, with Rabi-frequency $\Omega\equiv E_{\rm
AC} {}_j\bra{1,q}d_{q;j}\ket{0,0}_j / \hbar=dE_{\rm
  AC}/\sqrt{3}\hbar$.\\

The effects of the AC field on the two-particle scattering can be
summarized as: (a) Inducing {\em oscillating} dipole-moments in each
molecule, which determine long-range dipole-dipole interactions
whose sign and angular dependence are given by the polarization $q$
and the orientation in space, ${\bf e}_r$; (b) Inducing a coupling
of the ground and excited state manifolds of the {\em two-particle}
spectrum at a resonant (Condon) point $r_{\rm C}=(d^2/3h
\Delta)^{1/3}$, where the dipole-dipole interaction becomes
comparable to the detuning $\Delta$. This coupling is responsible
for an avoided crossing, whose properties depend crucially on the
polarization $q$. We show below that the character of the (3D)
ground-state effective interaction potential is very different at
distances larger and smaller than $r_{\rm C}$.

\begin{figure}[htb]
 \begin{center}
 \includegraphics[width=\columnwidth]{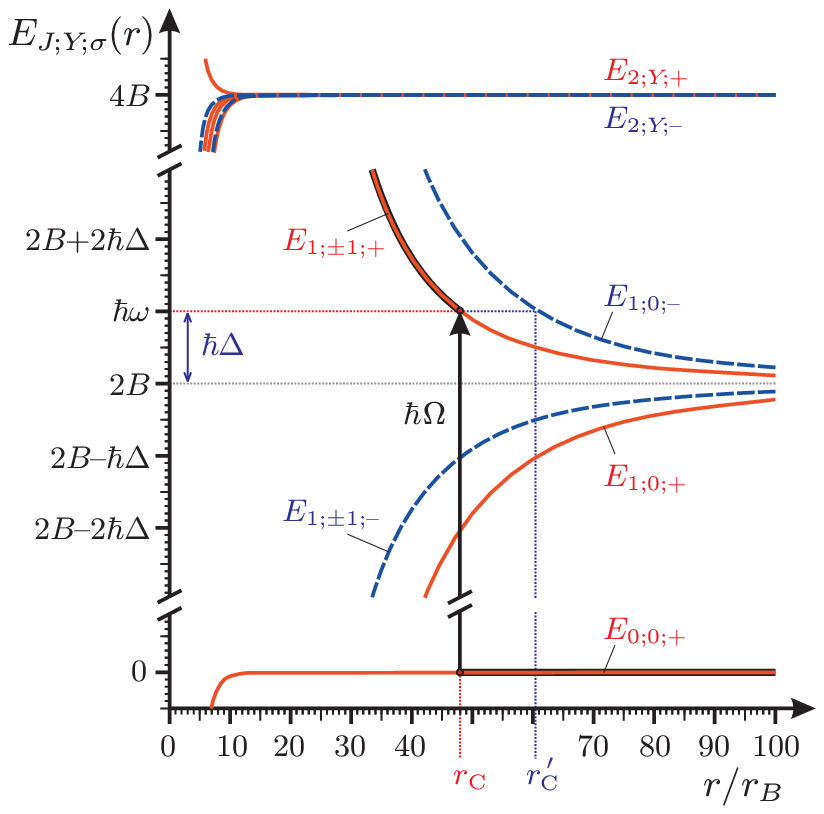}
  \caption{\label{fig:fig9} (color online) Schematic representation of the effects
    of an AC microwave field on the interaction of two molecules.
    The solid and dashed lines are the bare ($E_{\rm AC}=0$)-potentials
    $E_n(r)\equiv E_{J;Y;\sigma}(r)$ of Sec.~\ref{sec:secNoField}
    for the symmetric ($\sigma=+$) and antisymmetric ($\sigma=-$) states, respectively.
    An AC field of frequency $\omega=2B+\Delta$ is blue
    detuned by $\Delta=3 B/10^6\hbar$ from the single-particle rotational spacing
    $2B$, with Rabi-frequency $\Omega$.
    The dipole-dipole interaction splits the excited-state
    manifold, making the detuning position-dependent. Eventually, the
    combined energy of the bare ground-state potential $E_{0;0;+}(r)$
    and of an AC photon (vertical arrow) becomes degenerate with the energies of the
    bare symmetric $E_{1;\pm1;+}(r)$ and antisymmetric $E_{1;0;-}(r)$
    potentials. The corresponding resonant points are denoted as
    $r_{\rm C}=(d^2/3\hbar\Delta)^{1/3}$ (circles) and $r_{\rm C}'=(2d^2/3\hbar\Delta)^{1/3}$, respectively. The resulting {\em
    dressed} ground-state potential is sketched by a thick solid
    line. For molecular parameters of SrO ($B=h~10~{\rm GHz}$ and $d\approx 8.9~{\rm Debye}$) the detuning corresponds to $\Delta/2\pi=30~{\rm kHz}$, and the lengths $r_B$ and $r_{\rm C}$ are given by $r_B\approx11{\rm nm}$ and $r_{\rm C}\approx 0.5~\mu{\rm
    m}$, respectively.}
 \end{center}
 \end{figure}

\begin{figure}[htbp]
 \begin{center}
 \includegraphics[width=\columnwidth]{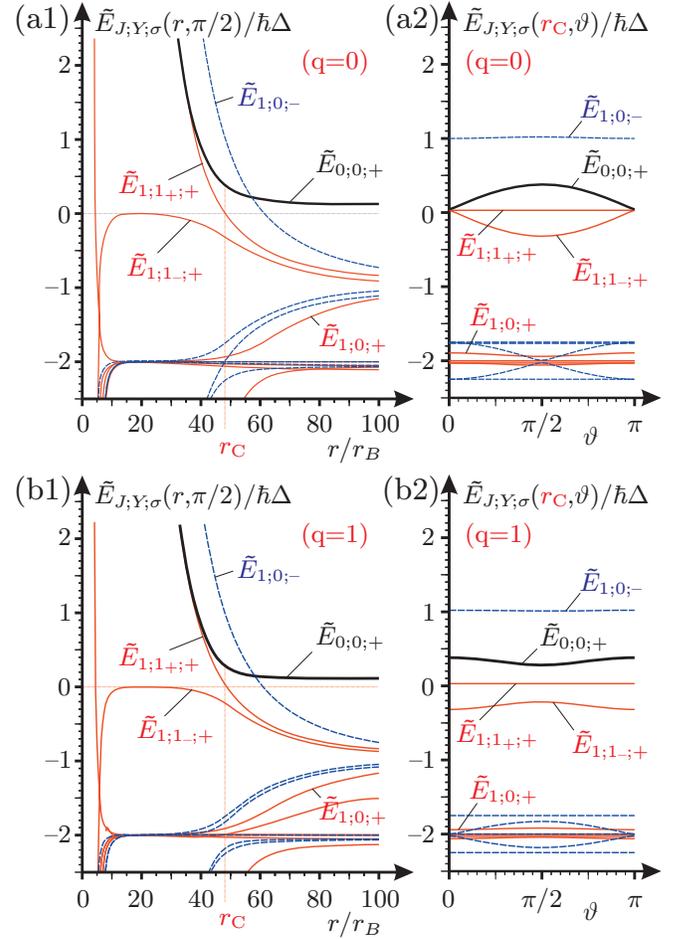}
  \caption{\label{fig:fig10}(color online) The dressed adiabatic potentials $\tilde
  E_n(r,\vartheta)$ of Eq.~\eqref{eq:dressed} for two molecules interacting in an AC
  field. The setup is the same as in Fig.~\ref{fig:fig9}.
  The field polarization is linear
    ($q=0$) in panels (a1) and (a2), while it is circular ($q=1$) in panels (b1) and (b2).
    The solid (red) and dashed (blue) lines correspond to the potentials
  for the symmetric ($\sigma=+$) and antisymmetric ($\sigma=-$)
  states, respectively. The thick continuous (black) line is the adiabatic dressed
  ground-state potential $\tilde{E}_{1;0;+}(r)$. Panels (a1) and (b1)
  show the potentials as a function of the
    separation $r$ for interactions in the ($z=0$)-plane
    ($\vartheta=\pi/2$). The position of the resonant Condon-point $r_{\rm C}$ is indicated
    by a vertical line. Panels (a2) and (b2) show the angular
    dependence of the potentials at $r=r_{\rm C}$.
    Note that for $q=0$ (panel (a2)) $\tilde{E}_{0;0;+}(r)$ becomes degenerate with
    $\tilde{E}_{1;1_{\pm};+}(r_{\rm C},\vartheta)$ at $\vartheta=0,\pi$,
    while it is non-degenerate at all angles for $q=1$ (panel (b2)),
    suggesting better shielding.
    The potential $\tilde{E}_{1;0;-}(r_{\rm C}, \vartheta)$
    has an energy larger than $\tilde{E}_{1;0;+}(r_{\rm C}, \vartheta)$
    for all angles $\vartheta$, indicating a level crossing ar $r>r_{\rm C}$ (see text).}
 \end{center}
 \end{figure}

The basic features of the scattering in the presence of the AC field
are depicted in Fig.~\ref{fig:fig9}. In the figure, the solid
(dashed) lines are the bare $(E_{\rm AC}=0)$-eigenvalues
$E_{J;Y;\sigma}({\bf r})$ of Eq.~\eqref{eq:eqInt} for $\sigma=+$
($\sigma=-$), plotted as a function of $r$. The color conventions
are the same as in Fig.~\ref{fig:fig3}. The microwave field, which
is detuned from the single-particle rotational spacing $2B$ by an
amount $\hbar\Delta>0$, is represented by a black arrow. Analogous
to Fig.~\ref{fig:fig3}, the excited-state manifold is split by the
dipole-dipole interaction. This splitting has the effect to render
the detuning position-dependent, so that eventually the combined
energy of the bare ground-state plus a microwave photon becomes
degenerate with the energies of some bare excited states. The
resonant points are denoted as $r_{\rm C}$ and $r_{\rm C}'$ for the
resonance with two symmetric ($\ket{\Phi_{1;1_{\pm};+}({\bf
r})}\equiv \ket{\Phi_{1;+1;+}({\bf r})}\pm\ket{\Phi_{1;-1;+}({\bf
r})}$) and an anti-symmetric state ($\ket{\Phi_{1;0;-}({\bf r})}$),
respectively. The symmetric bare ground-state is coupled by the AC
field to the symmetric bare excited-state $\ket{\Phi_{1;1_-;+}({\bf
r})}$ only, while the state $\ket{\Phi_{1;1_+;+}({\bf r})}$ is dark.
As it is explained below, this coupling induces a splitting of the
field-dressed energy levels at $r_{\rm C}$. Due to this coupling,
the 3D effective {\em dressed adiabatic} ground-state interaction
potential inherits the character of the bare ground and excited
potentials for $r \gg r_{\rm C}$ and $r \ll r_{\rm C}$, respectively
(thick solid line in the figure). Since the symmetric excited-state
potential is {\em repulsive}, during the collision the dynamics of
the particle is confined to the region $r \geq r_{\rm C}$, that is,
the AC coupling can determine an effective ``shielding'' of the
inner part of the molecular interaction potentials (the molecular
core of Fig.~\ref{fig:fig3}). This shielding is three-dimensional
and it is analogous to the optical shielding of Napolitano, Weiner
and Julienne developed in the context of ultra-cold atomic
collisions \cite{Napolitano97,Weiner99}. In particular, we show
below that the shielding efficiency depends strongly on the chosen
polarization $q$ of the AC field (see Fig.~\ref{fig:fig10} and
text), a characteristic which was found both in theory and in
experiments with cold atoms \cite{Napolitano97,Zilio96}.

Analogous to the optical shielding case, one expects that diabatic
couplings among symmetric states provide for a loss mechanism in the
3D ground-state collision, for any finite collisional energies. In
particular diabatic couplings, and therefore losses, are expected to
be particularly relevant in the region $r\approx r_{\rm C}$ and $r<
r_{\rm C}$, where the ground-state energy shows an avoided crossing
with another potential, and the ground-state energy becomes doubly
degenerate, respectively (see Figs.~\ref{fig:fig9} and
\ref{fig:fig10}). When a harmonic confinement in the $z$-direction
is considered, other loss channels may arise due to residual
non-compensated tensor-shifts Eq.~\eqref{eq:Tensor}, coupling the
ground-state to the anti-symmetric state $\ket{\Phi_{1;0;-}({\bf
r})}$, whose energy $E_{1;0;-}({\bf r})$ crosses the ground-state
potential at $r_{\rm C}'=(2d^2/3\hbar\Delta)^{1/3}$ (see
Fig.~\ref{fig:fig9}). When more particles are considered, three-body
interactions are expected to generate similar couplings to the
anti-symmetric state. Three-body interactions are of concern since,
as noted in Sec.~\ref{sec:secIntro}, we are interested in designing
effective two-dimensional interaction potentials for pairs of
molecules, which can lead to the realization of interesting phases
for an ensemble of polar molecules in the {\em strongly interacting}
regime (see Fig.~\ref{fig:fig1} and Ref.~\cite{Buechler07}).

Because of all these loss mechanisms, {\em two-dimensional}
shielding is not expected to be very efficient in the case of
interactions in an AC field. However, in Sec.~\ref{Sec:CouplingDCAC}
we show that most of these losses can be avoided, and an efficient
2D shielding recovered, by utilizing a properly chosen combination
of static and microwave fields, and a tight harmonic
confinement in the $z$-direction.\\

In the remainder of this section we further detail the interaction
processes. This analysis is instrumental to the discussion of the
collisions of two particles in the presence of both static and
microwave fields, which is addressed in Sec.~\ref{Sec:CouplingDCAC}.

In Sec.~\ref{sec:secAC1} we derive the dressed adiabatic potentials
for the interaction of two particles in an AC field. There, we show
that the shielding is strongly dependent on the chosen polarization
of the AC field. In fact, for linear polarization ($q=0$) the width
of the avoided crossing at the Condon point $r_{\rm C}$ between the
ground-state potential $E_{0;0;+}({\bf r})$ and the potential
$E_{1;+1;+}({\bf r})$ is dependent on the value of the polar angle
$\vartheta$, and it vanishes for $\vartheta=0$ and $\vartheta=\pi$.
This vanishing of the width of the avoided crossing entails the
existence of ``holes'' in the three-dimensional shielding, which
allow for reaching the molecular-core region [see
Fig.~\ref{fig:fig10}(a1) and Fig.~\ref{fig:fig10}(a2)]. On the other
hand, for circular polarization ($q=1$) the potential is repulsive
in three-dimensions [see Fig.~\ref{fig:fig10}~(b1) and
Fig.~\ref{fig:fig10}~(b2)]. Diabatic losses are most likely to occur
at the Condon point $r_{\rm C}$, and for $r<r_{\rm C}$ due to
couplings to the dark state $\ket{\Phi_{1;1_+;+}({\bf r})}$, which
becomes degenerate with the ground-state. In Sec.~\ref{sec:secAC2}
and Sec.~\ref{sec:secAC3} the interaction is further analyzed by
deriving a perturbative expansion for the ground-state potential
valid to second order in $\Omega/\Delta$, and by analyzing a reduced
model Hamiltonian valid in the vicinity of $r_{\rm C}$,
respectively. There, it is argued that couplings to the
antisymmetric manifold due to three-body interactions (and to the
possible existence of residual tensor shifts for a harmonic
confinement) reduce the efficiency of the shielding.

\subsubsection{Adiabatic potentials}\label{sec:secAC1}

The total Hamiltonian for the collision of two particles in the
presence of an AC field is
\begin{eqnarray}
H(t) =\sum_{j=1}^2 \left[\frac{{\bf p}_j^2}{2m} +\frac{1}{2}
m\omega_\perp^2z_j^2\right] + H_{\rm int}({\bf r},t),
\end{eqnarray}
with
\begin{eqnarray}
H_{\rm int}({\bf r},t)=\sum_{j=1}^2
\left[B{\bf J}^2_j - \left(E_{\rm
    AC}e^{-i\omega t}d_{q,j}+{\rm h.c.}\right)\right] + V_{\rm
dd}({\bf r}).\nonumber\\
\end{eqnarray}
Analogous to the discussion of Sec.~\ref{sec:secDC},  $H_{\rm
  int}({\bf r},t)$ entirely determines the non-trivial dynamics of the system, since
  the harmonic motion of the center of mass is decoupled from the
  relative motion. The permutation symmetry  $\sigma=\pm$ is conserved during the
  collision, since $H_{\rm int}({\bf r},t)$ is
invariant under the exchange of the position of the two molecules
$(j=1)\leftrightarrow(j=2)$, i.e.~${\bf r}\rightarrow -{\bf r}$.
Thus, $H_{\rm int}({\bf r},t)$ can be conveniently rewritten as
$H_{\rm int}({\bf r},t)=\sum_{\sigma=\pm}P_\sigma H^{(\sigma)}_{\rm
int}({\bf r},t)P_\sigma$, where $P_+$ and $P_-$ denote the projector
onto the symmetric and antisymmetric manifolds, respectively.

We obtain the solution of the time-dependent problem is obtained
analogous to Sec.~\ref{bla} by diagonalizing the Hamiltonian $H_{\rm
int}({\bf r},t)$ in a Floquet picture and proceeds as follows: First
we diagonalize the Hamiltonian in the absence of the AC field,
$E_{\rm AC}=0$, as $H_{\rm int}({\bf r})=\sum_n\ket{\Phi_n({\bf
r})}E_n({\bf
  r})\bra{\Phi_n({\bf r})}$ with $n=(J;Y;\sigma)$, which is the same as
  Eq.~\eqref{eq:eqInt} and in particular is {\em time-independent} (see also Table~\ref{tab:tab2}).
  Then, we consider the effect of the AC field, ${\bf E}_{\rm
  AC}(t)$, via a transformation to the Floquet picture, which is
  obtained by expanding the time-dependent wave-function in a
  Fourier series in the AC frequency $\omega$. After applying a rotating wave approximation, i.e. keeping
only the energy conserving terms, we obtain the {\em
time-independent} Hamiltonian $\tilde H({\bf r})$, which describes
the driven system. The Hamiltonian preserves the permutation
symmetry, $\sigma=\pm$, i.e. $\tilde{H}_{\rm
  int}({\bf r})=\sum_\sigma P_\sigma \tilde{H}_{\rm int}^{(\sigma)}({\bf r})
  P_\sigma$. Analogous to the zero-field case of Sec.~\ref{sec:secDC},
  we restrict the basis set to the 16 states belonging to
  the three lowest-energy manifolds. This is obtained
  by choosing a detuning much smaller than
  the rotational spacing $\Delta \ll B$, and working in the
  regime of weak saturation $\Omega \ll \Delta$. In fact, in this
  limit the anharmonicity of the single-particle rotational spectrum
  ensures that the population of high-energy rotational states is negligible.
  Finally, we solve for $\tilde H_{\rm int}({\bf r})$ by diagonalizing
  the Hamiltonian in the symmetric and antisymmetric subspaces separately,
  e.g. $\tilde H^{(\sigma)}_{\rm int}({\bf
  r})$.\\

The Hamiltonian $\tilde{H}^{(+)}_{\rm int}({\bf  r})$ for the
symmetric subspace expressed in the basis $\ket{\Phi_{J;Y;+}({\bf
r})}$ with $(J;Y)=\left\{(0;0),(1;Y)|_{Y=-1,0,1},
(2;Y)|_{Y=-2,-1,0_+,0_-,1,2}\right\}$ reads
\begin{widetext}
\begin{eqnarray}\label{eq:eq27}
\tilde{H}_{\rm int}^{(+)}({\bf r}) =
-\hbar\left[\begin{array}{c|ccc|ccccccc}
\Delta_{0;0}^{(+)}&\sqrt{2}\Omega_-^*&\sqrt{2}\Omega_0^*&\sqrt{2}\Omega_+^*&
0&0&0&0&0&0\\\hline \sqrt{2}\Omega_-&\Delta_{1;-1}^{(+)}&0&0&
\sqrt{2}\Omega_-^*&\Omega_0^*&c_+\Omega_+^*&-c_-\Omega_+^*&0&0\\
\sqrt{2}\Omega_0&0&\Delta_{1;0}^{(+)}&0&
0&-\Omega_-^*&c_-\sqrt{2}\Omega_0^*&c_+\sqrt{2}\Omega_0^*&\Omega_+^*&0\\
\sqrt{2}\Omega_+&0&0&\Delta_{1;1}^{(+)}&
0&0&c_+\Omega_-^*&-c_-\Omega_-^*&\Omega_0^*&\sqrt{2}\Omega_+^*\\\hline
0&\sqrt{2}\Omega_-&0&0&\Delta_{2;+2}^{(+)}&0&0&0&0&0\\
0&\Omega_0&\Omega_-&0&0&\Delta_{2;+1}^{(+)}&0&0&0&0\\
0&c_+\Omega_+&\sqrt{2}c_-\Omega_0&c_+\Omega_-&0&0&\Delta_{2;0_+}^{(+)}&0&0&0\\
0&-c_-\Omega_+&\sqrt{2}c_+\Omega_0&-c_-\Omega_-&0&0&0&\Delta_{2;0_-}^{(+)}&0&0\\
0&0&\Omega_+&\Omega_0&0&0&0&0&\Delta_{2;+1}^{(+)}&0\\
0&0&0&\sqrt{2}\Omega_+&0&0&0&0&0&\Delta_{2;+2}^{(+)}
\end{array}\right].
\end{eqnarray}
The Hamiltonian $\tilde{H}^{(-)}_{\rm int}({\bf  r})$ for the
antisymmetric subspace on the basis $\ket{\Phi_{J;Y;-}({\bf r})}$
with $(J;Y)=\left\{(1;Y)|_{Y=-1,0,1},(2;Y)|_{Y=-1,0,1}\right\}$)
reads
\begin{eqnarray}\label{eq:eq28}
\tilde{H}_{\rm int}^{(-)}({\bf r}) =
-\hbar\left[\begin{array}{ccc|ccc}
\Delta_{1;-1}^{(-)}&0&0&-\Omega_0^*&-\Omega_+^*&0\\
0&\Delta_{1;0}^{(-)}&0&\Omega_-^*&0&-\Omega_+^*\\
0&0&\Delta_{1;+1}^{(-)}&0&\Omega_-^*&-\Omega_0^*\\\hline
-\Omega_0&\Omega_-&0&\Delta_{2,-1}^{(-)}&0&0\\
-\Omega_+&0&\Omega_-&0&\Delta_{2,0}^{(-)}&0\\
0&-\Omega_-&\Omega_0&0&0&\Delta_{2,+1}^{(-)}
\end{array}\right].
\end{eqnarray}
\end{widetext}
In Eqs.~\eqref{eq:eq27} and \eqref{eq:eq28} off-resonant couplings
of order ${\cal O}(\Omega d^2/r^3B)$ between the various
$J$-manifolds have been neglected. The detunings
$\Delta_{J;Y}^{(\sigma)}$ and couplings $\Omega_Y$ in
Eq.~\eqref{eq:eq27} and Eq.~\eqref{eq:eq28} depend on the separation
${\bf r}$ of the two molecules as
\begin{eqnarray}
\Delta_{J;Y}^{(\sigma)}&\equiv&\Delta_{J;Y}^{(\sigma)}(r)
=J\Delta-E_{J;Y;\sigma}(r)/\hbar\\
\Omega_Y&\equiv&\Omega_{Y}(\vartheta,\varphi)
=\Omega D_{q,Y}^1(\varphi,\vartheta,0)^*.\label{eq:OmegaY}
\end{eqnarray}
Here $D_{q,Y}^1(\varphi,\vartheta,0)\equiv\bra{1,q}\exp(-i\varphi
J_z)\exp(-i\vartheta J_y)\ket{1,Y}$ are matrix-elements of
rotation-operator, which rotates the lab-frame onto the frame where
the collision axis is fixed along ${\bf e}_0$. The coefficients
$c_{\pm}$ are
$c_\pm=[(1\pm1/\sqrt{3})/2]^{1/2}$.\\

As said above, a set of dressed BO-potentials $\tilde{E}_n({\bf r})$
and of adiabatic eigenstates $\ket{\tilde{\Phi}_n({\bf r})}$ is
obtained by diagonalizing the Hamiltonian $\tilde{H}_{\rm int}$ as
\begin{eqnarray}\label{eq:dressed}
\tilde{H}_{\rm int}({\bf r}) = \sum_{n} \ket{\tilde\Phi_n({\bf r})}
\tilde E_{n}({\bf r}) \bra{\tilde\Phi_n({\bf r})},
\end{eqnarray}
with $n=(J;Y;\sigma)$. The tilde refers to the implicit dependence
of the dressed potentials and eigenstates on the Rabi-frequency
$\Omega$, the polarization $q$ and the detuning $\Delta$ of the
external AC-field. As mentioned above, we focus on blue detunings
$\Delta=\omega-2B/\hbar>0$, since we are interested in repulsive
potentials which can ``shield'' the short-range molecular-core
interaction.

\subsubsection{Asymptotic expansion: $r \gg r_{\rm C}$}\label{sec:secAC2}
An insight into the nature of the dressed ground state potential can
be obtained by deriving an expression for $\tilde E_0({\bf r})\equiv
\tilde{E}_{0;0;+}({\bf r})$ perturbatively in the small parameter
$\Omega/\Delta$. The perturbative expansion is valid at separations
$r\gg r_{\rm C}\equiv (d^2/\hbar\Delta)^{1/3}$, where the
dipole-dipole interaction in the first excited manifold
$n=(1;Y;\sigma)$ is smaller than the detuning of the AC field. Then,
to second order in $\Omega/\Delta$ the dressed ground-state
potential reads
\begin{eqnarray}\label{eq:dressedACgroundstatepotential}
  \tilde E_0({\bf r}) &\approx& -\hbar\Delta_{0;0}^{(+)}(r)
   + \hbar\sum_{Y=-1}^{+1}
  \frac{2|\Omega_\Lambda(\vartheta,\varphi)|^2}{\Delta_{1;Y}^{(+)}(r)
  -\Delta_{0;0}^{(+)}(r)}\nonumber\\
&\approx& +\frac{2\hbar|\Omega|^2}{\Delta} -
  \frac{2\hbar|\Omega|^2}{\Delta^2}\frac{d^2(2-3q^2)}{3r^3}
  (1-3\cos^2\vartheta),\nonumber\\
\end{eqnarray}
where terms of order ${\cal O}(1/r^6)$ have been neglected. The
first term in Eq.~\eqref{eq:dressedACgroundstatepotential} describes
a quadratic single-molecule AC-Stark shift, which is positive for
blue detunings. The second term is understood as follows: The
AC-field induces in each molecule an oscillating dipole-moment of
magnitude $\langle {\bf d}_j\rangle\sim d\Omega{\bf e}_q/\Delta$,
and on average the oscillating dipoles give rise to an effective
dipole-dipole interaction in the ground state, which is proportional
to $\propto\langle {\bf d}_j\rangle^2/r^3$.
Equation~\eqref{eq:dressedACgroundstatepotential} shows that the
overall sign of the induced interaction can be changed by varying
the polarization $q$.

The perturbative expression for the ground-state potential breaks
down at $r\sim r_{\rm C}=(d^2/3 \hbar \Delta)^{1/3}$, where two of
the bare ($J=1$)-excited potentials ($E_{1;\pm1;+}({\bf r})$) become
degenerate with the energy of the ground-state plus a photon of
frequency $\omega$.

While the validity of perturbation theory ceases at $r \sim r_{\rm
C}$, further insight into the solution of the adiabatic scattering
problem can be obtained by direct inspection of a specific example.
Figure~\ref{fig:fig10} shows the dressed BO-potentials
$\tilde{E}_n({\bf r})$ of Eq.~\eqref{eq:dressed} for
$\Delta=3B/10^6\hbar$ and $\Omega=\Delta/4$. The polarization is
linear, $q=0$, in panels (a1, a2), while it is circular in panels
(b1, b2), with $q=+1$. Panels (a1) and (b1) show $\tilde{E}_n({\bf
r})$ as a function of the separation $r$ for collisions in the plane
$\vartheta={\rm arccos}(z/r)=\pi/2$. Panels (a2) and (b2) depict the
angular dependence of $\tilde{E}_n({\bf r})$ at the Condon point
$r=r_{\rm C}$ for the two polarizations $q=0$ and $q=1$,
respectively. In all the panels, the solid and dashed
 lines denote symmetric and antisymmetric potentials,
respectively, while the dressed ground-state potential
$\tilde{E}_0({\bf r})=\tilde{E}_{0;0;+}({\bf r})$ is represented by
a thick solid line. Since we are interested in ground-state
collisions, the figure suggests the two following observations:
First, while the potentials $\tilde{E}_0({\bf r})$ is strongly
repulsive for $r<r_{\rm C}$ and $\vartheta = \pi/2$, both for $q=0$
and $1$ [panels (a1) and (b1)], the angular dependence at $r=r_{\rm
C}$ is very different [panels (a2) and (b2)]. In particular, panel
(a2) shows that for $q=0$ the repulsive potential is a maximum at
$\vartheta = \pi/2$, while it vanishes at $\vartheta = 0$ and
$\vartheta=\pi$. This vanishing of the repulsion allows for the
molecules to approach the molecular-core region, and thus the
polarization $q=0$ does not provide for an efficient
three-dimensional shielding of the molecular-core region. On the
other hand, panel (b2) shows that the shielding may in principle
work for $q=1$, since the ground-state potential is repulsive for
any angles. The second observation is that a level crossing of the
ground-state potential with the antisymmetric potential
$\tilde{E}_{1,0,-}({\bf r})$ appears at $r_{\rm C}'=2^{1/3}r_{\rm
C}$ for all polarizations [panels (a1) and (b1)]. Couplings to this
state can arise due to non-compensated residual tensor-shifts, when
a harmonic confinement in the $z$-direction is considered, or due to
three-body interactions, when an ensemble of polar molecules is
considered. These couplings will induce losses in the ground-state
interaction. In Sec.~\ref{Sec:CouplingDCAC} we show that the
position of this level crossing can be shifted to distances $r\ll
r_{\rm C}$, and the associated losses can be avoided, by
superimposing a weak static electric field to the AC field. In this
way, an efficient (2D) shielding of the
molecular-core region can be recovered.\\

\subsubsection{Resonant Condon-point: $r\sim r_{\rm C}$ }\label{sec:secAC3}

In the remainder of this section we analyze further the scattering
process at the resonance point $r_{\rm C}$. We restrict the
discussion to the three relevant states $\{\ket{\Phi_{0;0;+}({\bf
r})},\ket{\Phi_{1;+1;+}({\bf r})},\ket{\Phi_{1;-1;+}({\bf r})}\}$,
since all other symmetric states of the ($J=1$)-manifold are detuned
by $\Delta_{J;Y}^{(\sigma)}(r)\geq\Delta\gg\Omega$. In this subspace
the Hamiltonian Eq.~\eqref{eq:eq27} reads
\begin{eqnarray}
\tilde{H}_{\rm int}({\bf r}) = -\hbar\left[\begin{array}{ccc}
\Delta_{0;0}^{(+)}(r)&\Omega_-(\vartheta,\varphi)^*&\Omega_+(\vartheta,\varphi)^*\\
\Omega_-(\vartheta,\varphi)&\Delta_{1;1}^{(+)}(r)&0\\
\Omega_+(\vartheta,\varphi)&0&\Delta_{1;-1}^{(+)}(r)
\end{array}\right].
\end{eqnarray}

For $q=0$ we have
$\Omega_{\pm}(\vartheta,\varphi)=\mp\Omega\sin\vartheta/\sqrt{2}$
and the ground-state couples only to the bright superposition state,
$\ket{\Phi_{1;-1;+}({\bf r})}-\ket{\Phi_{1;+1;-}({\bf r})}$, with
coupling $\sqrt{2}\Omega\sin\vartheta$. The orthogonal state,
$\ket{\Phi_{1;-1;+}({\bf r})}+\ket{\Phi_{1;+1;-}({\bf r})}$, is dark
with respect to the AC coupling. The dressed ground-state is then a
(position-dependent) superposition of bare ground and excited states
and the corresponding dressed potential is
\begin{eqnarray}
\tilde{E}_{0;0;+}({\bf r})/\hbar=-\Delta_+(r)
+\sqrt{\Delta_-(r)^2+2|\Omega|^2\sin^2\vartheta},\label{eq:linearcase1}
\end{eqnarray}
with $\Delta_\pm(r)\equiv[\Delta_{1;1}^{(+)}(r)
\pm\Delta_{0;0}^{(+)}(r)]/2\approx\Delta-d^2/3\hbar r^3$. We notice
that for $q=0$ and $\vartheta=\pi/2$ the potential is repulsive with
a $\sim 1/r^3$ radial dependence, due to the avoided crossing at
$r=r_{\rm C}$, see Fig~\ref{fig:fig10}(a1). For $\vartheta \neq
\pi/2$ the splitting of the avoided crossing decreases as
$\sin\vartheta$ and vanishes at $r=r_{\rm
  C}$ for $\vartheta=0$ and $\vartheta=\pi$, see Fig.~\ref{fig:fig10}(a2).
Thus, close to the point ${\bf r}=r_{\rm C}{\bf e}_0$ the molecules
can penetrate the 3D ``shield'' provided by the AC field and
approach the short-range molecular-core region, $r \ll r_{\rm C}$.
This behavior resembles the one encountered in Sec.~\ref{sec:secDC}
for the collision of two dipoles polarized by a DC field, when the
intermolecular axis is parallel to the direction of the DC field,
see Fig.~\ref{fig:fig4}~(e). However, this unstable region now
appears at distances $r\sim r_{\rm
C}\approx(d^2/3\hbar\Delta)^{1/3}$, which are larger than the
short-distance length $r_B=(d^2/B)^{1/3}$ by a factor
$\sim(B/\hbar\Delta)^{1/3}$. For a detuning on the order of tens of
kHz and a rotational spacing of tens of GHz, $r_{\rm C}$ is two
orders of magnitudes larger than $r_B$.

For a circularly polarized field $|q|=1$, we have
$\Omega_\pm(\vartheta,\varphi)=e^{iq\varphi}(1\mp q\cos\vartheta)/2$
and hence the ground-state couples to the bright superposition
state, $\cos^2(\vartheta/2)\ket{\Phi_{1;q;+}({\bf r})}
+\sin^2(\vartheta/2)\ket{\Phi_{1;q;+}({\bf r})}$, with an amplitude
$\sqrt{2}\Omega$ which is now independent of the angle $\vartheta$.
The orthogonal superposition,
$\cos^2(\vartheta/2)\ket{\Phi_{1;q;+}({\bf r})}
-\sin^2(\vartheta/2)\ket{\Phi_{1;q;+}({\bf r})}$, is dark with
respect to the AC-coupling. The dressed ground-state potential is
\begin{eqnarray}
\tilde{E}_{0;0;+}({\bf r})/\hbar=-\Delta_+(r)
+\sqrt{\Delta_-(r)^2+2|\Omega|^2},
\end{eqnarray}
with $\Delta_\pm(r)$ defined as in Eq.~\eqref{eq:linearcase1}. The
behavior of the ground-state potential in the ($\vartheta=0$)-plane
is analogous to the linearly polarized case, see
Fig.~\ref{fig:fig10}(a1,b1). However, in contrast to the $q=0$ case,
now the width of the avoided crossing remains finite at all angles
[see Fig.~\ref{fig:fig10}(b2)], and the AC shielding of the
molecular core is effective.

However, as noted above, the pure AC shielding mechanism has an
intrinsic flaw that limits its utility, once an ensemble of polar
molecules is considered. The anti-symmetric state
$\ket{\Phi_{1,0,-}({\bf r})}$ is strongly repulsive with energy
$-\hbar\Delta_{1,0}^{(-)}(r)\approx 2d^2/3r^3-\hbar\Delta$ and thus
gives rises to a real crossing at $r_{\rm
C}'=(2d^2/3\hbar\Delta)^{1/3}=2^{1/3}r_{\rm C}$ (see dotted lines in
Fig.~\ref{fig:fig10}). This crossing at distances {\em larger} than
$ r_{\rm C}$ is expected to give rise to (strong) collisional losses
when an ensemble of polar molecules is considered. In fact,
three-body interactions can couple the ground-state to the
antisymmetric $\ket{\Phi_{1,0,-}({\bf r})}$-state. In addition,
analogous couplings can be provided by residual non-compensated
tensor shifts, when a harmonic confinement in the $z$-direction is
considered. In the next section we explain how some of these
problems can be circumvented by introducing an additional static
electric field. In that case, an efficient and collisionally stable
$2D$ shielding of the inner part of the potential can be recovered.

\subsection{Effective interactions in the presence of both a DC and an AC
fields}\label{Sec:CouplingDCAC}

 \begin{figure*}[htb]
 \begin{center}
 \includegraphics[width=.9\textwidth]{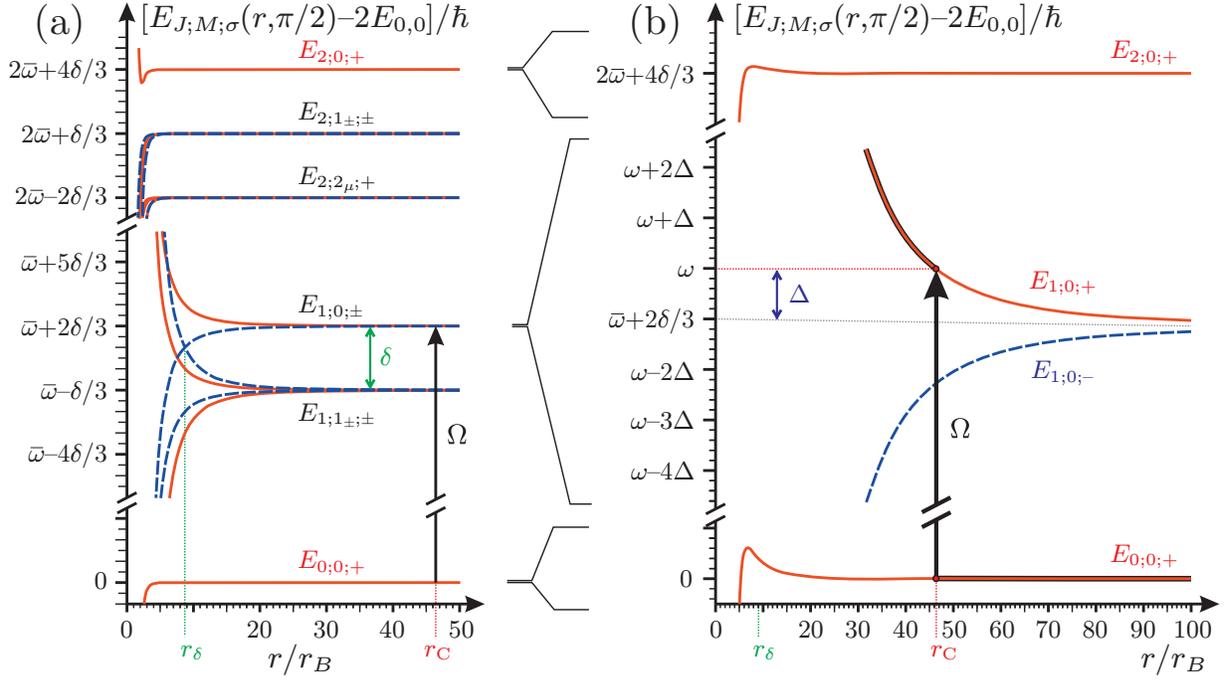}
  \caption{\label{fig:fig11}(color online) (a) Schematic representation of the effects
    of a DC and an AC microwave fields on the interaction of two molecules.
    The solid and dashed lines are the bare
    potentials $E_n({\bf r})\equiv E_{J;M;\sigma}(r,\vartheta)$ of
    Sec.~\ref{sec:secDCField} with $\vartheta=\pi/2$
    for interactions in the presence of the DC field only,
    for the symmetric ($\sigma=+$) and antisymmetric ($\sigma=-$)
    states,
    respectively. The DC field induces a splitting $\hbar\delta$
    of the first-excited manifold of the two-particle spectrum.
    A microwave-field of frequency $\omega=\overline{\omega}+2\delta/3+\Delta$ is blue
    detuned by $\Delta>0$ from the single-particle rotational
    resonance. The dipole-dipole interaction further splits the excited-state
    manifold, making the detuning space-dependent. Eventually, the
    combined energy of the bare ground-state potential $E_{0;0;+}({\bf r})$
    and of an AC photon (black arrow) becomes degenerate with the energy of the
    bare symmetric $E_{1;0;+}(r,\pi/2)$. The resonant point $r_{\rm C} =
(d^2/3 \hbar\Delta)^{1/3}$ occurs at $r\approx46~r_B$. A second
    resonant Condon point occurs at (much) shorter distances
    $r_{\rm C'}\lesssim r_{\delta}=(d^2/\hbar\delta)^{1/3}$
    with an anti-symmetric potential (not shown). (b) Blow-up of the
    potentials of panel (a) with $M=0$ (see text in Sec.~\ref{sec:secACDC2}).
    The {\em dressed} ground-state potential is sketched by a thick solid
    line.}
 \end{center}
 \end{figure*}

In this section we consider the interactions of two polar molecules
in the presence of {\em both} a weak DC field ${\bf E}_{\rm
DC}=E_{\rm DC}{\bf
  e}_0$ with $\beta\equiv dE_{\rm DC}/B\ll 1$ {\em and} of an AC
microwave field ${\bf E}_{\rm AC}(t)=E_{\rm
  AC}e^{-i\omega t}{\bf e}_q+{\rm c.c.}$, where the polarization $q$ is defined
  with respect to the $z$-direction, $\omega$
  is the frequency, $\Delta $ is the detuning from the
  single-particle resonance, and $\Omega$ is the Rabi frequency.

As explained in Sec.~\ref{Sec:CouplingDC}, the DC field partially
splits the three-fold degeneracy of the $(J_j=1)$-manifold of each
molecule by an amount $\sim\hbar\delta=3d^2E_{\rm DC}^2/20B$ (the
modulus of the projection $|M|$ is conserved). When the AC field is
superimposed to the weak DC field, this splitting can yield
significative advantages regarding the stability of the ground-state
collision: (a) The ground-state can couple to a single
non-degenerate excited state of the {\em two-particle} spectrum,
thus avoiding diabatic losses due to the presence of symmetric
(dark) states close to the ground-state for $r\lesssim r_{\rm C}
\sim (d^2/3\hbar\Delta)^{1/3}$ (see discussion in
Sec.~\ref{sec:secAC}). In fact, because of the splitting $\hbar
\delta$, the energies of other symmetric states become comparable to
the {\em dressed} ground-state energy only at distances $r \sim
r_\delta \equiv (d^2/\hbar\delta)^{1/3} \ll r_{\rm C}$, where the
dipole-dipole interaction becomes of the order of the splitting
$\hbar \delta $. (b) The location $r_{C}'$ of the real crossing of
Sec.~\ref{sec:secAC} is also shifted to small distances $r \lesssim
r_\delta \ll r_{\rm C}$, thus suppressing losses due to
three-body-induced (or residual-tensor-shift-induced due to
transverse confinement) couplings to the ground-state.

Both of the outlined processes are shown in Fig.~\ref{fig:fig11}(a)
and Fig.~\ref{fig:fig11}(b), which depict the bare ($E_{\rm AC}=0$)
energy levels of the two-particle eigenstates with
 $J\leq 2$ as a function of the distance $r$, for $\beta=1/10$. The
polarization of the AC-field is $q=0$ and its frequency $\omega$ is
blue-detuned from the ($\ket{\phi_{0,0}} \rightarrow
\ket{\phi_{1,0}}$)-transition of the single-particle spectrum by an
amount $\Delta=\omega-(\overline{\omega}+2\delta/3)>0$. In the
figure, the continuous (red) and dashed (blue) curves are the bare
BO-potentials for the symmetric and antisymmetric states,
respectively. The  presence of the AC-field is signaled by a black
arrow at the resonant (Condon) point $r_{\rm C} \sim
(d^2/3\hbar\Delta)^{1/3}$.
 Analogous to the case of Fig.~\ref{fig:fig4}(b),
Fig.~\ref{fig:fig11}(a) shows that the ($J=1$)-manifold is
asymptotically split by the DC-Stark-shift $\hbar\delta$. As said
above, for $r\gg r_\delta$ this splitting suppresses the coupling
among states of the $(J=1)$-manifold due to dipole-dipole
interactions. Moreover, we note that since the characteristic length
$r_\delta$ is such that $r_\delta\ll r_{\rm C}$ and $r_{\rm C}'
\lesssim r_\delta$ (not shown in the figure), the presence of the
splitting ensures that $r_{\rm C}\gg r_{\rm C}'$, as opposed to the
($E_{\rm DC}=0$)-case of the previous section Sec.~\ref{sec:secAC}.
As a consequence, for $r \gg r_\delta> r_{\rm C}'$ we expect
diabatic and three-body-induced losses to be largely suppressed.
Then, we show below that a strong optical confinement in the
$z$-direction allows for the realization of stable 2D collision
setups, analogous to Sec.~\ref{sec:secDC}. However, at variance with
the DC case of Sec.~\ref{sec:secDC}, utilizing a combination of DC
and AC fields allows for much greater flexibility in designing
interparticle interactions. In particular, we here focus on the
realization of a  2D potential achievable with a single AC field,
whose character is very
different at distances larger and smaller than $r_{\rm C}$.\\

In the remainder of this section, we discuss further the
above-mentioned processes. In Sec.~\ref{sec:secACDC1} we derive the
3D dressed adiabatic potentials for interactions in the presence of
combined DC and AC fields. In Sects.~\ref{sec:secACDC2} and
\ref{sec:secACDC3} we illustrate the main features of the
two-particle interaction, by specializing to the case where the
AC-field polarization is linear ($q=0$), and by solving a model
Hamiltonian comprising only a limited number of states, whose
energies are close to the one of the ground-state. An expression for
the ground-state interaction potential is obtained which shows that
at large distances $r\gg r_{\rm C}$ the potential has a behavior
$\sim 1/r^3$ similar to the one obtained for two molecules in a DC
field. However now the effective dipolar strength is given by the
combination of both the DC {\em and} the AC fields, and it can be
much weaker than for $r<r_{\rm C}$. Thus, 3D interaction potentials
can be engineered that have a marked ``step-like'' character, being
strongly and weakly repulsive at distances smaller and larger than
$r_{\rm C}$, respectively. Analogous to the DC case of
Sec.~\ref{sec:secDC}, an effective 2D interaction potential
shielding of the short-range region $r\ll r_\delta \ll r_{\rm C}$ is
obtained by adding a harmonic confinement in the $z$-direction and
tracing over the fast particle motion along $z$, (see
Sec.~\ref{sec:secACDC4} and Sec.~\ref{sec:secACDC5}).

\subsubsection{Adiabatic potentials}\label{sec:secACDC1}
The total Hamiltonian including the couplings to DC and AC fields
reads
\begin{eqnarray}\label{eq:ancora}
H(t) = \sum_j\left[\frac{{\bf
p}_j^2}{2m}+\frac{1}{2}m\omega_\perp^2z_j^2\right]+H_{\rm int}({\bf
r},t),
\end{eqnarray}
with
\begin{eqnarray}
H_{\rm int}({\bf r},t)&=& \sum_j \left[B{\bf J}_j^2-E_{\rm DC}d_{0;j}\right.\nonumber\\
&&-\left.\left(E_{\rm AC}e^{-i\omega t}d_{q;j}+{\rm
h.c.}\right)\right]+V_{\rm dd}({\bf r}).\nonumber\\
\end{eqnarray}

Similar to the discussion following Eq.~\eqref{eq:Hamilt}, the
non-trivial system dynamics is determined by the dynamics of the
relative degrees of freedom, which decouple from the harmonic motion
of the center of mass. In this section we set $\omega_\perp=0$ in
Eq.~\eqref{eq:ancora}, and thus diagonalizing the Hamiltonian for
the relative coordinates in the adiabatic limit corresponds to
diagonalizing $H_{\rm int}({\bf r},t)$. The case $\omega_\perp \neq
0$ is treated in the next sections.

The Hamiltonian $H_{\rm int}({\bf r},t)$ is invariant under the
permutation of the two molecules, $(j=1)\leftrightarrow(j=2)$, and
thus it can be conveniently rewritten as $H_{\rm
  int}({\bf r},t)=\sum_{\sigma=\pm}P_\sigma H_{\rm int}^{(\sigma)}({\bf
  r},t)P_\sigma$. Here $P_+$ and $P_-$ are the projectors onto the
manifold of symmetric and antisymmetric states, respectively. Since
(several) external fields are present, parity is not conserved.

In Sec.~\ref{sec:secDCField} we have already diagonalized $H_{\rm
int}({\bf r},t)$ in the absence of the AC field, that is $H_{\rm
int}({\bf r})=\sum_n\ket{\Phi_n({\bf r})}E_n({\bf
r})\bra{\Phi_n({\bf
    r})}$. The respective adiabatic potentials $E_n({\bf r})$ are
shown in Fig.~\ref{fig:fig4}, together with the corresponding
quantum numbers $n=(J;M;\sigma)$. We remark that $J$ is not a good
quantum-number since the electric field breaks the parity for each
molecule; Thus $J=J_1+J_2$ merely indicates the asymptotic manifold.
The corresponding adiabatic potentials and eigenstates for the
ground-state $n=(0;0;+)$ [valid for $r\gg r_B=(d^2/B)^{1/3}$] and
the lowest excited states [valid for $r\gg
r_\delta=(d^2/\hbar\delta)^{1/3}$] are given in Tab.~\ref{tab:tab3}.
Our goal in this section is to extend that treatment to account for
the driving by the AC microwave field, which we assume to be
near-resonant with the transition from the ground to the
first-excited manifold, i.e.
$\omega\sim\overline{\omega}+(2/3-q^2)\delta$ for polarizations
$q=0,\pm1$, see Fig.~\ref{fig:fig2}. The average energy separation
$\hbar\overline{\omega}$ is defined in Eq.~\eqref{eq:eqomega}.\\

Since for $\beta\lesssim1$ the single-particle rotor spectrum is
strongly anharmonic and the AC-field is near-resonant with the
$(J_j=0 \leftrightarrow 1)$-transition, we restrict our discussion
to the rotor states with $J_j=0,1$ for each molecule, that is we
consider $16$ two-particle states. Moreover, we focus on the region
$r\gg r_\delta$, where the dipole-dipole  interaction is (much)
weaker than the DC-field-induced splitting $\hbar\delta$ in the
excited states. Therefore, (up to corrections of order
$\sim{d^2/\hbar\delta{r^3}}$ and $\sim{d^2/Br^3}$) the states
$\ket{\Phi_{J;M;\sigma}({\bf r})}$ are given by the states
$\ket{\Phi_{J;M;\sigma}^{(0)}(\vartheta,\varphi)}
\equiv\ket{\Phi_{J;M_\mu;\sigma}(r\rightarrow\infty,\vartheta,\varphi)}$,
which are reported in Tab.~\ref{tab:tab3}. These states are
independent of $r$, that is, they depend only on the orientation of
the two molecules. In analogy to the treatment of
Sec.~\ref{sec:secAC}, we utilize the approximate states
$\ket{\Phi_{J;M;\sigma}({\bf r})}$ to diagonalize the {\em
time-dependent} Hamiltonian $H_{\rm int}({\bf r},t)$ in a Floquet
picture: We expand the time-dependent wave-function in a Fourier
series in the AC frequency $\omega$. After applying a rotating wave
approximation, i.e. keeping only the energy conserving terms, we
obtain the {\em time-independent} Hamiltonian $\tilde H({\bf r})$,
which again preserves the permutation symmetry, $\sigma=\pm$. The
Hamiltonian $\tilde{H}_{\rm int}^{(+)}({\bf r})$ for the symmetric
manifold ($\sigma=+$) is expressed on the basis
$\{\ket{\Phi_{J;M_\mu;\sigma=+}({\bf r})}\}$ with
$(J;M_\mu)=\{(0;0),(1;1_\mp)|;(1;0);(2;2_{\mu})|_{\mu=-,0,+};(2;1_\mp);(2;0)\}$
as
\begin{widetext}
\begin{eqnarray}
\tilde{H}_{\rm int}^{(+)}({\bf
r})&=&-\hbar\left[\begin{array}{c||cc|c||ccc|cc|c}
\Delta_{0;0}^{(+)}&\sqrt{2}\Omega_-^*&\sqrt{2}\Omega_+&\sqrt{2}\Omega_0&0&0&0&0&0&0\\\hline\hline
\sqrt{2}\Omega_-&\Delta_{1;1_-}^{(+)}&0&0
&\Omega_+^*&-c_+\Omega_-^*&c_-\Omega_-^*&\Omega_0^*&0&0\\
\sqrt{2}\Omega_+&0&\Delta_{1;1_+}^{(+)}&0
&\Omega_-^*&-c_-\Omega_+^*&c_+\Omega_-^*&0&\Omega_0^*&0\\\hline
\sqrt{2}\Omega_0&0&0&\Delta_{1;0}^{(+)}&0&0&0&\Omega_-^*&-\Omega_+^*&\sqrt{2}\Omega_0\\\hline\hline
0&\Omega_+&\Omega_-&0&\Delta_{2;2_-}^{(+)}&0&0&0&0&0\\
0&-c_+\Omega_-&c_-\Omega_+&0&0&\Delta_{2;2_0}^{(+)}&0&0&0&0\\
0&c_-\Omega_-&c_+\Omega_+&0&0&0&\Delta_{2;2_+}^{(+)}&0&0&0\\\hline
0&\Omega_0&0&\Omega_-&0&0&0&\Delta_{2;1_-}^{(+)}&0&0\\
0&0&\Omega_0&\Omega_+&0&0&0&0&\Delta_{2;1_+}^{(+)}&0\\\hline
0&0&0&\sqrt{2}\Omega_0&0&0&0&0&0&\Delta_{2;0}^{(+)}
\end{array}\right],\label{eq:eq38a}
\end{eqnarray}
$\Delta_{J;M_\mu}^{(\sigma)}\equiv\Delta_{J;M_\mu}^{(\sigma)}(r,\vartheta)=J\omega-E_{J;M_\mu;\sigma}({r,\vartheta})/\hbar$
denote position-dependent detunings (for each rotational
excitation), $\Omega_\pm\equiv\Omega_\pm(\varphi)$ are
orientation-dependent couplings, which are detailed below, and
$c_\pm\equiv c_\pm(\vartheta)=\cos(\xi/2)\pm\sin(\xi/2)$ depends on
the polar angle $\vartheta$. The parameter $\xi$ is defined in the
caption of Tab.~\ref{tab:tab3}.

The Hamiltonian $\tilde{H}_{\rm
  int}^{(-)}({\bf r})$ for the antisymmetric manifold $(\sigma=-)$ expressed on
the basis $\{\ket{\Phi_{J;M_\mu;\sigma=-}({\bf r})}\}$ with
$(J;M_\mu)=\{(1;1_\mp);(1;0);(2;2);(2;1_\mp)\}$ reads
\begin{eqnarray}\label{eq:eq38b}
\tilde{H}_{\rm int}^{(-)}({\bf
r})&=&-\hbar\left[\begin{array}{cc|c||c|cc}
\Delta_{1;1_-}^{(-)}&0&0&\Omega_+^*&\Omega_0^*&0\\
0&\Delta_{1;1_+}^{(-)}&0&-\Omega_-^*&0&\Omega_0^*\\\hline
0&0&\Delta_{1;0}^{(-)}&0&-\Omega_-^*&-\Omega_+^*\\\hline\hline
\Omega_+&-\Omega_-&0&\Delta_{2;2}^{(-)}&0&0\\\hline
\Omega_0&0&-\Omega_+&0&\Delta_{2;1_-}^{(-)}&0\\
0&\Omega_0&-\Omega_+&0&0&\Delta_{2;1_+}^{(-)}\\
\end{array}\right].
\end{eqnarray}
\end{widetext}
In Eq.~\eqref{eq:eq38a} and Eq.~\eqref{eq:eq38b} we neglected
off-resonant (second-order) corrections
$\sim\Omega_M(\varphi)d^2/\hbar\delta{r^3}$ to the Rabi-frequency.
The couplings $\Omega_-(\varphi),\Omega_0,\Omega_+(\varphi)$ are
given by $\Omega_-(\varphi)=qf_1E_{\rm AC}e^{iq\varphi}/\sqrt{2}$,
$\Omega_0=(1-q^2)f_0E_{\rm AC}$, $\Omega_+(\varphi)=qf_1E_{\rm
AC}e^{iq\varphi}/\sqrt{2}$, respectively. Thus for linear
polarization ($q=0$) one has $\Omega_0\equiv\Omega$ and
$\Omega_\pm=0$, while for circular polarization $(|q|=1)$
$\Omega_0=0$ and $\Omega_+(\varphi)=\pm\Omega_-(\varphi)=\Omega
e^{\pm i\varphi}/sqrt{2}$ for $q=\pm1$, respectively.

\subsubsection{Model Hamiltonian for $\Delta\ll\delta$}\label{sec:secACDC2}

\begin{figure*}[htb]
 \begin{center}
 \includegraphics[width=\textwidth]{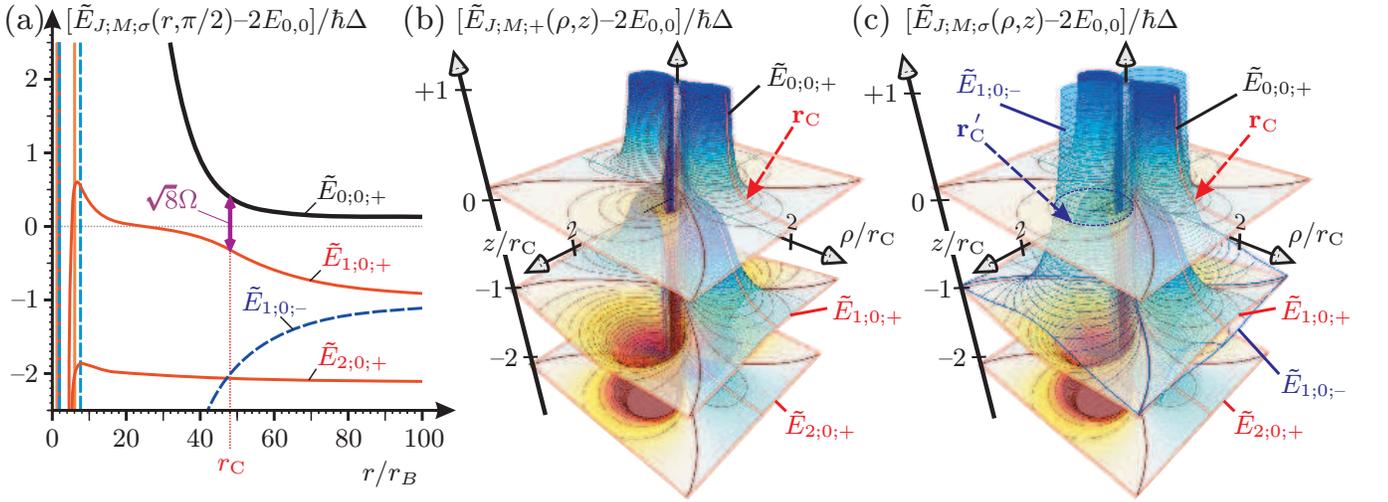}
  \caption{\label{fig:fig12}(color online) Dressed adiabatic potentials $\tilde{E}_{J;M;\sigma}({\bf r})$ of Eq.~\eqref{eq:eq40True}
  for the interaction of two molecules polarized by a (weak)
  DC field ${\bf E}_{\rm DC}=E_{\rm DC}{\bf e}_z$ with $\beta\equiv dE_{\rm DC}/B=1/10$
  and dressed by an AC with linear polarization $q=0$.
  The AC-field detuning and Rabi-frequency are $\Delta=3B/10^6\hbar$ and $\Omega=\Delta/4$,
  respectively. For a typical rotational spacing of $B\sim h~10~{\rm GHz}$
  these numbers entail $\Delta/2\pi= 30{\rm kHz}$ and $\Omega/2\pi= 7.5{\rm
  kHz}$. Panel (a): Dressed adiabatic potentials $\tilde{E}_n({\bf r})$ of
  Eq.~\eqref{eq:eq40True} plotted as a function of $r$ for $z=0$ ($\vartheta=\pi/2$). The
  potentials corresponding to symmetric (antisymmetric) states are given by solid (dashed) lines, and
  indicated by $\tilde{E}_{J;M;\sigma}({\bf r})$ for $J=0,1,2$, $M=0$, $\sigma=+$ ($J=1$, $m=0$, $\sigma=-$). The
  dressed ground-state potential $\tilde{E}_{0;0;+}({\bf
  r})$ has the highest energy (thick solid line). The other
  potentials are asymptotically detuned by a multiple of $\Delta$.
  The ground-state potential $\tilde{E}_{0;0;+}({\bf r})$ shows an avoided
  crossing with the symmetric potential $\tilde{E}_{1;0;+}({\bf r})$ at $r=r_{\rm C}$.
  (b) Dressed adiabatic potentials $\tilde{E}_{J;M;\sigma}({\bf r})\equiv\tilde{E}_{J;M;\sigma}(\rho,z)$ for the symmetric states ($\sigma=+$) of
  Eq.~\eqref{eq:eq40True} plotted as a function of $\rho=r\sin\vartheta$ {\em and}
  $z=r\cos\vartheta$. Blue regions correspond to a repulsive potential, red regions to an attractive potential.
  For $z=0$ ($\rho$-axis) we recognize the case of panel (a), where the
  symmetric ground-state potential has the largest energy. The
  position of the Condon point $r_{\rm C}$ is indicated by a arrow.
  Accordingly, the avoided crossing with the potential
  $\tilde{E}_{1;0;+}({\bf r})$ observed at $r=r_{\rm C}$ in panel (a) is now visible in
  transparency, below the upper layer. For $|z|>0$ the potential
  $\tilde{E}_{1;0;+}$ becomes less and less repulsive.
    For $|z|>\rho/\sqrt{2}$ we have $1-3\cos^2\vartheta<0$ and
    thus $\tilde{E}_{1;0;+}({\bf r})$ is attractive and the Condon-point vanishes
    since the two states are off-resonant. (c) Dressed adiabatic potential as in panel (b), but also showing the antisymmetric states ($\sigma=-$).
    We see that the dressed potential $\tilde{E}_{1;0;-}({\bf r})$ for the antisymmetric
    state with $(1;0;-)$ is strongly attractive in the plane, i.e.for $z=0$, which corresponds to the profile shown in panel (a).
    With increasingly separation $|z|/r>0$ the potential become less and less attractive.
    For $|z|/\rho>1/\sqrt{2}$ we have $3\cos^2\vartheta-1>0$ and the potential becomes repulsive. Thereby
    a crossing between the asymmetric state and the ground-state appears at a (second) Condon-``point''
    ${\bf r}_{\rm C}'$ (dashed line),
    as the two states are resonant, however due to the permutation symmetry do not couple.}
 \end{center}
 \end{figure*}

In the following we illustrate the main features of the scattering
in the combined DC and AC fields, using the example of
Fig.~\ref{fig:fig11}. The Rabi-frequency $\Omega$ is chosen real and
positive and in particular smaller than the detuning,
$\Omega\ll\Delta$. Moreover, we choose $\Delta\ll\delta$ since we
want to address the potentials in regions where the two molecules in
the first-excited two-particle manifold are aligned by the DC field
and not by the dipole-dipole interaction, see Fig.~\ref{fig:fig11}.
The figure shows that for $r\gg r_\delta$ the bare ($E_{\rm
AC}=0$)-states $\ket{\Phi_{J;M;\sigma}({\bf r})}$ with $M\neq0$ are
largely detuned from resonance by an amount of order
$\delta\gg\Delta$. Thus, the discussion of Eq.~\eqref{eq:eq38a} and
Eq.~\eqref{eq:eq38b} can be simplified by restricting the Hilbert
space to the four states with $M=0$ only. For our basis-set, these
are: The three symmetric states $\ket{\Phi_{J;0;+}({\bf r})}$ with
$J=0,1,2$ and the antisymmetric state $\ket{\Phi_{1;0;-}({\bf r})}$.
Then, Eq.~\eqref{eq:eq38a} and Eq.~\eqref{eq:eq38b} reduce to
\begin{subequations}\label{eq:eq40}
\begin{eqnarray}
\tilde{H}_{\rm int}^{(+)}({\bf r})&=&-\hbar\left[\begin{array}{ccc}
\Delta_{0;0}^{(+)}&\sqrt{2}\Omega&0\\
\sqrt{2}\Omega&\Delta_{1;0}^{(+)}&\sqrt{2}\Omega\\
0&\sqrt{2}\Omega&\Delta_{2;0}^{(+)}
\end{array}\right],\\
\tilde{H}_{\rm int}^{(-)}({\bf
r})&=&-\hbar\left[\Delta_{1,0}^{(-)}\right].
\end{eqnarray}
\end{subequations}
\noindent The position-dependence of the detunings
$\Delta_{J;0}^{(\sigma)}\equiv\Delta_{J;0}^{(\sigma)}(r,\vartheta)$
has the usual dipolar form $\Delta_{J;0}^{(\sigma)}(r,\vartheta)\sim
\Upsilon/r^3$, with $\Upsilon\equiv 1-3\cos^2\vartheta$. Explicitly,
we have $\Delta_{J;0}^{(\sigma)}(r,\vartheta)=
J\Delta-[C_{3;(J;0;\sigma)}\Upsilon/r^3-C_{6;(J;0;\sigma)}(\vartheta)/r^6]/\hbar$.
The coefficients $C_{3;n}$ and $C_{6;n}(\vartheta)$ are given in
Tab.~\ref{tab:tab3} for $n=0,5,15$ (symmetric states) and $n=6$
(antisymmetric state), respectively. For the following discussion,
it is important to notice that for a weak DC electric field
$\beta\ll1$, the $C_{3;n}$-coefficients $C_{3;(0;0;+)}\equiv
g_0^2\approx(d\beta/3)^2$ and $C_{3;(2;0;+)}\equiv
g_2^2\approx(d\beta/5)^2$ are quite small, since they are suppressed
by a factor $\sim\beta^2$. On the other hand, the coefficients
$C_{3;(1;0;\pm)}\equiv g_0g_2\pm f_1^2\approx \pm d^2/3$ for states
belonging to the first-excited manifold are as large as the bare
dipolar coefficients, see Tab.~\ref{tab:tab1} and
Tab.~\ref{tab:tab3}.

By diagonalizing Eq.~\eqref{eq:eq40}~(a) we obtain
  the three dressed symmetric potentials $\tilde{E}_{J;0;+}({\bf r})$ (with
  $J=0,1,2$) in terms of complex cubic roots by
\begin{eqnarray}\label{eq:eq40True}
  \tilde{E}_{J;0;+}({\bf r})&=&\sum_\pm e^{\pm2\pi
    i J/3}\left[-\frac{Q}{2}\pm
    i\sqrt{\frac{P^3}{27}-\frac{Q^2}{4}}\right]^{1/3}\nonumber\\
&&-\hbar\overline{\Delta}(r,\vartheta),
\end{eqnarray}
where  $P\equiv
4\Omega^2+\sum_J[\Delta_{J;0}^{(+)}(r,\vartheta)-\overline{\Delta}(r,\vartheta)]^2/2$,
  $Q=2\Omega^2[\Delta_{1;0}^{(+)}(r,\vartheta)-\overline{\Delta}(r,\vartheta)]
  -\prod_J[\Delta_{J;0}^{(+)}(r,\vartheta)-\overline{\Delta}(r,\vartheta)]$ and $\overline{\Delta}(r,\vartheta)=\sum_J\Delta_{J;0}^{(+)}(r,\vartheta)/3$.
The dressed potential for the antisymmetric state,
$\tilde{E}_{1,0,-}({\bf r})=-\hbar\Delta_{1;0}^{(-)}(r,\vartheta)$,
is the same as the bare one.\\

  The dressed
  potentials $\tilde{E}_{J;0,\sigma}(r,\vartheta)$ are plotted in
  Fig.~\ref{fig:fig12}, for $\Delta=4\Omega=3B/10^6\hbar$, $B=h~10~{\rm GHz}$, linear polarization
  ($q=0$) and $\beta=1/10$. These parameters are the same as in
  Fig~\ref{fig:fig11}.
  In particular,
  Fig.~\ref{fig:fig12}(a) shows $\tilde{E}_{J;0;\sigma}(r,\vartheta=\pi/2)$ as
  a function of the distance $r$, for molecules on the plane
  $z=r\cos\vartheta=0$ ($\vartheta=\pi/2$). Figure~\ref{fig:fig12}(b)
  is a three-dimensional representation of
  the potential-energy surfaces
  $\tilde{E}_{J;0;+}(r,\vartheta)\equiv\tilde{E}_{J;0;+}(\rho,z)$
  for the three symmetric states $J=0,1,2$,
  for finite transverse displacements $z=r\cos\theta$
  ($\vartheta\neq\pi/2$), while Fig.~\ref{fig:fig12}(c)
  is the same as Fig.~\ref{fig:fig12}(b), with the addition of the
  potential $\tilde{E}_{1;0;-}(\rho,z)$ for the antisymmetric
  state.

  In Fig.~\ref{fig:fig12}~(a) the dressed ground-state potential
  $\tilde{E}_{0;0;+}(r,\pi/2)$ is the thick solid curve with
largest energy, which undergoes an avoided crossing with the
potential $\tilde{E}_{1;0;+}(r,\pi/2)$ at a distance $r_{\rm C} \sim
(d^2/3\hbar\Delta)^{1/3}$. The precise value of $r_{\rm C}$ is
derived below. The figure shows that the Condon point $r_{\rm C}$
separates an inner region $r<r_{\rm C}$ where the ground-state
potential is strongly repulsive $\tilde{E}_{0;0;+}(r<
r_{C},\pi/2)\sim\tilde{C}_3(r<r_{\rm C})/r^3$, from an outer region
$r>r_{\rm C}$ where the potential is only weakly repulsive
$\tilde{E}_{0;0;+}(r>r_{C},\pi/2)\sim\tilde{C}_3(r>r_{\rm C})/r^3$
with $\tilde{C}_3(r>r_{\rm C})\ll\tilde{C}_3(r<r_{\rm C})$. This
marked dependence of the potential strength on $r$ is the
realization of the ``step-like'' potential of
Fig.~\ref{fig:fig3added}, and it is due to the fact that the dressed
ground-state inherits the character of the bare ground-state and of
the bare state $\ket{\Phi_{1;0;+}({\bf r})}$ for $r
> r_{\rm C}$ and $r<r_{\rm C}$, respectively. Thus, we have
$\tilde{C}_3(r> r_{C}) \sim C_{3;(0;0;+)}\approx(d\beta/3)^2$ and
$\tilde{C}_3(r< r_{C})\sim C_{3;(1;0;+)}\approx d^2/3$. A harmonic
confinement in the $z$-direction will be added in
Sec.~\ref{sec:secACDC4} to ensure the stability of the 2D
interaction.

Figure~\ref{fig:fig12}(b) is a three-dimensional representation of
the dressed adiabatic potentials $\tilde{E}_{J;M;+}({\bf r})$ for
the symmetric states, plotted as a function of $\rho=r\sin\vartheta$
{\em and} $z=r\cos\vartheta$. The blue and red regions correspond to
repulsive and attractive potentials, respectively. The thin gray
lines are equipotential energy contours. For $z=0$ ($\rho$-axis) we
recognize the case of Figure~\ref{fig:fig12}~(a), where the
symmetric ground-state potential has the largest energy. The
position of the Condon point $r_{\rm C}$ is indicated by an arrow.
The avoided crossing between the ground-state potential
$\tilde{E}_{0;0;+}({\bf r})$ and the potential
$\tilde{E}_{1;0;+}({\bf r})$ observed at $r=r_{\rm C}$ for
$\vartheta=\pi/2$ in panel (a) is now visible in transparency, below
the upper layer. The figure shows that for $|z|>0$ the potential
$\tilde{E}_{1;0;+}(\rho,z)$ becomes less and less repulsive, and
thus the Condon point ${\bf r}_{\rm C}=r_{\rm C}(\vartheta){\bf
e}_r$ occurs at shorter distances (see below,
Eq.~\eqref{eq:Condon}). For $|z|>\rho/\sqrt{2}$, we have
$1-3\cos^2\vartheta<0$ and therefore $\tilde{E}_{1;0;+}({\bf r})$
becomes attractive. Thus, the Condon-point vanishes since the
combined energy of the bare ground-state plus a photon and the
energy $E_{1;0;+}({\bf r})$ of the bare state
$\ket{\Phi_{1;0;+}({\bf r})}$ are not resonant [for the dependence
of the bare potential $E_{1;0;+}({\bf r})$ on the angle $\vartheta$,
see also Fig.~\ref{fig:fig4}(b) and Fig.~\ref{fig:fig4}(d)]. This
vanishing of the avoided crossing for $|z|>\rho/\sqrt{2}$
corresponds to the formation of a ``hole'' in the 3D potential
shielding the molecular-core region, and it allows for the familiar
attraction of dipole-dipole interactions. The presence of this
"hole" is reminiscent of the vanishing of the Condon-point at
$\vartheta=0$ and $\vartheta=\pi$ for the case of a
linearly-polarized AC field in the absence of a DC field [see
Fig.~\ref{fig:fig10}(a2) in Sec.~\ref{sec:secAC}]. However, here
there are no dark states present at $r_{\rm C}$, due to the
DC-field-induced splitting $\hbar\delta$ of the ($J=1$)-manifold.
This fact eliminates a significant non-adiabatic loss channel for
ground-state interactions.

From Fig.~\ref{fig:fig12}(b) we see that a real crossing with the
anti-symmetric state $\ket{\Phi_{1;0;-}({\bf r})}$ takes place for
$|z|>\rho/\sqrt{2}$ at a second Condon point, which we denote as
${\bf r}_{\rm C}'$. This is at variance with the case of
Sec.~\ref{sec:secAC} [see Fig.~\ref{fig:fig10}~(a1)], where $r_{\rm
C}'$ was $r_{\rm C}'>r_{\rm C}$ for all angles, thus opening loss
channels due to three-body-induced (or tensor-shift-induced, when a
harmonic confinement along $z$ is considered) couplings to the
symmetric ground-state for any $\vartheta$. The exact position of
the point ${\bf r}_{\rm C}'$ is obtained in the next section.\\

\subsubsection{Effective 3-D interaction potential }\label{sec:secACDC3}

In the following we are interested in the effective 3D potential
$V_{\rm eff}^{\rm 3D}({\bf r})$ for two molecules in their
ground-state dressed by the external fields. In the absence of a
trap ($\omega_\perp=0$) $V_{\rm eff}^{\rm 3D}({\bf r})$ reads
\begin{eqnarray}
V_{\rm eff}^{\rm 3D}({\bf r}) \equiv \tilde{E}_{0;0;+}(r,\vartheta)
- 2\left[E_{0,0}+E_{0,0}'\right],
\end{eqnarray}
where the terms in brackets are the Stark-shifts $E_{0,0}\approx
-d^2\beta^2/6$ and $E_{0,0}'\approx \hbar\Omega^2/\Delta$ induced by
the DC and AC electric field, respectively.

At separations $r \gg r_{\rm C}$ the effective potential resembles
the dipolar potential for two dipoles aligned along ${\bf e}_z$ and
in second order in the saturation amplitude $\Omega/\Delta$ is given
by
\begin{eqnarray}\label{eq:eq4X}
V_{\rm eff}^{\rm 3D}({\bf r}) \approx
\frac{C_{3;(0;0;+)}\Upsilon}{r^3}
+\frac{2\Omega^2}{\Delta^2}\frac{(C_{3;(1;0;+)}-C_{3;(0;0;+)})\Upsilon}{r^3},\nonumber\\
\end{eqnarray}
where $\Upsilon\equiv 1-3\cos^2\vartheta$, and terms of order ${\cal
O}(\Omega^4)$ and ${\cal O}(1/r^6)$ have been neglected. The first
term in Eq.~\eqref{eq:eq4X} is the dipole-dipole interaction for the
two weakly polarized molecules induced by the DC field (see
Sec.~\ref{sec:secDCField}), while the second term is the familiar
dipole-dipole interaction induced by the coupling to the AC field.
The proportionality factor
$(C_{3;(1;0;+)}-C_{3;(0;0;+)})2\Omega^2/\Delta^2$ appears due to the
competition of the oscillating dipole-moment ($\sim d\Omega/\Delta$)
induced by the AC field with the permanent dipole-moment already
present because of the DC field.

The perturbative expression Eq.~\eqref{eq:eq4X} breaks down when the
level spacing becomes comparable to the coupling, that is for
$|\Delta_{1;0}^{(+)}({\bf
  r})-\Delta_{0;0}^{(+)}|\sim\Omega$. In particular,
for $\Delta_{1;0}^{(+)}({\bf r}_{\rm C})=\Delta_{0;0}^{(+)}({\bf
r}_{\rm C})$
  an avoided crossing occurs between the potentials
  $\tilde{E}_{0;0;+}({\bf r})$ and $\tilde{E}_{1;0;+}({\bf r})$,
  which defines the resonant Condon point, ${\bf r}_{\rm C} \equiv r_{\rm C}(\vartheta) {\bf
e}_r$, where ${\bf e}_r$ is the intermolecular axis. The Condon
distance $r_{\rm C}(\vartheta)$ is parameterized in terms of the
polar angle $\vartheta$ as
\begin{eqnarray}
r_{\rm
C}(\vartheta)=\left[\frac{C_{3;(1;0;+)}-C_{3;(0;0;+)}}{\hbar\Delta(1-3\cos^2\vartheta)}\right]^{1/3}.\label{eq:Condon}
\end{eqnarray}
For $\vartheta=\pi/2$ ($z/r=0$) the Condon point is attained at
$r_{\rm C}=r_{\rm
C}(\pi/2)=[(C_{3;(1;0;+)}-C_{3;(0;0;+)})/\hbar\Delta]^{1/3}\approx[d^2/3\hbar\Delta]^{1/3}$,
see Fig.~\ref{fig:fig11}~(a), which depends on the detuning $\Delta$
and the difference in the $C_{3;n}$ coefficients of the first
excited state and the ground state
($C_{3;(1;0;+)}-C_{3;(0;0;+)}\approx d^2/3$ for a weak DC-field
$\beta\ll 1$). For $\vartheta\neq\pi/2$ ($z/r\neq 0$) the avoided
crossing occurs at smaller separations $r_{\rm C}(\vartheta)<r_{\rm
C}$ until it vanishes for $\cos^2\vartheta=(z/r)^2=1/3$, see
Fig.~\ref{fig:fig11}(b).

The position of the point ${\bf r}_{\rm C}'\equiv r_{\rm
C}'(\vartheta) {\bf e}_r$ is determined by the crossing between the
dressed ground-state potential $\tilde{E}_{0;0;+}({\bf r})$ and the
potential for the antisymmetric state $\tilde{E}_{1;0;-}({\bf r})$.
As mentioned above [and shown in Fig.~\ref{fig:fig12}(b)], this
crossing occurs in the region $|z|>\rho/2$ ($\cos^2\vartheta>1/3$).
The distance $r_{\rm C}'(\vartheta)$ is given by $r_{\rm
C}'(\vartheta)\approx
[(C_{3;(0;0;+)}-C_{3;(1;0;-)})(3\cos^2\vartheta-1)/\hbar\Delta]^{1/3}$.\\%

The discussion above suggests that an effective 2D interaction
potential $V_{\rm eff}^{\rm 2D}(\boldrho)$ with no losses due to
couplings of the ground-state to other symmetric or antisymmetric
states may be obtained for distances $r\gg r_\delta$, by introducing
a parabolic potential in the $z$-direction confining the particles
to the sector $(z/r)^2< 1/3$. This ``shielding'' of the loss
channels is analogous to the ``shielding'' of the attractive part of
the potential and of the molecular-core region of the DC case for
$r>\ell_\perp \gg r_\star\sim (d^2/B\beta^2)^{1/3}$ (see
Sec.~\ref{sec:secDC}). However, now $\ell_\perp$ is effectively
replaced by $r_{\rm C} \gtrsim \ell_\perp$, and $r_{\rm C}$ allows
for much greater flexibility in tuning by external fields.

In the next section we detail the requirements for obtaining a
stable effective interaction in 2D. In this way, it is possible to
realize the 2D potential with ``step-like'' character, as shown in
Fig.~\ref{fig:fig3added}.

\subsubsection{Parabolic confinement}\label{sec:secACDC4}

The presence of a finite trapping potential of frequency
$\omega_\perp$ in the $z$-direction provides for a
position-dependent energy shift of Eq.~\eqref{eq:eq4X}. Thus, the
new potential reads
\begin{eqnarray}
  V(\rho,z) \equiv V_{\rm eff}^{\rm 3D}({\bf r}) +
  \frac{1}{4}m\omega_\perp^2z^2. \label{eq:ACDCTRAP}
\end{eqnarray}
Analogous to the discussion of Sec.~\ref{sec:secDCTrap}, the
combination of the dipole-dipole interaction, which is repulsive for
$r \gg r_{\rm C}$, and of the harmonic confinement yields a
repulsive potential which provides for a three-dimensional barrier
separating the long-distance repulsive regime from the
short-distance regime, where collisional losses can occur. If the
collisional energy is much smaller than this barrier, the relative
motion of the particles is confined to the long-distance region,
where the potential is purely repulsive.

 \begin{figure}[htb]
 \begin{center}
 \includegraphics[width=\columnwidth]{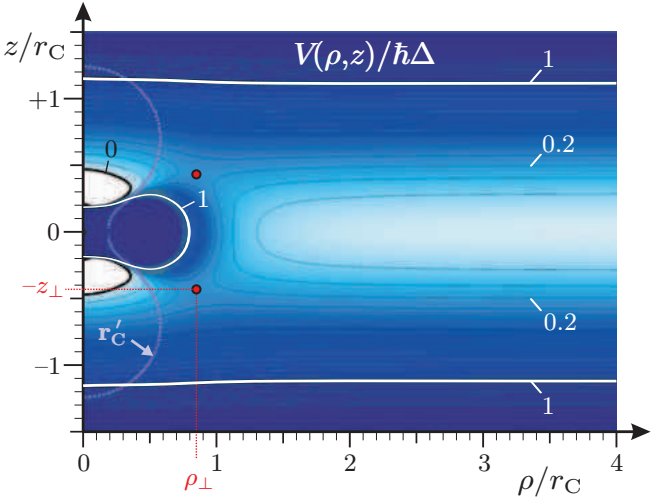}
  \caption{(color online) Contour plot of the effective potential $V(\rho,z)$
  of Eq.~\eqref{eq:ACDCTRAP} for two polar
    molecules interacting in the presence of a weak DC field and an
    AC field. The field parameters are the same as in
    Fig.~\ref{fig:fig12}. The frequency of the confining harmonic
    potential in the $z$-direction is $\omega_\perp=\Delta/5$.
    Darker regions represent stronger repulsive interactions. The white region for $\rho<1/2$ indicates a potential $V(\rho,z)<0$.
     The
    combination of the dipole-dipole interactions induced by the
    DC electric and AC (microwave) fields and of the quadratic confinement leads to
    realizing a 3D repulsive potential for $r \gg r_{\rm C}$. In
    particular, two saddle-points located at
    $(\rho_\perp,\pm z_\perp)$ (circles) separate the repulsive long-distance regime
    $r\gg r_{\rm C}$ from the short-distance regime $r<r_{\rm C}$ where diabatic
    losses occur. The dotted line signaled by ${\bf r}_{\rm C}'$ indicates the
    location of the crossing between the ground-state potential and
    the energy $\tilde{E}_{1;0;-1}({\bf r})$ of the antisymmetric state, see text and
    Fig~\ref{fig:fig12}.}\label{fig:fig13}
 \end{center}
 \end{figure}

Figure~\eqref{fig:fig13} is a contour plot of
Eq.~\eqref{eq:ACDCTRAP} for the same parameters as in
Fig.~\ref{fig:fig12}, i.e. $\Delta=4\Omega=3B/10^{6}\hbar$ and
$\beta=1/10$. The frequency $\omega_\perp$ for the harmonic
confinement is $\omega_\perp=\Delta/5$. In the figure, darker
regions correspond to a stronger repulsive potential, and the white
region for $\rho,z\lesssim r_{\rm C}/2$ corresponds to
$V(\rho,z)<0$. The repulsion due to the dipole-dipole and harmonic
potentials is clearly distinguishable at $z=0$ and $|z|/r_{\rm C}
\gtrsim 1$, respectively. Two saddle points located at
$(\rho_\perp,\pm z_\perp)$
 separate the repulsive long-distance
from the short-distance regions (circles in Fig.~\eqref{fig:fig13}).
The location of the saddle points approaches
$(\rho_\perp,|z_\perp|)\sim (r_{\rm C},r_{\rm C}/2)$ with increasing
confining potential $\omega_\perp$ in the $z$-direction. In the
figure, the dotted line signaled by ${\bf r}_{\rm C}'$ marks the
location of the crossing between the dressed ground-state potential
$\tilde{E}_{0;0;+}({\bf r})$ and the potential
$\tilde{E}_{1;0;-}({\bf r})$ for the antisymmetric state. The figure
shows that this crossing occurs in the short-distance region $r <
r_{\rm C}$ for all $z$, in agreement with previous discussions.
Thus, for $r\gg r_{\rm C}$ and collisional kinetic energies smaller
than the potential barrier at the saddle point the ground-state
interactions are stable and purely repulsive, consistent with the
discussion above.

\subsubsection{Effective 2-D Interaction}\label{sec:secACDC5}

\begin{figure*}[htb]
 \begin{center}
 \includegraphics[width=\textwidth]{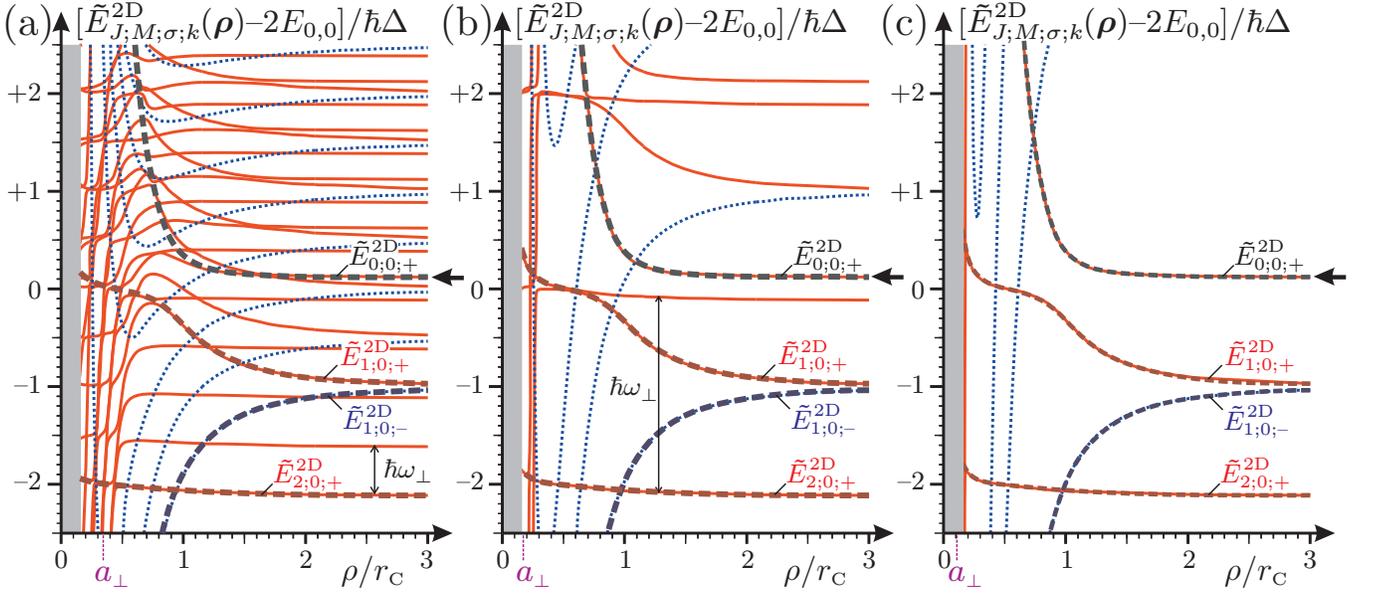}
  \caption{(color online) Effective 2D potentials $\tilde E_{J;M;\sigma;k}^{\rm
  2D}(\boldrho)$
  with $(J=0,1,2;M=0;\sigma=\pm1;k=0,1,2,\ldots)$ of
  Eq.~\eqref{eq:eq53} for the AC- and DC-field setups of Fig.~\ref{fig:fig13}.
  The strength of the (weak) DC field is $\beta=dE_{\rm DC}/B=1/10$.
  The AC field has polarization $q=0$, is blue detuned by $\Delta=3B/10^6\hbar$
  and the saturation amplitude is $\Omega/\Delta=1/4$.
  Panels (a), (b) and (c) correspond to a harmonic oscillator
  frequency of the confining potential $\omega_\perp/\Delta=1/2,2$ and $5$,
  respectively. The tick-dashed lines indicate the (single band)
  effective potentials $\tilde E_{J;0;\sigma}^{\rm 2D}(\boldrho)$ of Eq.~\eqref{eq:eq45}.
  The solid lines are the non-perturbative potentials
  $E_{J;0;\sigma;k}^{\rm 2D}(\boldrho)$ of Eq.~\eqref{eq:eq53},
  where $k=0,1,2,\ldots$ denotes the $k^{\rm th}$ transversal excitations in the $z$-direction.
  The ground-state potential approaches twice the single-particle
  Stark-shifts, $2(E_{0;0}+E_{0;0}')$ (see text), for $\rho\rightarrow\infty$ and it is
  indicated by arrow(s). The harmonic oscillator
  length $a_\perp$ for a typical mass $m\approx 100{\rm amu}$,
  a dipole moment $d\approx 8.9{\rm Debye}$ and a rotational constant
  $B\approx h~10~{\rm GHz}$ is indicated on the $\rho$-axis.
  The corresponding detuning is $\Delta\approx2\pi\times30~{\rm kHz}$ and the
  trapping frequency is $\omega_\perp/2\pi\approx 15{\rm kHz}, 60{\rm kHz}, 150{\rm kHz}$
  in Panel (a), (b), (c), respectively. The gray region corresponds to
  $r<r_\delta$.}\label{fig:fig14}
 \end{center}
 \end{figure*}

Analogous to the discussion for the DC-field case in
Sec.~\ref{sec:secEff2D}, in the limit of tight optical confinement
it is possible to derive effective two-dimensional ground
interaction potentials $V_{\rm eff}^{\rm 2D}(\boldrho)$ by
integrating over the fast transverse degrees of freedom, $z_1$ and
$z_2$. For $r
> r_{\rm C} \gg a_\perp$, the two-particle eigenfunctions in the
$z$-direction approximately factorize into products of
single-particle harmonic oscillator wave-functions, and thus the
integration is conveniently carried out in the harmonic oscillator
basis. In the adiabatic approximation we find to first order in
$V_{\rm eff}^{\rm 3D}({\bf r})/\hbar\omega_\perp$ the 2D effective
ground-state potential as
\begin{eqnarray}
V_{\rm eff}^{\rm 2D}(\boldrho)&\approx&\int dz_1 dz_2
|\psi_0(z_1)|^2
|\psi_0(z_2)|^2 V_{\rm eff}^{\rm 3D}(\boldrho,z_2-z_1)\nonumber\\
&=&\frac{1}{\sqrt{2\pi}a_\perp}  \int dz e^{-z^2/2a_\perp^2} V_{\rm
eff}^{\rm 3D}(\rho,z),\label{eq:eff2dpotentialACDC}
\end{eqnarray}
where $\psi_k(z_j)$ is the $k$-th harmonic oscillator wave-function
for the transverse confinement. In an analogous way, effective 2D
potentials can be derived for all the dressed potentials
$\tilde{E}_{J;M;\sigma}({\bf r})$, as (up to a constant shift)
\begin{eqnarray}
\tilde{E}_{J;M;\sigma}^{\rm 2D}(\boldrho) \approx
\frac{1}{\sqrt{2\pi}a_\perp} \int dz e^{-z^2/2a_\perp^2}
\tilde{E}_{J;M;\sigma}(\rho,z),\label{eq:eq45}
\end{eqnarray}
with $\tilde{E}_{0;0;+}^{\rm 2D}(\boldrho)= V_{\rm eff}^{\rm
2D}(\boldrho)$. In the following we discuss the validity of the
adiabatic approximation in the case when both the DC and the AC
fields are present. We focus only on the four above-mentioned
potentials, since the remaining states of the $(J=1)$-manifold are
detuned by a large amount $\sim\delta/\omega_\perp\sim 10^3$. Thus,
we neglect non-adiabatic couplings from the ground-state to the
continuum corresponding to high-energy transverse excitations of the
far-detuned states, since these couplings are expected to vanish at
large inter-particle separations.\\

At variance with the DC-field case of Sec.~\ref{sec:secEff2D},
satisfying the adiabatic approximation in the presence of both DC
and AC fields is non as trivial. In fact, for a blue-detuned
($\Delta > 0$) AC field the dressed ground-state potential
$\tilde{E}_{0;0;+}({\bf r})$ has the largest energy, see
Fig~\ref{fig:fig12}(a). Thus, it can happen that
$\tilde{E}_{0;0;+}({\bf r})$ becomes degenerate with the energy of
one of the other states plus some multiple $k$ of the harmonic
oscillator energy in the transverse direction $\hbar \omega_\perp$.
When these degeneracies happen, avoided and real crossings occur
with the energies of the symmetric and the anti-symmetric states, so
that satisfying the adiabatic requirement becomes in general much
harder than in the DC-field case of Sec.~\ref{sec:secEff2D}. In
fact, there the ground-state is the lowest-energy state and the
lowest-energy excitations are the ones of the harmonic oscillator
along $z$. In that case the adiabaticity condition is satisfied for
$V_{\rm eff}^{\rm 2D}(\boldrho)\ll\hbar \omega_\perp$. On the other
hand, it is still possible to derive an expression analogous to the
latter even for the case when the AC field is present, if the
trapping potential is large enough so that $\omega_\perp \gtrsim 2
\Delta$. In fact, then for large distances $\rho \gg r_{\rm C}$ the
energy difference between $\tilde{E}_{0;0;+}({\bf r})$ and the one
of the first-excited state is approximately
$\hbar\omega_\perp-2\hbar\Delta$, where $-2\hbar\Delta$ is the
energy $\tilde{E}_{2;0;+}({\bf r})=-2\hbar\Delta$ (see
Fig.~\ref{fig:fig12}). In this case, the adiabatic approximation is
still valid provided
\begin{eqnarray}
V_{\rm eff}^{\rm 2D}({\bf\rho})\ll\hbar\omega_\perp
-2\hbar\Delta.\label{eq:eq46}
\end{eqnarray}

The perturbative expressions Eq.~\eqref{eq:eq45} for the dressed
effective 2D potentials $\tilde{E}_{J;0;\sigma}^{\rm 2D}(\boldrho)$
are shown as thick dashed lines in Fig.~\ref{fig:fig14} for the
combination of a weak DC field with $\beta=1/10$ and an AC field
with linear polarization $q=0$ and detuning
$\Delta=4\Omega=3B/10^6\hbar$. The panels (a,b,c) of
Fig.~\ref{fig:fig14} represent different transverse trapping
frequencies given by $\omega_\perp/\Delta=1/2,2,5$, respectively.
The effective potential for the ground-state $\tilde{E}_{0;0;+}^{\rm
2D}$ is indicated at large separation $\rho\gg r_{\rm C}$, where it
approaches the value $\tilde{E}_{0;0;+}^{\rm
2D}(\rho\rightarrow\infty)=2E_{0,0}+2E_{0,0}'$, corresponding to the
DC and AC Stark-shift of the separated molecules, with
$2E_{0,0}'/\hbar\Delta\approx +2(\Omega/\Delta)^2=1/8$. The thin
solid and dotted lines in Fig.~\ref{fig:fig14} show the potentials
$\tilde{E}_{J;M;\sigma;k}^{\rm 2D}(\boldrho)$ for $\sigma=+$ and
$\sigma=-$, respectively. The potentials
$\tilde{E}_{J;M;\sigma;k}^{\rm 2D}(\boldrho)$ have been obtained by
diagonalizing numerically
\begin{eqnarray}
\tilde{H}_{\rm
rel}=\frac{p_z^2}{m}+\frac{1}{4}m\omega_\perp^2z^2-\frac{\hbar\omega_\perp}{2}+\tilde{H}_{\rm
int}({\bf r}),\label{eq:eq53}
\end{eqnarray}
with $\tilde{H}_{\rm int}({\bf r})$ given in Eq.~\eqref{eq:eq40}.
Here, the index $k=0,1,2,\ldots$ labels the transverse excitations
and at large separations $\rho\gg r_{\rm C}$ the corresponding
potentials approach $\tilde{E}_{J;M;\sigma;k}^{\rm
2D}(\rho\rightarrow\infty)\approx
\tilde{E}_{J;M;\sigma}(r\rightarrow\infty)+k\hbar\omega_\perp$.

In Fig.~\ref{fig:fig14}(a) we observe a series of avoided crossings
involving the ground-state potential in the region $\rho\lesssim
r_{\rm C}$. For $\rho \gg  r_{\rm C}$ the ground-state potential has
the characteristic $1/\rho^3$-dependence. Fig.~\ref{fig:fig14}(b)
shows that for $\omega_\perp=2 \Delta$ the ground-state potential
$V_{\rm eff}^{\rm 2D}(\boldrho)$ is already well separated from the
energy of the first-excited state with $k>0$ in a region
$\rho>r_{\rm C}$. Finally, Fig.~\ref{fig:fig14}(c) shows that for a
tight trapping, $\omega_\perp=5\Delta$, the ground-state potential
$V_{\rm eff}^{\rm 2D}(\boldrho)$ for $\rho>r_{\rm C}/2$ is well
separated by $~\hbar\Delta$ from all the excited state with $k>0$ .
The adiabatic approximation is valid for $V_{\rm eff}^{\rm
2D}(\boldrho)\ll\hbar\Delta$, consistent with Eq.~\eqref{eq:eq46}.

Remarkably, we find that since the spontaneous emission rates in the
excited rotational levels of polar molecules, $\Gamma_{\rm SE}$, are
negligible compared to achievable optical confinements $\sim
\omega_\perp\sim 2\pi\times 150~{\rm kHz}$, the regime where
$\Delta<\omega_\perp/2$ is widely accessible. In fact one can
achieve the limit of weak saturation and strong confinement while at
the same time be sufficiently detuned to not suffer of spontaneous
emission. That is, it is possible to fulfil all of the inequalities
$\Gamma_{\rm SE}\ll\Omega<2\Delta<\omega_\perp$.\\

\section{Conclusions}

In this work, we have shown how to engineer 2D interaction
potentials for optically-trapped polar molecules in their electronic
and vibrational ground-state. In particular, we have shown how to
modify the {\em shape} as well as the {\em strength} of the
inter-particle interaction potentials, by manipulating the
rotational dynamics using external DC and AC microwave fields in
combination with dipole-dipole interactions. In Sec.~\ref{sec:secDC}
we have shown that in the presence of a DC field and of a tight
optical confinement it is possible to realize effective 2D
potentials, where particles interact {\em via} a purely repulsive
$\sim 1/r^3$. A potential barrier {\em shields} the attractive
inner-region of the interaction potential, thus providing for the
stability of the collisional setup. In Sec.~\ref{sec:secAC} we have
analyzed the interactions in the presence of an AC field. We have
derived the 3D adiabatic potentials for the molecular interactions,
finding several degeneracies in the two-particle spectrum at
distances of the order of the resonant Condon point $r_{\rm C}$
between the energy of the ground-state plus one photon and states of
the first-excited manifold. The presence of these degeneracies opens
(diabatic, three-body-induced, and, for the case of transverse
confinement, residual tensor-shift-induced) loss channels for the
ground-state collisions, which make the case of interactions in a
pure AC-field less appealing for realizing stable collisional setups
in two-dimensions. In Sec.~\ref{Sec:CouplingDCAC} we show that it is
possible to realize stable 2D interaction setups with considerable
flexibility in potential-designing by combining DC and AC fields, in
the presence of strong transverse confinement. In fact, the DC field
helps to greatly suppress the presence of loss channels at large
distances, while the AC field allows for realizing potentials whose
shape can vary markedly between the long and short distance regimes.

\section{Acknowledgements}

The Authors thank Roman V. Krems and Paul S. Julienne for
stimulating and helpful discussions.\\

This work was supported by the Austrian Science Foundation (FWF),
the European Union projects OLAQUI (FP6-013501-OLAQUI), CONQUEST
(MRTN-CT-2003-505089), the SCALA network (IST-15714) and the
Institute for Quantum Information.

\end{document}